\documentclass[conference]{IEEEtran}

\usepackage{booktabs} 
\usepackage[colorinlistoftodos, textwidth=2.0cm]{todonotes}
\usepackage{verbatim}
\usepackage{subcaption}
\usepackage{pifont}
\usepackage[linesnumbered,ruled,vlined]{algorithm2e}
\SetKwInput{KwInput}{Input}
\usepackage{adjustbox}
\usepackage{amsmath}
\usepackage{hyperref}
\usepackage{tabularx}
\usepackage{multirow}
\SetKwInput{KwInput}{Input}
\usepackage{xurl}

\usepackage[linesnumbered,ruled,vlined]{algorithm2e}

\usepackage{listings}

\newcommand\YAMLcolonstyle{\color{red}\mdseries}
\newcommand\YAMLkeystyle{\color{black}\bfseries}
\newcommand\YAMLvaluestyle{\color{blue}\mdseries}

\makeatletter

\newcommand\language@yaml{yaml}

\expandafter\expandafter\expandafter\lstdefinelanguage
\expandafter{\language@yaml}
{
  keywords={true,false,null,y,n},
  keywordstyle=\color{darkgray}\bfseries,
  basicstyle=\YAMLkeystyle,                                 
  sensitive=false,
  comment=[l]{\#},
  morecomment=[s]{/*}{*/},
  commentstyle=\color{purple}\ttfamily,
  stringstyle=\YAMLvaluestyle\ttfamily,
  moredelim=[l][\color{orange}]{\&},
  moredelim=[l][\color{magenta}]{*},
  moredelim=**[il][\YAMLcolonstyle{:}\YAMLvaluestyle]{:},   
  morestring=[b]',
  morestring=[b]",
  literate =    {---}{{\ProcessThreeDashes}}3
                {>}{{\textcolor{red}\textgreater}}1     
                {|}{{\textcolor{red}\textbar}}1 
                {\ -\ }{{\mdseries\ -\ }}3,
}

\lst@AddToHook{EveryLine}{\ifx\lst@language\language@yaml\YAMLkeystyle\fi}
\makeatother

\newcommand\ProcessThreeDashes{\llap{\color{cyan}\mdseries-{-}-}}

\usepackage{csquotes}
\usepackage[noabbrev,capitalise]{cleveref}
\newcount\colveccount
\newcommand*\colvec[1]{
        \global\colveccount#1
        \begin{pmatrix}
        \colvecnext
}
\def\colvecnext#1{
        #1
        \global\advance\colveccount-1
        \ifnum\colveccount>0
                \\
                \expandafter\colvecnext
        \else
                \end{pmatrix}
        \fi
}

\def\BibTeX{{\rm B\kern-.05em{\sc i\kern-.025em b}\kern-.08em
    T\kern-.1667em\lower.7ex\hbox{E}\kern-.125emX}}
\begin{document}

\title{A Case Study on Optimization of Platooning Coordination}

\author{\IEEEauthorblockN{Veronika Lesch, Marius Hadry, Samuel Kounev}
\IEEEauthorblockA{\textit{University of Würzburg}\\
Würzburg, Germany \\
\{firstname.lastname\}@uni-wuerzburg.de}
\and
\IEEEauthorblockN{Christian Krupitzer}
\IEEEauthorblockA{\textit{University of Hohenheim}\\
Hohenheim, Germany \\
christian.krupitzer@uni-hohenheim.de}
}

\maketitle

\begin{abstract}
In today's world, circumstances, processes, and requirements for software systems are becoming increasingly complex.
In order to operate properly in such dynamic environments, software systems must adapt to these changes, which has led to the research area of Self-Adaptive Systems~(SAS).
Platooning is one example of adaptive systems in Intelligent Transportation Systems, which is the ability of vehicles to travel with close inter-vehicle distances.
This technology leads to an increase in road throughput and safety, which directly addresses the increased infrastructure needs due to increased traffic on the roads.
However, the No-Free-Lunch theorem states that the performance of one platooning coordination strategy is not necessarily transferable to other problems.
Moreover, especially in the field of SAS, the selection of the most appropriate strategy depends on the current situation of the system.
In this paper, we address the problem of self-aware optimization of adaptation planning strategies by designing a framework that includes situation detection, strategy selection, and parameter optimization of the selected strategies.
We apply our approach on the case study platooning coordination and evaluate the performance of the proposed framework.
\end{abstract}


\section{Introduction}
In a world as dynamic as we find it today, where circumstances, processes, and requirements are becoming increasingly complex, the challenges for software systems to be able to work in these dynamic environments are also increasing. 
One of the most critical challenge for these systems is to analyze their environment and to adapt to changes accordingly. 
The Self-Aadaptive System~(SAS)~\cite{Cheng2009,Krupitzer2015} research area attempts to address these challenges. 
The SAS can change their behavior and deal with changes in their environment and the system itself~\cite{lesch2020toward}. 
In our daily lives, we are constantly in contact with SAS that aim to support and improve our way of life without us directly noticing it.
For example, the first electric traffic signals as part of the Intelligent Transportation Systems~(ITS) is one use case of SAS that has led to the development of real-time traffic control in urban areas~\cite{xie2014coping}.
Another promising example for ITS is platooning, which addresses increased infrastructure needs due to increased traffic on roads.
Due to advances in autonomous driving, an increased infrastructure need can be reduced through platooning, which is the ability of vehicles to travel with very close inter-vehicle distances, enabled by communication~\cite{robinson2010operating}.
The use of platooning increases road throughput~\cite{alam2011fuel} and safety~\cite{robinson2010operating}.
Platooning coordination is the process of assigning vehicles to platoons and controlling the platooning activities. 
The platooning coordination problem is a multi-objective problem with several dimensions, since objectives of the drivers, aspects of the platoon, and global traffic need to be considered~\cite{Sturm2020evaluation}.
Platoons are usually coordinated using platooning coordination strategies.
This coordination is an example of SAS in ITS, as these coordination strategies can be considered as adaptation planning strategies that adapt the system, in this case the platoons, to their current state and environment.

In line with the No-Free-Lunch theorem~\cite{Wolpert1997} the proper selection of adaptation planning strategies is a key factor in the success of any SAS, as the performance of one strategy may not necessarily be transferable to other application scenarios. 
In the year 1976, John R. Rice already defined the algorithm selection problem, which involves finding the best performing algorithm for the current problem~\cite{rice1976algorithm}. 
This leads to the idea of a mechanism that automatically selects the most promising algorithm that is also generalizable to be applied in a variety of applications.
The observation from~\cite{fredericks2019planning} that the choice of the strategy for adaptation planning in self-adaptive systems~\cite{Cheng2009,Krupitzer2015} depends on the situation of the system opens a wide area to which such a mechanism can be applied.
Gathered observations can be used to apply different strategies in different situations or to adjust the parameters of a strategy.
Furthermore, the knowledge can be used in combination with previous experiences to learn in which situation which strategy and which parameter configuration work best. 
This idea of combined reasoning and learning can be found in the SeAC research area, whose ideas and approaches will be applied in this work.
There are several approaches to situation detection~\cite{calinescu2020understanding,endsley2017toward,liu2015situation,rockl2007architecture,hardes2019dynamic,porter2016losing,kang2020far}, algorithm selection~\cite{smith2009cross,kerschke2019survey,kerschke2019automated,kotthoff2015improving,bischl2016aslib}, and parameter optimization~\cite{neumuller2012large,feurer2015initializing,zhang2014empirical,chis2013multi,vinctan2015improving} especially in the SAS literature. 
However, there is no integrated approach that combines these ideas into a mechanism that is generalizable and applicable to a variety of use cases.

In this paper, we propose self-aware optimization of adaptation planning strategies and optimization of systems-of-systems, with a particular focus on the field of ITS.
We address the problem of self-aware optimization of adaptation planning strategies by designing a framework that includes situation detection, strategy selection, and parameter optimization of the selected strategies on the case study of platooning coordination.
In addition, the framework applies concepts from SeAC research and is able to learn from previous decisions.

The remainder of this paper is organized as follows:
Section~\ref{relwork} presents and discusses related work.
Afterwards, Section~\ref{sec:framework:assumptions} proposes a self-aware approach on optimization of platooning coordination strategies.
Then, Section~\ref{eval} presents our platooning coordination case study.
Finally, Section~\ref{conclusion} summarizes the paper and outlines future work.

\section{Related Work}
\label{relwork}
A recent study by Calinescu~et~al.~\cite{calinescu2020understanding} has shown that situation-awareness is the main driver for the development of self-adaptive systems and is therefore still an important research topic with many open research challenges.
Endsley~\cite{endsley2017toward} presents a theoretical model of situation-awareness in relation to dynamic human decision making, building on research on naturalistic decision making.
Fredericks~et~al.~\cite{fredericks2019planning} present an approach that uses clustering to determine the current situation. 
They use this information for optimization techniques to discover the optimal configuration for black-box systems. 
Liu~et~al.~\cite{liu2015situation} propose an approach to situation-awareness in autonomous driving that aims to improve the decision-making process in an urban environment. 
Rockl~et~al.~\cite{rockl2007architecture} propose an architecture for driver assistance systems that uses increased environmental information to detect hazardous situations.
Hardes~et~al.~\cite{hardestowards} address communication problems in urban platooning scenarios by using the concept of situation-awareness. 
Porter~et~al.~\cite{porter2016losing} propose a software framework that learns optimal system assemblies in emergent software systems.
Kang~et~al.~\cite{kang2020far} analyze which history length and sensor range provide the best results for long-term situational awareness.

According to Lewis~et~al.~\cite{lewis2017towards}, meta-self-awareness \enquote{leads to the ability to model and reason about changing trade-offs during the system's lifetime}.
Cox~et~al.~\cite{cox2005metacognition} research on meta-cognition, which bridges psychology and computer science.
Agarwal~et~al.~\cite{agarwal2009self} provide an approach that allows computer systems to reason about their own knowledge.
Perrouin~et~al.~\cite{perrouin2012towards} propose a rule-based approach to meta-self-awareness. 
They use layered MAPE-K control loop to optimize adaptation decisions and make an adaptive system \enquote{resilient to a larger number of unexpected situations}~\cite{perrouin2012towards}.
Gerostathopoulos~et~al.~\cite{gerostathopoulos2017strengthening} propose the concept of meta-adaption for cyber-physical systems, which improves the adaptation of a cyber-physical system by generating new self-adaptation strategies at runtime.
Kinneer~et~al.~\cite{Kinneer2018uncertainty} propose the idea of re-using knowledge from previous plans for optimization.
They use a white-box approach with knowledge about the system combined with a genetic algorithm to respond to unexpected adaptation scenarios.

Kate Smith-Miles considers algorithm selection as learning problem~\cite{smith2009cross}.
She reviews the interdisciplinary literature dealing with algorithm selection and presents the developments in this research area.
Kerschke~et~al. provide a survey on automated algorithm selection~\cite{kerschke2019survey}.
The survey covers early and recent work in this area and discusses promising application areas.
Further, it includes an overview on related areas such as algorithm configuration and scheduling.
Pascal Kerschke and Heike Trautmann contribute an approach for automatic model construction for algorithm selection in continuous black-box optimization problems~\cite{kerschke2019automated}.
The goal of this approach is to reduce the required resources of the selected optimization algorithms.
Kotthoff~et~al. apply algorithm selection on the TSP problem~\cite{kotthoff2015improving}. 
They apply two existing TSP solvers and show that they perform complementary in different instances.
The authors design algorithm selectors based on existing TSP features from the literature as well as new features.
Bischl~et~al. propose a benchmark library for algorithm selection~\cite{bischl2016aslib}.
They define a standardized format for representing algorithm selection scenarios. 
Further, they provide a repository containing data sets from the literature to compare proposed approaches.

Neumüller~et~al.~\cite{neumuller2012large} present an implementation of parameter meta-optimization for the heuristic optimization environment \emph{HeuristicLab Hive}.
Their approach minimizes the expert knowledge required  to adapt the parameters of a meta-heuristic.
In their evaluation, Neumüller~et~al. showed that the obtained parameter combinations in some cases deviate strongly from the usual settings.
However, their approach mainly covers single-objective optimization, whereas a multi-objective problem can only be assessed using a normalized and weighted sum of objectives. 
Feurer~et~al.~\cite{feurer2015initializing} improve the Sequential Model-based Bayesian Optimization used for tuning the parameters of machine learning algorithms involving meta-learning.
Using the knowledge from past optimization runs, they showed significant improvement in the Sequential Model-based Bayesian Optimization algorithm.
Zhang~et~al.~\cite{zhang2014empirical} address the problem of release planning, which means the process of deciding which features to integrate into the next version of a software release.
The authors perform a study on various meta- and hyper-heuristics used for multi-objective release planning.
They use different hyper-heuristic algorithms to decide on search operators for meta-heuristics to improve solution quality and compare their performance.
Chis~et~al.~\cite{chis2013multi} use the Framework for Automatic Design Space Exploration to compare the performance of different multi-objective meta-heuristics.
The authors show that all algorithms find similar Pareto front approximations with good solution quality. 
Similarly, Vinctan~et~al.~\cite{vinctan2015improving} deal with design space exploration by implementing a meta-optimization layer for the tool Framework for Automatic Design Space Exploration. 
With this approach, it is possible to introduce a meta-optimization function that can use multiple meta-heuristics simultaneously by switching between them at simulation runtime.
In the evaluation, the authors show that their meta-optimization approach leads to better results than running two different meta-heuristics independently and combining their results.

Another research direction related to this work is the area of Auto-ML.
As the name suggests, automated machine learning focuses on automating machine learning mechanisms by using pipelines in combination with hyperparameter optimization to reduce manual effort.
Reinbo, for example, is an Auto-ML framework that uses task pipelines and implements reinforcement learning and Bayesian optimization to automatically determine the parameters~\cite{sun2019reinbo}.
A similar approach is used by Chai~et~al. who propose an Auto-ML framework that covers the common problem of data drift in machine learning~\cite{chai2019auto}. 
Thornton~et~al. propose a mechanism for hyper-parameters selection and optimization in the context of classification algorithms~\cite{thornton2013auto}.
Finally, Li~et~al. attempt to solve the problem of tuning hyper-parameters using a random search mechanism combined with adaptive resource allocation and early-stopping~\cite{li2017hyperband}.

This work delineates from the presented related work as follows:
All mentioned approaches already cover parts of our proposed framework, such as a rule-based meta-self-aware approach, situation-awareness, determining the optimal configuration of a system, or performance comparison of optimization techniques.
However, there is no other work that integrates all these aspects into one framework. 
The combination of a multi-layered framework with the LRA-M control loop and the integration of adaptation planning strategies, situation-awareness, strategy selection, learning approaches, and optimization techniques make the proposed approach unique and a valuable contribution to the research community.

\section{Self-Aware Optimization of Platooning Coordination}
\subsection{Assumptions}
\label{sec:framework:assumptions}
In this section, assumptions are made for the design of the framework to ensure broad applicability in various use cases.
The following assumptions ensure the proper operation of the framework as well as the use case and define the interactions between both systems.
At the same time, they point out limitations that can be addressed in future work.

First, we assume that the use case for which the framework is to be used consists of two parts.
One part is the environment in which entities operate based on their individual goals and actions.
The second part is an adaptation planning system that monitors the entities and decides upon adaptation actions based on global goals.
We assume that the operating entities adhere to the given plan of the adaptation planning system and execute all given adaptations.
If they cannot implement these instructions, they report this to the adaptation executor, who then decides on further actions that should be taken by the entities.
Further, we assume that the communication between entities in the use case and adaptation planning system is flawless and that the entities regularly report measurement and observation values to the adaptation planning system.
Additionally, we assume that the applied strategy of the adaptation planning system is interchangeable and has the possibility to change its parameters at runtime.

Second, we assume that the use case to which the framework is to be applied is digitized, meaning that performance and monitoring data are captured and stored digitally---typically centrally in the adaptation planning system.
Further, the adaptation planning system is able to transmit relevant data to a defined management entity---in this work the framework---where higher level optimizations take place.
We assume that the interaction between framework and adaptation planning system of the use case is always successful. 
Therefore, we exclude any case where the connection between the two systems fails or the computed changes cannot be transmitted to the adaptation planning system due to other failures - resilience management of both systems is part of the future work of this paper.

Third, we assume that the adaptation planning system works independently of a higher-level optimization, i.e., the framework, and can be used with a previously defined strategy algorithm and parameter set.
Thus, it remains functional regardless of whether the framework determines an optimization adjustment.
This is especially important in the startup phase of the framework, when optimization adjustments have not yet been determined.
We also assume that this adaptation planning algorithm works independently and flawlessly and does not need to be monitored for failures.

Finally, we assume that the framework provides optimized decisions to the adaptation planning system without explicit request.
Furthermore, we assume that the adaptation planning system regularly retrieves and successfully implements these changes.
We assume that the adaptation planning system reports its current configuration along with other monitoring data to the framework to execute the optimizations based on the current state at given time intervals.
Based on this data, future work can extend the framework to include a mechanism to ensure successful implementation of new strategies and parameter settings in the adaptation planning system.

\subsection{Terminology}
\label{sec:framework:terminology}
In this section, we define the terminology used in the following to avoid misunderstandings and imprecise expressions. 
We start with the definition of the use case as well as the entities and their actions in the use case, continue with the definition of an observation, a context, and a situation, and finally define the term framework.

\textbf{Use Case:}
We define a use case as a group of entities operating in a particular environment, pursuing their own goals.
Entities in the use case can be linked to an adaptation planning system that helps them achieve their goals more efficiently, or that adapts the entity's actions to achieve global, regional, or local goals.
The complexity and abstraction level of a particular use case are irrelevant as long as a adaptation planning system is in place.
This adaptation planning system must provide multiple adaptation planning strategies and can provide configuration options.
With respect to the running example platooning coordination, the use case could be defined at the regional level, e.g., as coordination of platooning of vehicles on a road segment with a central Platooning-Coordination-System fulfilling the role of the adaptation planning system.
The use case could also be defined at a lower level, such as optimizing the inner platoon structure, i.e., the order of vehicles within the platoon (cf. \cite{lesch2020comparison}).

\textbf{Entity in a Use Case:}
An entity within a use case is a machine, human, or other object that can receive and execute instructions from an adaptation planning system.
Entities may have the ability to make decisions for themselves according to their individual goals and do not necessarily need to receive instructions from the adaptation planning system.
Entities may also work with coarse-grained instructions or work toward individual goals.
The entities within the use case are expected to follow the adaptation actions they receive from an adaptation planning system, even if their individual goals dictate a different direction.
For a discussion of research challenges related to coordinating global, regional, and local goals, we refer the interested reader to our publication~\cite{LeKrTo2019}.

\textbf{Actions of an Entity:} 
Entities of a use case have a given set of possible actions that they can execute to accomplish certain tasks or achieve their individual goals.
The actions to be taken can either be specified in fine-grained terms by an adaptation planning system, or entities can work autonomously toward a coarser-grained goal.
The second case also means that entities can operate without an adaptation planning system if the entities' goals are defined and the available actions enable the entities to achieve that goal.

\textbf{Observation:} 
An observation contains information about the use case at a particular point in time.
This includes details about the entities, their sensed data from the environment as well as the current configuration of the adaptation planning system and its performance. 
These performance indicators must be defined individually for each use case.
Using expert knowledge, the performance of the adaptation planning system can be evaluated based on these indicators.
We define each observation as a triple $(\textit{context}, \textit{input}, \textit{metrics})$ at a given point in time that is sent to the system on a regular basis. 
The \textit{context} represents a set of values used to determine the current situation of the use case.
The \textit{input} parameters are the configuration parameters of the adaptation planning system.
The \textit{metrics} are a set of indicators that represent the current performance of the adaptation planning system.

\textbf{Adaptation Planning System:}
An adaptation planning system is a mechanism that uses the retrieved observations from a use case in order to plan adaptations within a Self-adaptive System.
These systems aim to identify changes in the environment and the system itself and to react accordingly and apply adaptation planning strategies to plan adaptations.
The strategies are exchangeable within a Self-adaptive System and require the configuration of parameters.
These properties can be used to tune the performance of Self-adaptive System by an optimized selection of adaptation planning strategies and parameter tuning.

\textbf{Situation:}
We define a situation as a set of observed contexts that have similar values.
This means that environmental factors, entities, and entity behavior occur in a similar combination to previous context data.
We use the term situation from a technical point of view, following Cámara's definition, \enquote{where a situation includes at least the elements of the situation [...], and environmental factors and their current states}~\cite[p.~38]{camara2017self}.
We use the context data to determine the situation the system is in.
The knowledge of the current situation is then used to adjust the adaptation planning strategy and its parameters to optimize the overall performance of the system.
We also use this information to learn good strategies for each situation and improve system performance for similar situations in the future.

\textbf{Framework:} 
We consider a framework as an abstract modular application that defines a specific process structure, pursues a specific goal, and provides generic functionality by combining components.
We assume that these components have well-defined interfaces through which they can communicate with other components to ensure smooth integration into the overall framework.
As part of this work, we have implemented all relevant components of our framework. 
In addition, we have designed the framework to provide the ability to extend it by adding new components or customizing existing components. 
Furthermore, the user of the framework has to define a configuration for the specific use case, which defines the composition, setup and configuration of the framework and the components used.

\subsection{System Model}
\label{sec:framework:systemmodel}
\begin{figure*}[htb]
    \centering
    \includegraphics[width=0.85\textwidth]{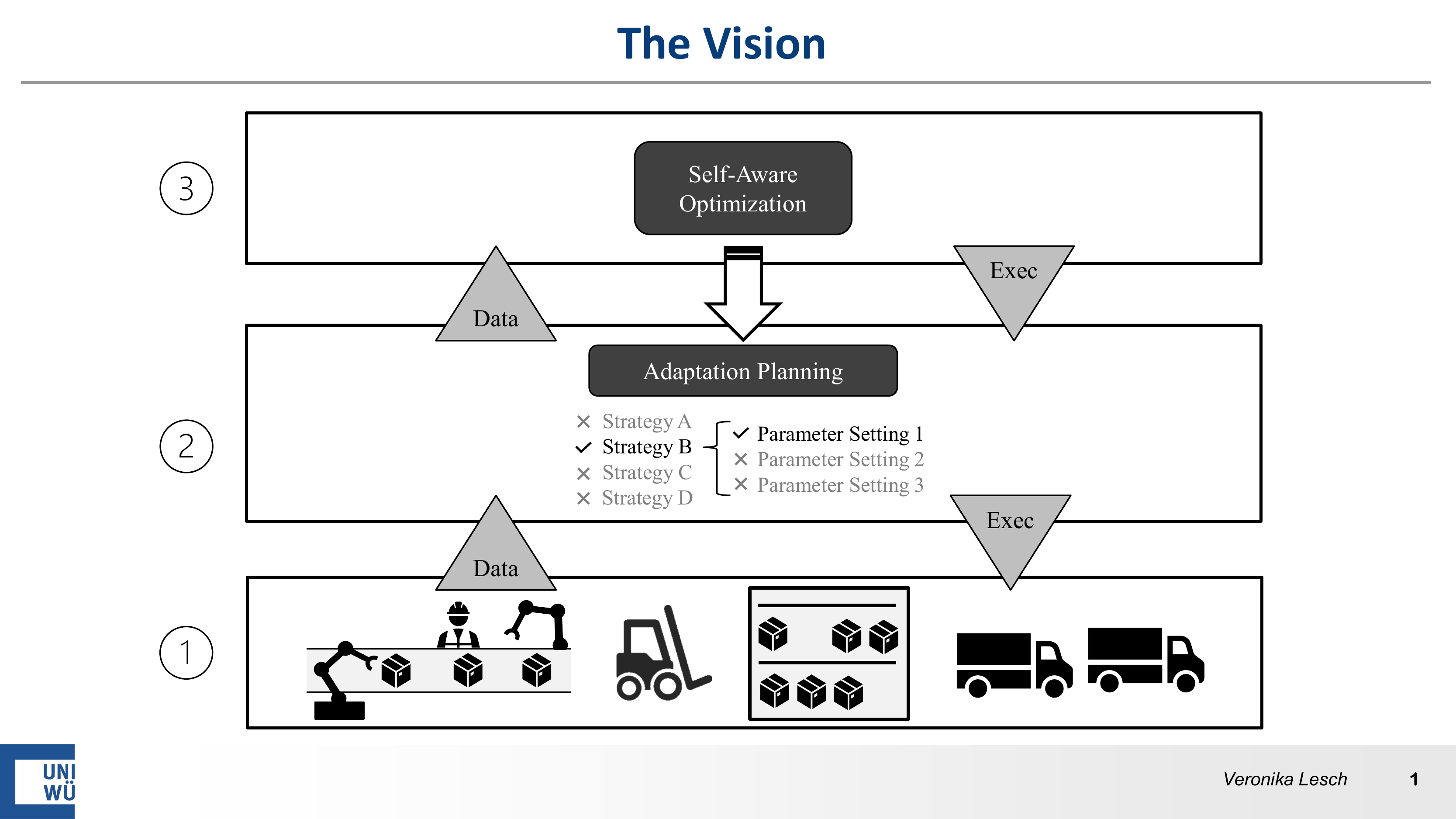}    
    \caption[Multi-layer architecture of the self-aware optimization framework.]{Multi-layer architecture of the self-aware optimization framework. Layer 1 represents an adaptive system, the adaptation planning system is shown in Layer 2, and Layer 3 shows the self-aware optimization.}
    \label{fig:framework:systemmodel}
\end{figure*}

This section introduces the system model we use for defining the self-aware optimization framework.
The system model is presented in \cref{fig:framework:systemmodel} and integrates three layers following the three layer architecture proposed by Kramer and Magee~\cite{kramer2007self} to incorporate the principles of maintainability and separation of concerns: (i)~Application, (ii)~Adaptation Planning, and (iii)~Self-Aware Optimization.
In the following, we explain the details of each layer: the self-aware optimization framework.

We refer to the bottom layer~\ding{192} of the system model as the application layer and consider real-world use cases from Intelligent Transportation Systems and logistics.
Entities of the use case monitor themselves and their environment.
The collected data is sent to the upper layer where the adaptation planning system receives the data.
After an adaptation planning cycle, the use case entities can receive adaptation actions to follow and execute. 
If the entities fail to carry out these instructions, we assume that they will report this to the adaptation planning system, which will decide on further action.

The middle layer~\ding{193}, called adaptation planning, includes the adaptation planning system, which receives observations from the use case.
The adaptation planning system applies a strategy that uses the received observations to plan adaptations for the managed system.
These strategies are selected from various existing strategies in the adaptation planning system.
The adaptation planning strongly depends on the use case and is therefore out of scope of this work.
The strategies can range from simple rule-based algorithms to complex (multi-objective) optimization algorithms.
Furthermore, we assume that the user of the framework will provide multiple strategies per use case, customized for the particular use case, to provide the possibility of strategy exchange when needed.
The performance data of the selected strategy is collected and---together with the use case's monitoring data---transferred to the next layer, which performs a self-aware optimization.
After one self-aware optimization cycle, the adaptation planning layer may receive instructions to  change the strategy parametrization or even to replace the strategy.
As mentioned earlier, we assume that the adaptation planning layer executes these commands without interference.

Finally, the third layer~\ding{194} is called self-aware optimization.
This layer is responsible for optimizing the parameters of the selected strategy in the adaptation planning layer as well as for the selection of strategies for the~\ding{193} layer and, therefore, integrates several components: (i)~situation detection, (ii)~algorithm selection, and (iii)~parameter optimization.
The situation detection component receives the monitoring data, that is, the use case observations, and the performance data from the adaptation planning system and categorizes the observation into a currently present situation.
The algorithm selection component uses the information about the current situation, combines it with experience from similar situations in the past and selects the most appropriate adaptation planning strategy.
Finally, the parameter optimization component also receives monitoring data and tunes the parameters of the adaptation planning strategy.
All decisions---including the situation, the selected strategy, and the parameter settings---are used in combination with monitoring and performance data to learn from previous decisions.
A knowledge base manages the set of known situations as well as corresponding decisions and continuously learns which parameter and algorithm combination fits best for the situations already experienced.
In addition, it is possible to develop another component that includes prediction and forecasting mechanisms to enable proactive adaptation of the system.
Finally, the third layer passes the decisions to the adaptation planning layer~\ding{193}, which executes them.

\subsection{LRA-M Loop Adoption}
\label{sec:framework:lram}
\begin{figure}[htb]
    \centering
    \includegraphics[width=\columnwidth]{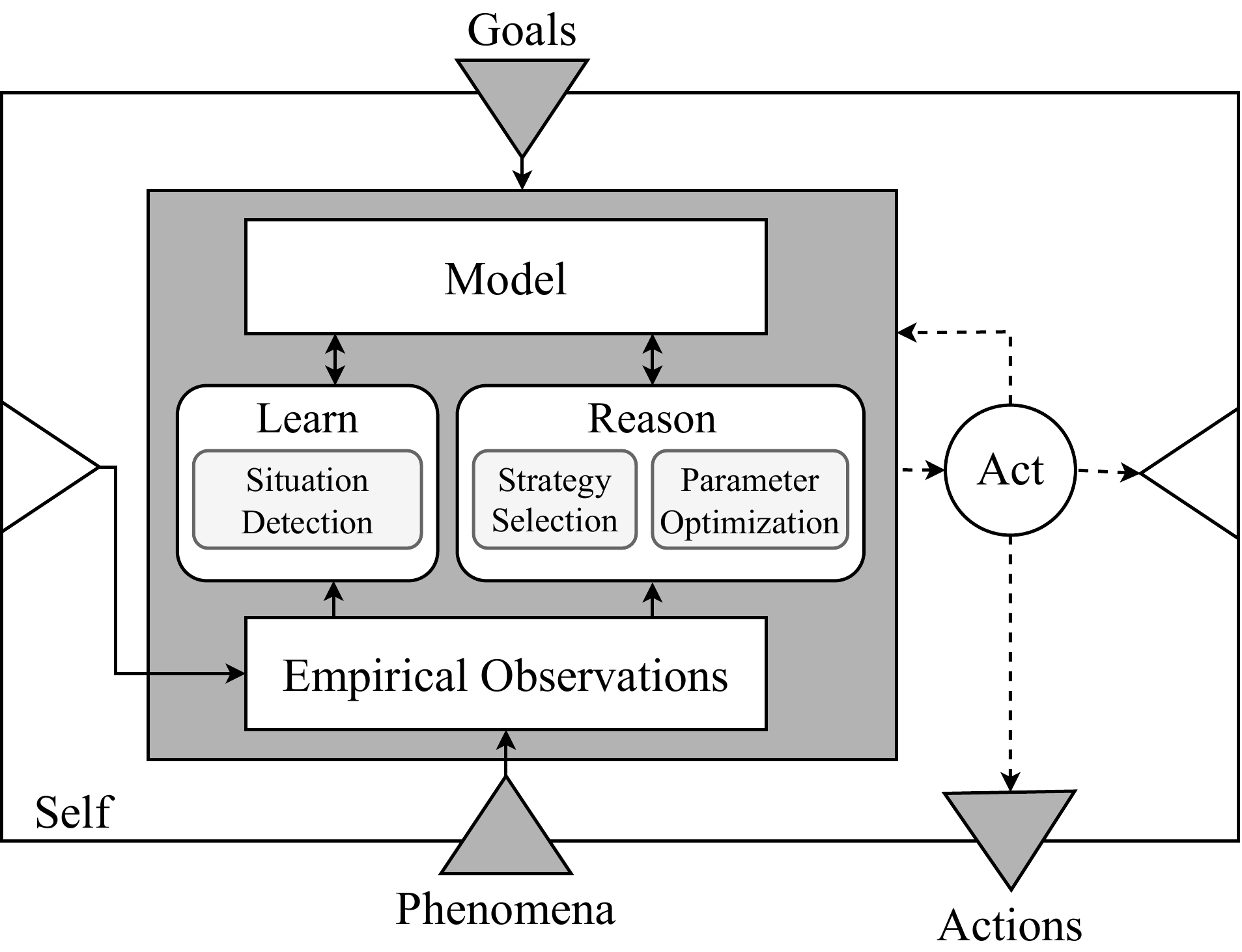}
    \caption[Modified LRA-M control loop.]{Modified LRA-M control loop based on Kounev~\textit{et~al.}~2017. The basic LRA-M control loop is extended to include analysis and the meta-optimization in the Learn module and planning through optimization in the Reason module.}
    \label{fig:framework:lram}
\end{figure}

In this section, we present our concept of a self-aware optimization framework from the control loop point of view. 
This perspective allows us to elaborate on the idea of the framework and explain the interplay between ongoing learning and reasoning in the framework.
Since we use the terminology of  self-awareness in this work, we focus this section on the corresponding LRA-M control loop.
The LRA-M control loop was first introduced by Kounev~\textit{et~al.} in 2017~\cite{kounev2017notion} in his work on Self-aware Computing Systems.
This loop is quite similar to other concepts like the MAPE-K control loop~\cite{Kephart2003} or the Observer/Controller concept~\cite{tomforde2011observation} and most of these concepts can be transformed into each other~\cite{LeKrTo2019}.
However, the LRA-M control loop explicitly includes a \emph{Learn} and a \emph{Reason} component.
Learning allows the system to learn models of the system itself and the environment, while reasoning uses these models to trigger adaptation actions that modify the system and affect the environment. 
These components are essential parts of the framework because learning enables the framework to form models of the environment, i.e., the two lower levels of the system model (c.f.~\cref{sec:framework:systemmodel}) and to recognize new situations.
Reasoning gives the framework the ability to consider which adaptation actions might be beneficial in a given situation based on the knowledge of recent decisions or decisions in similar situations. 
This combination of ongoing learning and model-based reasoning forms the basis for the proposed framework, which is why we chose to use the LRA-M control loop.
Since the LRA-M control loop is a general-purpose concept applicable to diverse systems, we modify the control loop to explicitly include the functionalities of our proposed framework, as shown in \cref{fig:framework:lram}.

The loop displays the system, also called the self, and its interfaces with the environment.
It interacts with the environment by (i)~perceiving \emph{Phenomena} and storing them as \emph{Empirical Observations}, (ii)~receiving \emph{Goals} to be achieved, and (iii)~executing \emph{Actions} based on the decisions made.
The Empirical Observations are captured in the use case, i.e., the application layer of the system model, and used in the \emph{Learn} and \emph{Reason} modules.
Furthermore, the decisions of the adaptation planning layer are part of the captured phenomena since they are needed as additional sources of information for the third layer.
In the ongoing learning process, the observations are abstracted into models that contain knowledge about the environment and the system itself.
We add the \emph{Situation Detection} component into the Learn module, which enables to interpret the observations and updates the models to persist all gathered information.
So far, we use clustering algorithms in the Situation Detection component to determine the current situation.
However we have built each component in a modular fashion so that it is easy to extend the techniques used.
Further, the learning component receives performance data of the managed use case with periodic observations and learns the impacts of the actions taken based on the current situation.
This enables the system to continuously improve its reasoning and acting, and to keep the system's models of itself and the environment up-to-date.
These models serve as the basis for the reasoning process that determines actions to be taken in response to a changing environment.
The reason module determines actions for the adaptation planning system to adapt to changes in the environment or to deteriorated performance values. 
Hence, we assign the two components (i)~\emph{Strategy Selection}, and (ii)~\emph{Parameter Optimization} to this module.
The Strategy Selection component combines the information from Situation Detection, the current use case performance with the learned models about the use case and determines whether to keep the current strategy or switch to another existing strategy.
The Parameter Optimization component applies optimization techniques to tune the parameters for the selected strategy.
So far, we use known, well-performing parameter settings as initial values for the optimization process to achieve a faster convergence of the optimization.
The Situation Detection, Strategy Selection, and Parameter Optimizations in the modified LRA-M control loop are newly introduced components and not part of the original definition of the LRA-M control loop.
These three components build the main contribution in terms of the proposed framework and are meant to be generically applicable to a wide range of suitable use cases.

\subsection{Framework Composition}
\label{sec:framework:architecture}
\begin{figure}[htb]
    \begin{center}
        \includegraphics[width=0.99\columnwidth]{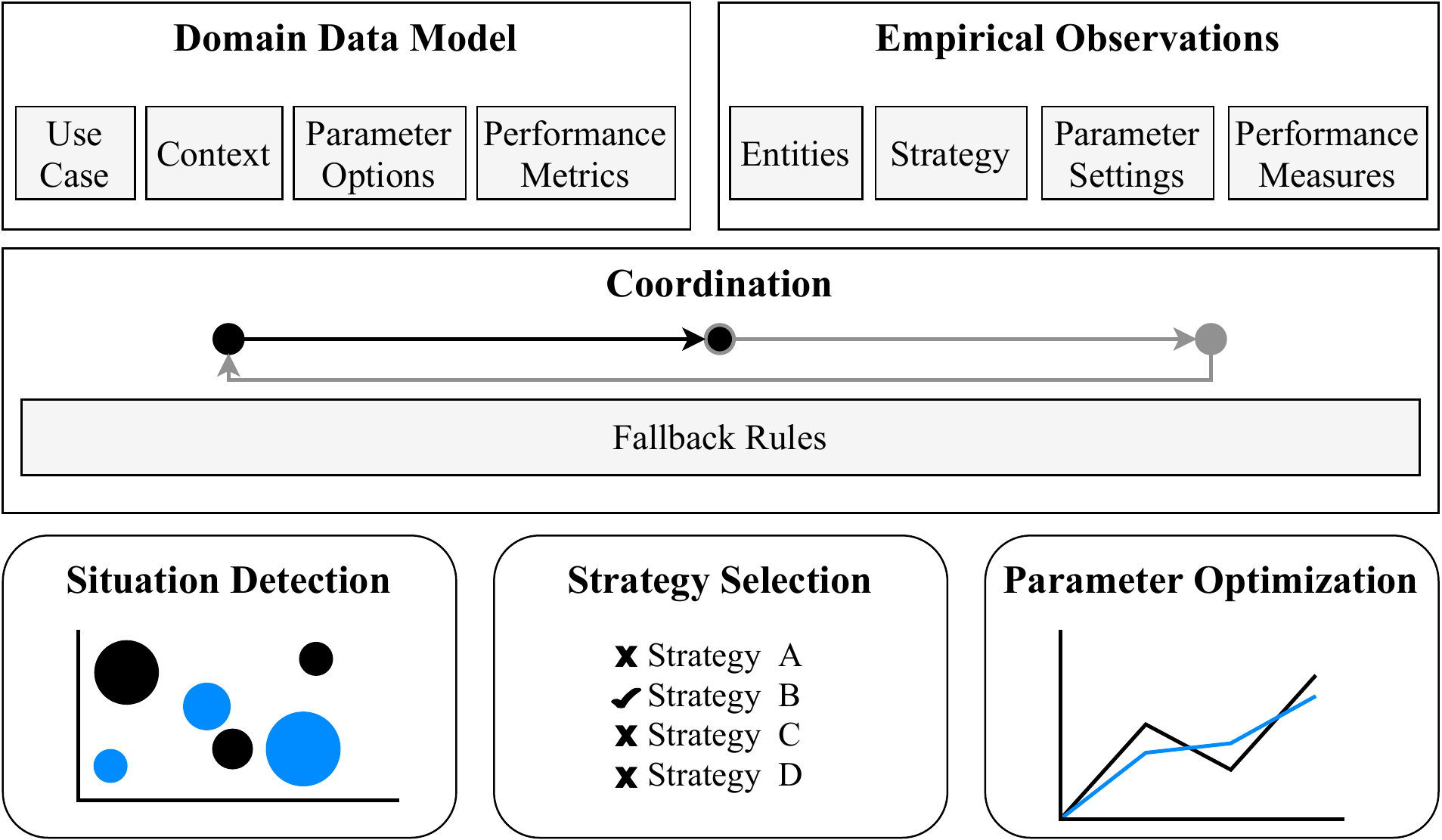}
        \caption[Composition of the self-aware optimization framework.]{Composition of the self-aware optimization framework. The framework contains the Domain-Data-Model for configuration, the Empirical Observations as a repository, a Coordination component that manages the workflow, and the three main components Situation Detection, Strategy Selection, and Parameter Optimization.}
        \label{fig:framework:architecture}
    \end{center}
\end{figure}

This section presents the composition of the generically applicable self-aware optimization framework.
The framework consists of several components that configure the framework, store its observations, and execute the desired functionality, that is, to determine which strategy algorithm and parameters to use in the adaptation planning system.
\cref{fig:framework:architecture} provides a comprehensive overview of the framework's structure. 
In the following, we briefly introduce each component and state its main contribution to the framework.
All details of the components can be found in the following sections.

First of all, the user of the framework can use the \emph{Domain-Data-Model} to configure the entire framework and all its components. 
The Domain-Data-Model is the only part of the framework that the user needs to configure with use case specific information.
Therefore, the framework considers the two lower layers from \cref{fig:framework:systemmodel} as a black box, of which it only knows the information defined in the Domain-Data-Model.
In the Domain-Data-Model, relevant information about the use case such as the name and existing strategies in the adaptation planning system are defined. 
The context part of the Domain-Data-Model defines what sensor data the adaptation planning system sends to the framework with respect to the context of the system.
With regards to the platooning coordination use case, this sensor data could be the number of cars and trucks on the road, the platooning percentage, or the average speed of the vehicles. 
The parameter options of the Domain-Data-Model specify which configuration parameters exist for the adaptation planning strategy of the second layer and which values they can accept. 
Finally, the Domain-Data-Model contains a definition of the performance metrics used to assess the performance of the use case. 

The second component of the framework is responsible for managing all sensor data received from the use case and is called \emph{Empirical Observations}.
This component processes incoming data from the use case and provides an interface for the other components to retrieve the relevant data for the according computation step of the framework.
For example, it maintains information about the entities of the framework such as the number of vehicles, the platooning percentage, and the vehicle speed. 
It also obtains information about the currently executed adaptation planning strategy and its parameter settings in combination with performance metrics. 
This enables the framework to reflect on previous adaptation decisions and learn which combination of strategy and parameter settings works best in a given situation.

The central component of the framework is the \emph{Coordination}, which is responsible for retrieving the required data from the observation storage and passing them to the next component whose execution it triggers. 
This component is constantly active and regularly invokes the other components of the framework, namely the \emph{Situation Detection}, the \emph{Strategy Selection}, and the \emph{Parameter Optimization}, in this predefined order.
In the event that one of the other components fails, the coordination component can fall back to user-defined fallback rules from the Domain-Data-Model to remain functional. 
These fallback rules can be simple if-then-else rules, but since we provide the possibility to load arbitrary Python code into the fallback rules, the user could also extend the framework with a more sophisticated fallback mechanism.

The \emph{Situation Detection} component of the framework receives the observation data of the use case, such as the entities and their current state, and determines the situation the use case is currently in. 
So far, we only use clustering algorithms for this purpose.
However, it is easy to extend the component with other approaches, as we have designed the framework to be modular and the approach used is configured in the Domain-Data-Model.
The identified situation is then returned to the Coordination component, which forwards this information to the Empirical Observations component.

After the Situation Detection completes its computation, the Coordination invokes the \emph{Strategy Selection} component.
This component combines knowledge about the current situation with experience from previous decisions in similar situations and determines which adaptation planning strategy is most appropriate for this situation.
This decision is returned to the Coordination component that triggers the next component.

The last component of the framework is the \emph{Parameter Optimization} component. 
This component receives the current parameter settings as starting point, historical data for the current situation, the corresponding adaptation planning algorithm, and performance measures.
It then performs an optimization process to tune the parameter setting for this adaptation planning strategy to the current situation.
It then returns the settings to the Coordination component, which stores all the collected information of this round of execution from the components, updates the system models, and sends adaptation actions to the adaptation planning system in layer two.

\begin{figure}[htb]
    \begin{center}
        \includegraphics[width=0.99\columnwidth]{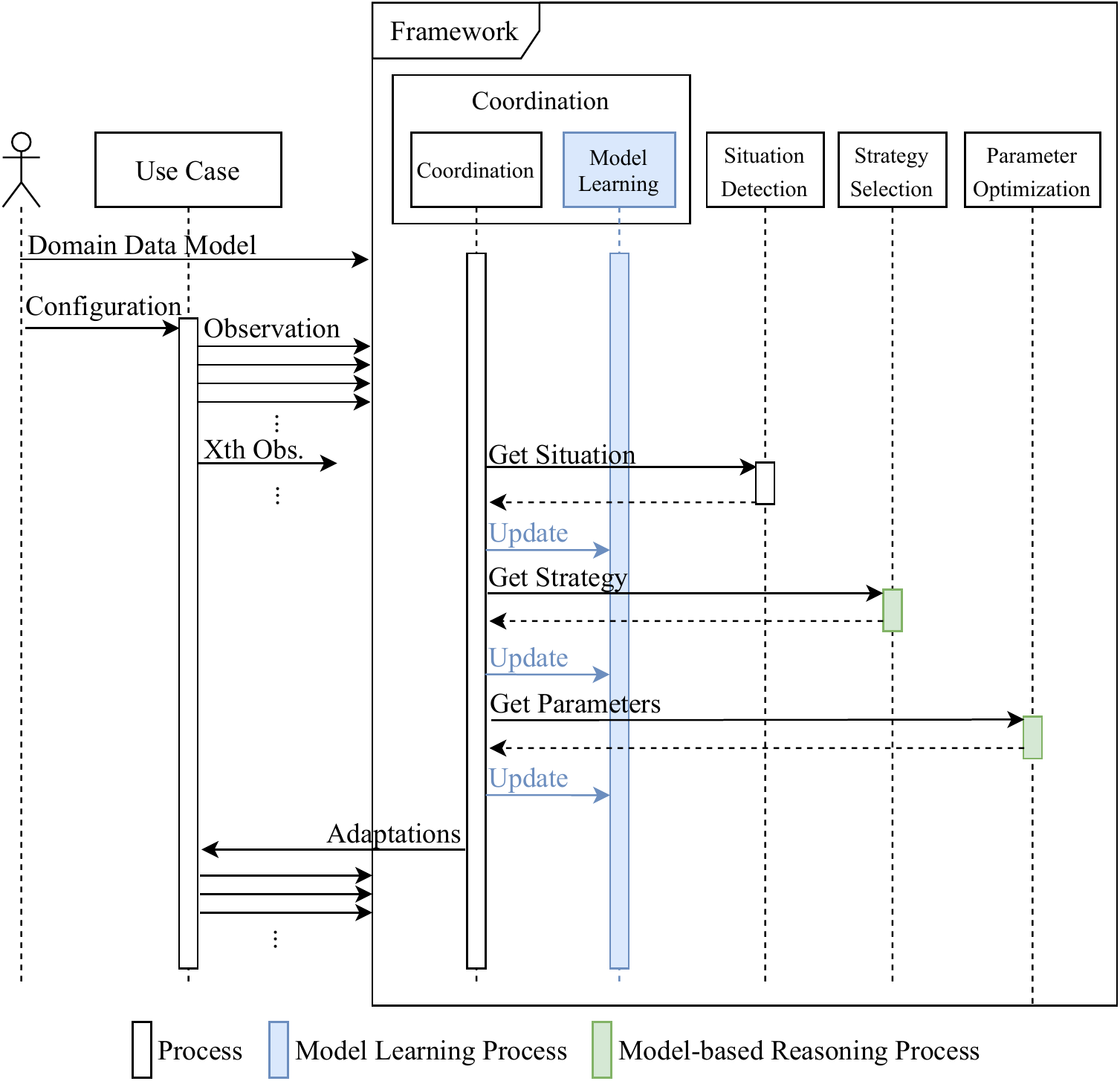}
        \caption[Sequence diagram of the self-aware optimization framework.]{Sequence diagram of the workflow of the self-aware optimization framework. The user configures the framework and the use case sends observations. The framework processes the observations, identifies the current situation, selects the strategy and parameter setting, and continuously learns and updates its models.}
        \label{fig:framework:seqdiagram}
    \end{center}
\end{figure}

In addition to the general composition of the framework, we illustrate the workflow of the framework as a sequence diagram in \cref{fig:framework:seqdiagram}.
The user is shown on the left side of the sequence diagram.
He configures and starts the framework using the Domain-Data-Model, sets up the use case and configures it. 
The use case then starts its execution and sends the defined observations to the framework in regular intervals, regardless of the current computational state of the framework.
The Coordination component of the framework processes the incoming observations and forwards them to the Empirical Observations.
After a certain number of received observations, the Controller component triggers the first execution of the Situation Detection component and forwards relevant observation data to this component. 
In the meantime, the Coordination component receives further observations from the use case, which are stored but not used until the next round of execution.
After the situation is detected, this component returns the situation ID to the Coordination, which updates the system model of the environment.
Then, the Coordination component triggers the Strategy Selection component with filtered observation data containing only observations of the identified situation.
This component applies model-based reasoning based on this data to determine the most promising adaptation planning strategy.
Again, this decision is fed back to the Coordination component which again updates the system model.
Finally, the observed data is filtered again to include only data for the current situation and the adaptation planning strategy determined by the Strategy Selection. 
With this data, the Coordination triggers the model-based reasoning of the Parameter Optimization, which performs an optimization process to find the best parameter setting for the current situation and the selected strategy. 
After the Coordination component obtains this parameter setting, it updates the system model and sends adaptation tasks to the adaptation planning system, which executes them.
This step completes one round of execution in the framework and after a predefined waiting time, the Coordination starts the next round.

\subsubsection{Coordination}
\label{sec:framework:coordination}
This section provides a more technical view of the Coordination component depicted in~\cref{fig:framework:architecture} and extends the descriptions of the previous sections. 
We further summarize the workflow of the Coordination component using Pseudocode in \cref{algo:framework:coordination}.
The Coordination is responsible for initializing and invoking all other components of the framework.
It also processes incoming observations and updates the system models based on observations and the framework's adaptation decisions.
It is triggered at the start of the framework and instantiates all components of the framework~(lines~1-2).
To do so, the Coordination receives the Domain-Data-Model specified by the user, in which he defines the configuration of all components.
It parses the Domain-Data-Model and instantiates the other components. 
After that, the framework is fully set up and ready to start its execution. 
\begin{algorithm}
	\caption{Pseudocode workflow of the Coordination component.}
	\label{algo:framework:coordination}
    \SetKwInOut{Input}{Input}
    \SetKwInOut{Data}{Data}
    \Input{Domain-Data-Model, new observation, existing observations}
    \BlankLine
    \If{start of framework}{
        initialize components defined in the Domain-Data-Model\;
    }
    derive additional information from the observation\;
    save new observation\;
    situation $\leftarrow$ invoke Situation Detection on all observations\;
    \eIf{situation could not be determined}
        {adaptations $\leftarrow$ apply fallback rules to all observations\;
        update system model with current adaptation decision\;
        send adaptations\;
        }
        {update system model with current situation\;
        \If{waiting time after previous adaptation action is over}{
            \eIf{same situation as before AND number of optimization attempts not met}{
                    parameter setting $\leftarrow$ invoke Parameter Optimization on observations of current situation and strategy\;
                }{
                    strategy $\leftarrow$ invoke Strategy Selection on observations of current situation\;
                    parameter setting $\leftarrow$ invoke Parameter Optimization on observations of current situation and strategy\;
                }
            update system model with current adaptation decision\;
            send adaptation decision to use case\;
            }
        }
\end{algorithm}

The use case that the framework is intended to optimize is responsible for sending observations on a regular basis. 
Each observation consists of the use case entities, the currently active adaptation planning strategy, its parameter settings, and the use case performance metrics.
Each new observation received triggers a new round of execution in the Coordination component. 
As a first step, the component uses the received data to compute additional important information relevant to subsequent processing~(line~3): the time that the currently active parameter setting was active and the Hypervolume of the use case performance metrics.
We require a user-defined waiting time in the Domain-Data-Model to allow adjustments to take effect.
Thus, the framework calculates the time that the current configuration is active in the use case and waits a predefined amount of time before evaluating the performance of the latest adaptation decisions to reduce unstable effects after recent changes.
This also prevents too many adaptation actions from being sent to the use case without enough time for implementation.
The Hypervolume measure is a widely used quality indicator for multi-objective optimization, especially in evolutionary optimization~(c.f. \cite{Wang2016}).
We use the Hypervolume to reduce the observed performance indicators of the use case to a single performance value.
This allows us to use any single-objective optimization technique in the Parameter Optimization component without requiring of multi-objectiveness for this technique.
Afterwards, the component forwards the observation combined with the derived information to the Empirical Observations component (c.f. \cref{fig:framework:architecture}) that stores the incoming data~(line~4).

Then, the Coordination passes the new observation to the Situation Detection component~(line~5).
Since the Situation Detection component applies clustering algorithms for identifying the current situation, it needs all the observation data collected from the use case for each execution. 
Therefore, we decided to implement an additional internal data management for the Situation Detection component to reduce the communication and data transfer between the components. 
After the Situation Detection identified the current situation, it returns the situation to the Coordination. 
If the available observation data is not sufficient for the clustering algorithm or the current situation is clustered as noise, the Situation Detection does not return a situation.

The Coordination component checks whether the situation detection was successful and returned a situation~(line~6). 
If the situation detection did not return a situation due to insufficient data or classification as noise, the Coordination component applies the fallback rules to the current observations~(line~7).
Then, the Coordination updates the system model with the most recent adaptation decision and sends the adaptations to the use case~(lines~8-9). 
In case the Situation Detection returned a valid situation~(line~10), the Coordination adds information about the current situation to the system model. 
Since we apply a clustering algorithm in the Situation Detection that always clusters all observation data, it could restructure the whole data and find different clusters compared to the clustering of previous executions. 
In this case, the Coordination updates the system model and reclassifies the already clustered observation data to match the latest clustering~(line~11).

After successfully updating the system model with respect to the current situation, the Coordination checks whether the waiting time after a previous adaptation action has expired~(line~12). 
This waiting time is defined by the user in the Domain-Data-Model and serves as a cool-down period for use case adaptations to take effect. 
By doing this, we ensure that the transient phase of the use case is waited for and performance measures are retrieved that evaluate only the most recent adaptations.
If the waiting time is still active, the current round of execution has ended and the Coordination waits for the next observations of the use case.
When the waiting time has expired, new adaptation decisions can be send to the use case.
In the next step, the Coordination requires another user-defined parameter from the Domain-Data-Model: the number of optimization attempts for the Parameter Optimization.
This parameter specifies how many optimization cycles are performed per situation before a change in strategy is considered. 
This definition of optimization attempts per situation provides sufficient time to tune the parameters and avoids a hasty change of the selected strategy.
The Coordination first checks if the current situation is the same as in the previous execution.
Then, based on the user-defined parameter, it checks whether the necessary number of optimization attempts for this situation has already been executed~(line~13). 
If this is the case, the Coordination requests all observations of the current situation and strategy combination and passes them to the Parameter Optimization.
The Parameter Optimization computes a new set of parameters and returns it to the Coordination~(line~14).
However, if the number of optimization attempts has been exceeded this indicates poor performance of the currently used strategy which the framework uses to search for a new, better fitting strategy.
In this case, or whenever the situation changed~(line~15), the Coordination requests all observations of the current situation and passes them to the Strategy Selection component~(line~16).
This component uses this information to reason about the most promising strategy for adaptation planning.
After the computation, this component returns the selected strategy to the Coordination. 
Then, the Coordination requests all observations of the current situation and the selected strategy to pass them to the Parameter Optimization~(line~17). 
Using this information, this component performs an optimization task to select the most promising parameter settings for this strategy and returns the results to the Coordination.
The Coordination, in turn, uses the strategy decision and its parameterization to update the system model~(line~18).
Finally, it sends the adaptation decisions including the strategy and the parameter setting to the use case~(line~19).

\begin{figure}[htb]
    \begin{center}
        \includegraphics[width=0.99\columnwidth]{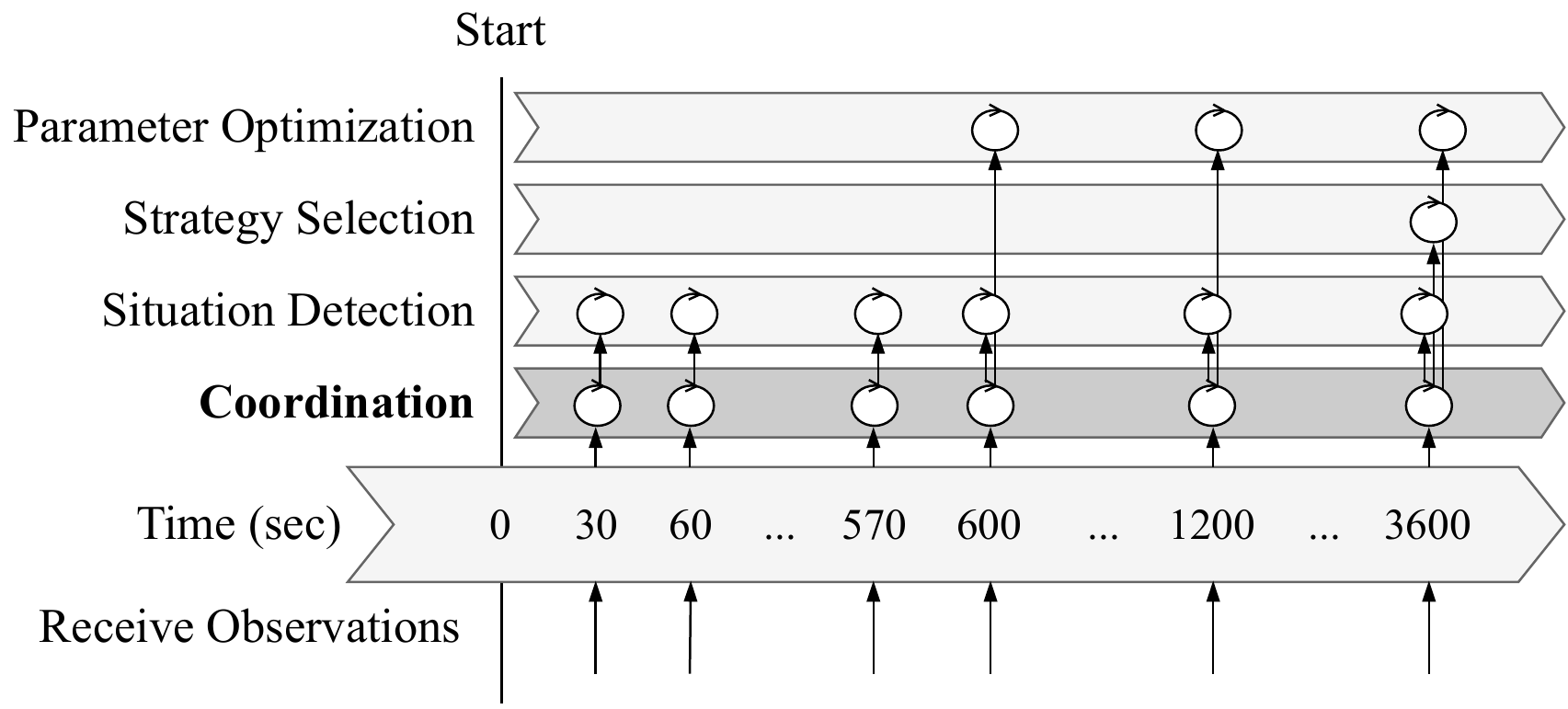}
        \caption[Timescale of the components the Coordination invokes.]{Timescale of the components and their computations the Coordination invokes. Illustrated is a time scale of 3600 seconds where observations arrive every 30 seconds. Each observation triggers an execution of the Coordination which then decides which other components to invoke.}
        \label{fig:framework:timescale}
    \end{center}
\end{figure}

To better understand the timing within the framework, we present an example timescale for invoking the three components Situation Detection, Strategy Selection, and Parameter Optimization in \cref{fig:framework:timescale}.
All timing values can be defined by the user with respect to the use case.
Therefore, the timing presented here should only be considered as an example for demonstration and not as the fixed timing of the framework for all use cases.
For simplicity, we assume that no situation changes occur in this example.
The figure shows time in seconds along the x-axis as a time scale, arranges the components above the time scale, and received observations are shown as arrows pointing to a specific time on the time scale. 
The use case in this example is configured to send current observations at a regular interval of 30~seconds.
Each incoming observation triggers the Coordination that decides which other components are required at that time. 
At the beginning of the framework execution, the Coordination stores the received observations and forwards them to the Situation Detection.
However, since there is not enough data at the beginning of the execution, the Situation Detection does not provide a situation and the Coordination applies the fallback rules.
Once there is enough data (at second 600), the Coordination component triggers the Situation Detection that returns a specific situation ID.
The situation identification then triggers the Parameter Optimization for the first time. 
Strategy Selection is omitted at this point because we decided to first optimize the parameters of the current strategy to see if the performance of the strategy can be sufficiently improved by an optimized parameter setting.
Therefore, the user defines a number of optimization attempts that must be computed before the Strategy Selection can be triggered.
This parameter is situation dependent and the number of optimization attempts is executed as long as the situation remains the same.
If the Situation Detection component identifies a different situation than the last one, the Coordination triggers the Strategy Selection and Parameter Optimization regardless of whether the required number of optimization attempts is reached.
In the presented example, this number of optimization attempts is set to five.
Thus, after 3600~seconds execution time, the Coordination has already triggered five optimization attempts and now additionally triggers the Strategy Selection. 

\subsubsection{Domain Data Model}
The Domain-Data-Model is a representation of the use case for the framework and serves as configuration file for the framework as depicted in~\cref{fig:framework:architecture}.
It contains all use case-specific information the framework needs to optimize the use case and thus enables the generic applicability of the framework for a variety of use cases.
This means that these settings strongly depend on the chosen use case and can individually be enriched by use case specific parameters.
Further, the Domain-Data-Model provides configuration information for the components of the framework that the Coordination component uses to instantiate the components.
The Domain-Data-Model is defined using YAML and consists of four main parts: (i)~use case, (ii)~context, (iii)~parameter\_options, and (iv)~performance\_measures.
In the following, we describe each of these parts separately and provide a short example YAML file for this part.

We name the first part of the Domain-Data-Model \emph{use case}~(\cref{listing:ddm:usecase}, line~1) which contains general information about the use case.
The \emph{name}~(\cref{listing:ddm:usecase}, line~2) of the use case is the first key of this part, which is used to identify all information collected during the execution. 
The second key is called \emph{available\_strategies}~(\cref{listing:ddm:usecase}, line~3) and consists of a list of available adaptation planning strategies in the use case.
The Strategy Selection component of the framework uses this list to determine the most promising strategy for the current situation. 
The framework refers to them as black-box strategies and sends the name to the second layer which is able to select the appropriate strategy identified by its name.
This list of possible strategies does not need to be exhaustive and the user can omit strategies he does not want to be executed.
The last key of this part is the \emph{fallback\_rules} key~(\cref{listing:ddm:usecase}, line~4), which defines a path to a Python file that contains fallback rules for the framework. 
These fallback rules should reflect expert knowledge from the use case and are used by the framework in case the situation detection is not possible due to insufficient data or the current situation is identified as noise.
\cref{listing:ddm:usecase} presents the first part of the YAML file of an example use case called \texttt{platooning\_coordination}. 
In this use case, two adaptation planning strategies \texttt{s\_1} and \texttt{s\_2} are available. 
Finally, the path to the predefined fallback rules is defined as \texttt{Path.To.Rules}.

\begin{lstlisting}[language=yaml,firstnumber=1,stepnumber=1,numbers=left,
    frame=single,xleftmargin=2em,xrightmargin=2em,framexrightmargin=1.5em,
    label=listing:ddm:usecase,basicstyle=\small,
    caption={Example for the use case part of the Domain-Data-Model YAML.}]
use_case:
  name: platooning_coordination
  available_strategies: ["s_1", "s_2"]
  fallback_rules: "Path.To.Rules"
\end{lstlisting}

The second part of the Domain-Data-Model is called \emph{context}~(\cref{listing:ddm:context}, line~5) and specifies what context \emph{data}~(\cref{listing:ddm:context}, line~6), i.e., observations, the use case sends to the framework.
Furthermore, this part defines the configuration of the Situation Detection component with the key \emph{situation\_detection\_settings}~(\cref{listing:ddm:context}, line~13). 
The data key of this part contains any number of context parameters from the use case, which can be named arbitrarily, but must be unique~(\cref{listing:ddm:context}, line~9,11).
The framework will use these keys as identifiers when logging information to a database. 
Further, each context parameter requires a \emph{data\_type} specification~(\cref{listing:ddm:context},  line~10,12) and we currently accept \texttt{int} and \texttt{double} values.
The situation\_detection\_settings key describes the configuration of the Situation Detection component and consists of the two keys \emph{algorithm} and \emph{settings}~(\cref{listing:ddm:context}, line~16,17).
The algorithm key expects the definition of an available situation detection algorithm. 
So far, four algorithms are available which we describe in more detail in the next section: \texttt{RuleBased}, \texttt{K-Means}, \texttt{DBSCAN}, and \texttt{OPTICS}.
We limit our contribution to them as these are the most common algorithms, however, this list can easily be extended whenever another algorithm might perform better.
Each algorithm requires additional configuration parameters that are part of the settings key.
\cref{listing:ddm:context} provides a short YAML example for the context part.
It defines two context parameters \texttt{context1} and \texttt{context2} for the data key with data\_type \texttt{int} and \texttt{double}.
For the situation\_detection\_settings it is specified that the algorithm \texttt{DBSCAN} should be used and the required settings for this algorithm \texttt{min\_samples = 120} and \texttt{eps = 34} are defined~(\cref{listing:ddm:context}, line~18,19). 

The third part of the Domain-Data-Model is called \emph{parameter\_options}~(\cref{sec:framework:strategyselection}, line~20).
It defines input parameters of the adaptation planning strategy that can be tuned by the framework and provides configuration information for the Strategy Selection component.
This part consists of the \emph{options} for the input parameters and the \emph{strategy\_selection\_settings}~(\cref{sec:framework:strategyselection}, line~21,34). 
The options key contains an arbitrary number of input parameter options for strategies and the key is in turn used as identifier for this parameter~(\cref{sec:framework:strategyselection}, line~24,28).
Thus, it can be named arbitrarily but must be unique within this Domain-Data-Model.
\begin{minipage}{0.99\columnwidth}
\begin{lstlisting}[language=yaml,stepnumber=1,frame=single,
    firstnumber=5,stepnumber=1,numbers=left,
    frame=single,xleftmargin=2em,xrightmargin=2em,framexrightmargin=1.5em,
    label=listing:ddm:context,basicstyle=\small,
    caption={Context part of the YAML definition of the Domain-Data-Model.}]
context:
  data:
    # any number of context parameters 
    # with unique names
    context1:
      data_type: int
    context2:
      data_type: double
  situation_detection_settings:
    # available algorithms: RuleBased,
    # kMeans, DBSCAN, OPTICS
    algorithm: "DBSCAN"
    settings:
      min_samples: 120
      eps: 34
\end{lstlisting}
\end{minipage}
Each input parameter option further consists of three mandatory keys: \emph{data\_type}, \emph{min}, and \emph{max} and an optional key \emph{strategies}.
The \emph{data\_type} key defines the data type of the input parameter option, where we accept \texttt{int} and \texttt{double}~(\cref{sec:framework:strategyselection}, line~25,29).
The \emph{min} and \emph{max} keys allow the user to specify the value range the input parameter can take~(\cref{sec:framework:strategyselection}, line~26,27,30,31). 
Finally, the strategies key allows the user to define for which adaptation planning strategy this input parameter is meaningful by defining a list of strategies~(\cref{sec:framework:strategyselection}, line~33).
This key is optional and the absence of this key leads to the conclusion that this parameter applies to all strategies.
The second key of this part is the \emph{strategy\_selection\_settings} key, which configures the Strategy Selection component. 
This key consists of five mandatory keys: \emph{observations\_between\_adaptations}, \emph{min\_optimization\_attempts}, \emph{window\_size}, \emph{threshold\_exceeds}, and \emph{method} and one optional key called \emph{hypervolume\_threshold}. 
The key \emph{observations\_between\_adaptations} defines the number of observations the framework must receive before new adaptation actions can be performed~(\cref{sec:framework:strategyselection}, line~35). 
This property allows the user to define the transient phase for the use case where measurement data might be unreliable due to recent changes in the system. 
The \emph{min\_optimization\_attempts} key defines the number of parameter optimization attempts for a situation before the Coordination component considers computing a new adaptation planning strategy~(\cref{sec:framework:strategyselection}, line~36).
The \emph{window\_size} and \emph{threshold\_exceeds} keys determine whether a new strategy should be chosen~(\cref{sec:framework:strategyselection}, line~37,38).  
For a detailed explanation of these keys, please refer to \cref{sec:framework:strategyselection}.

\cref{listing:ddm:paramopt} provides a short YAML example for the parameter\_options part of the Domain-Data-Model.
It defines two parameter options \texttt{param1} and \texttt{param2}, where the first one is of type \texttt{int}, can accept values in the interval $[0,100]$, and applies to all defined strategies.
The second parameter option is of type \texttt{double}, can take values in the range $[0.0,2.0]$, and is only applicable for the strategy \texttt{s\_1}.
Furthermore, it specifies that the minimum number of optimization attempts is set to five, as well as other required settings for the Strategy Selection component.

\begin{minipage}[tb]{0.99\columnwidth}
\begin{lstlisting}[language=yaml,stepnumber=1,frame=single,
    firstnumber=20,stepnumber=1,numbers=left,
    frame=single,xleftmargin=2em,xrightmargin=2em,framexrightmargin=1.5em,
    label=listing:ddm:paramopt,basicstyle=\small,
    caption={Parameter options part of the YAML definition of the Domain-Data-Model.}]
parameter_options:
  options:
    # any number of context parameters 
    # with unique names
    param1:
      data_type: int
      min: 0
      max: 100
    param2:
      data_type: double
      min: 0.0
      max: 2.0
      # optional definition of 
      # relevant strategies
      strategies: ["s_1"]
  strategy_selection_settings:
    observations_between_adaptations: 1
    min_optimization_attempts: 5
    window_size: 5
    threshold_exceeds: 3
    # available methods: 
    # hypervolume, threshold
    method: "hypervolume"
    hypervolume_threshold: 3.4
\end{lstlisting}
\end{minipage}

The last part of the Domain-Data-Model is called \emph{performance\_measures}~(\cref{listing:ddm:perfmeasures}, line~42) and defines indicators of the performance of the defined use case.
This part contains any number of performance measures from the use case, which can be named arbitrarily~(\cref{listing:ddm:perfmeasures}, line~43,47).
Since these names are used as identifiers in the framework, they need to be unique.
Each performance measure consists of three mandatory keys \emph{data\_type}, \emph{higher\_is\_better}, and \emph{reference\_value}, and an optional key called \emph{threshold\_value}. 
The \emph{data\_type} specifies the performance measurement data type, which can be either \texttt{int} or \texttt{double}~(\cref{listing:ddm:perfmeasures}, line~44,48). 
The \emph{higher\_is\_better} key defines whether a higher or a lower value of this metric is better for this use case, and is of type Boolean~(\cref{listing:ddm:perfmeasures}, line~45,49).
The \emph{reference\_value} key specifies a reference value for the calculation of the Hypervolume, which needs to be of the same type as specified in \emph{data\_type}~(\cref{listing:ddm:perfmeasures}, line~46,50).
Finally, the \emph{threshold\_value} key is only required if the \texttt{threshold} method is selected in the \emph{strategy\_selection\_settings} of the \emph{parameter\_options} part and defines a threshold value that cause the Strategy Selection component to compute a new strategy. 
\cref{listing:ddm:perfmeasures} provides a YAML example for the \emph{performance\_measures} part of the Domain-Data-Model and defines two performance measures \texttt{pm1} and \texttt{pm2}. 
The first is of type \texttt{int}, where higher values represent a better use case performance and a reference value of \texttt{-1}. 
The second performance measure is of type \texttt{double}, with lower values representing better use case performance and a reference value of \texttt{100.0}.
\begin{minipage}[htb]{\columnwidth}
\begin{lstlisting}[language=yaml,stepnumber=1,frame=single,
    firstnumber=42,stepnumber=1,numbers=left,
    frame=single,xleftmargin=2em,xrightmargin=2em,framexrightmargin=1.5em,
    label=listing:ddm:perfmeasures,basicstyle=\small,
    caption={Performance measures part of the YAML definition of the Domain-Data-Model.}]
performance_measures:
  pm1:
    data_type: int
    higher_is_better: True
    reference_value: -1
  pm2:
    data_type: double
    higher_is_better: False
    reference_value: 100.0
\end{lstlisting}
\end{minipage}

\subsubsection{Situation Detection}
The Situation Detection component is responsible for identifying the current situation the use case is currently experiencing as depicted in~\cref{fig:framework:architecture}.
The use case periodically sends observation data to the framework, as defined in the context part of the Domain-Data-Model.
The frameworks' Coordination component forwards this data to the Situation detection.
So far, this component provides four methods for detecting the current situation: (i)~rule-based, (ii)~K-Means, (iii)~DBSCAN,  and (iv)~OPTICS.
We limit the available methods for this work but provide the possibility to easily integrate other methods in the framework.
All methods operate on all context data available in the system. 
To reduce the communication overhead within the framework, the Situation Detection contains a duplicated set of received observation data within the component, and the Coordination only needs to forward the current observation.
The Situation Detection component computes the current situation and returns a situation ID to the Coordination component.
This ID is further used in the Strategy Selection and Parameter Optimization components to find appropriate adaptation decisions for this specific situation and to learn from previous decisions in this situation.

The situation detection process can be defined as a mathematical function that maps observation data from the use case to an integer value.
This value represents the situation ID as defined in \cref{eq:framework:situationdetection}.
We define the value interval of this function as $[-1,\infty)$, where the value $-1$ indicates that the situation could not be detected.
This could be the case for two reasons: 
First, the amount of available data is insufficient to determine the situation.
Second, the observation data is classified as noise, meaning that the currently observed values cannot be classified as a specific situation.
This could be due to a novel situation for which these is not enough data, or measurement inaccuracies in the use case.
In the case that the Situation Detection classified the current situation as $-1$, the framework does not invoke any further computational processes, such as Strategy Selection or the Parameter Optimization. 
However, the Coordination component uses the user-defined fallback rules from the Domain-Data-Model~(\cref{listing:ddm:usecase}, line~4) to determine any adaptation actions that may be required.
If the returned situation ID is equal to or greater than zero, the Situation Detection component has determined a valid situation. 
Therefore, the Coordination component can invoke the Strategy Selection and Parameter Optimization components.
The actual value of the situation ID does not allow for further interpretation regarding the similarity of situations.
For example, if the component identified three situations $s_1=0, s_2=1, s_3=10$, it means that these three situations exist and are all different from each other. 
Moreover, the proximity of the values $0$ and $1$ does not mean that the situations $s_1$ and $s_2$ are more similar to each other than the situation $s_3$.
\begin{equation}
\label{eq:framework:situationdetection}
\textit{sit\_det}(\textit{context}) = \begin{cases}
	-1, 		  	& \textit{if} \text{  situation is classified as noise} \\
	>= 0, 			& \text{otherwise}
\end{cases}
\end{equation}

Due to the ongoing nature of the framework, the use case regularly sends new observation data.  
Therefore, the amount of observation data grows as the framework is executed and the Situation Detection component receives more and more data to improve decision making. 
However, this could lead to a changed in the assignment of context data to situations during the execution time. 
This means, the situations identified during the last Situation Detection process may not be the same as those identified in the current process.
Completely new situations or a change in assignment from an already assigned observation could lead to inconsistencies in the data.
For example, a context observation classified as situation $s_1$ in the last process could now be classified as $s_2$ when more data is available. 
Therefore, the Situation Detection component updates its learned models after each execution to match the latest findings to the observation data.

We provide two types of situation detection mechanisms, one rule-based mechanism and four clustering algorithms that can be selected and configured by the user in the Domain-Data-Model. 
Since we designed the framework to be modular, it is easy to extend the framework with additional components or to further develop individual components with additional techniques. 
The following \cref{algo:framework:situationdetection} summarizes the workflow behavior of the Situation Detection component.
The component receives the Domain-Data-Model and the new observation and selects the configured algorithm for the Situation Detection. 
In all cases, the component retrieves required parameters for the selected technique from the Domain-Data-Model and invokes the configured technique.
All techniques return the \texttt{situationID}s for all observations, that is, the cluster to which each observation in the data set is assigned. 
The component then update its situation model of all observed data with the latest classification and returns the \texttt{situationID} of the new observation to the Coordination component.

\begin{algorithm}
	\caption{Pseudocode workflow of the Situation Detection component.}
	\label{algo:framework:situationdetection}
    \SetKwInOut{Input}{Input}
    \SetKwInOut{Data}{Data}
    \Input{Domain-Data-Model, new observation}
    \BlankLine
    \Switch{Domain-Data-Model.situation\_detection\_settings.algorithm}{
        \Case{RuleBased}{
            retrieve path to fallback rules from Domain-Data-Model\;
            situationID $\leftarrow$ execute fallback rules\;
        }
        \Case{kMeans}{
            retrieve K-Means parameters from Domain-Data-Model\;
            situationID $\leftarrow$ invoke K-Means\;
        }
        \Case{DBSCAN}{
            retrieve DBSCAN parameters from Domain-Data-Model\;
            situationID $\leftarrow$ invoke DBSCAN\;
        }
        \Case{OPTICS}{
            retrieve OPTICS parameters from Domain-Data-Model\;
            situationID $\leftarrow$ invoke OPTICS\; 
        }
    }
    update situation model with latest classifications\;
    return situationID of new observation\;
\end{algorithm}

The rule-based situation detection offers the possibility to integrate domain knowledge in the identification process of this component. 
For example, in the platooning use case, the user could specify frequent traffic volumes for which he knows the best performing configuration of the adaptation planning system.
The user defines the rules in form of a Python file that is loaded and executed by the component. 
As the simplest option, the user can define Event-Condition-Action rules to specify known, well-performing configurations. 
However, since the user describes the fallback rules in a Python file, he can also construct arbitrarily complex functions to identify situations. 
Still, the user must provide a script that matches our definition of the situation detection function in \cref{eq:framework:situationdetection}.
The user can adapt these rules for future executions of the framework as he gains new domain knowledge from running the framework and analyzing its decisions.
In the context of this paper, we omit updating the user-provided rule set with new knowledge from previous executions of the Situation Detection.
This also results in the framework being unable to react to new situations in the fallback case, since they are not present in the rules. 
In this case, the new situation must be classified as noise.
However, there are several approaches to automatically update rule sets during execution~\cite{nguyen2012computational,chand2018use,ghandar2009computational}.

In addition to the static rule-based situation detection, we provide three clustering-based situation detection methods.
Advantage of these methods are that they can automatically detect new situations due to their unsupervised learning approach, and that they do not require domain knowledge~\cite{alelyani2013feature,fredericks2019planning}.
One clustering algorithm we integrate into our framework is K-Means in two versions.
The first version works with a predefined parameter $k$ that specifies the number of clusters to identify. 
In the second version, the algorithm can determine the parameter $k$ automatically by applying the concept of gap statistics~\cite{tibshirani2001estimating}. 
This method requires the definition of a minimum and a maximum value for $k$ but no further interaction with the user is required. 
The gap statistics estimates the best value for $k$ by applying K-Means to different values of $k$ and analyzing the quality of the clustering.
Another method for automatically defining $k$ could be the elbow method~\cite{elbowMethodBlogWebsite}.
In this method, the user must plot various possible values of $k$ and their performance with regards to the resulting clustering. 
Then, the user identifies the elbow of the resulting line that represents the best value for $k$.
Due to the mandatory user interaction, we decided to omit the elbow technique.
The performance of the K-Means algorithm depends heavily on the definition of $k$ and the user may not have the expertise to determine the number of distinct situations a priori. 
Further, the K-Means algorithm always assigns all observations to an existing cluster and cannot identify noise, which could negatively affect the performance of the framework.
Therefore, we additionally integrate two density-based clustering algorithms into the Situation Detection component to reduce these drawbacks.

We select DBSCAN and OPTICS as density-based clustering approaches.
Neither method requires a number of clusters as input. 
Instead, DBSCAN requires the definition of \texttt{min\_samples}, which specifies the minimum number of observation samples to form a cluster. 
Additionally, an $\epsilon$ (\texttt{eps}) value is required that defines the neighborhood of a data point in which at least \texttt{min\_samples} must be found to classify that data point as core-point.
For the definition of $\epsilon$ the user needs domain knowledge and it has a great impact on the identified cluster structure.
OPTICS needs the parameter \texttt{min\_samples} which is the number of data points in a neighborhood, to consider this point as core-point.
Also required is the parameter \texttt{min\_cluster\_size}, which is the minimum number of data points required to form a cluster.
The user can determine both values by considering how long a situation is usually active in the use case and how many observations are sent to the framework. 
Both density-based clustering algorithms can classify observations as noise, which could happen when the use case observes a new situation for a short time.

One important point that the user of the framework must keep in mind is data management. 
Since the use case continuously sends observation data, the amount of data is constantly increasing. 
So far, we have not implemented any feature to reduce the amount of considered data in the decision making, which may lead to errors due to memory limitations.
To reduce the amount of considered and stored data, the framework needs to determine what information will be important in the future and what information can be omitted without negatively impacting the future performance of the framework.
One option is to set a maximum number of data points considered, but this could result in sparse situations being forgotten.
Techniques for reducing an ever-increasing amount of observation data can be found in the literature.
For example, Kang~et~al.~\cite{kang2020far} research on the required knowledge of robots about their environment to reduce the probability of collisions due to estimation errors regarding other robots.
Such a technique could be easily added to the Situation Detection component as future work to prepare the framework for long-term executions as well.

\subsubsection{Strategy Selection}
\label{sec:framework:strategyselection}
The Strategy Selection is the second component of the framework, that is invoked by the Coordination component as depicted in~\cref{fig:framework:architecture}.
This component is responsible for selecting the most promising adaptation planning strategy for the use case with respect to the current situation.
This assumes, of course, that the use case supports different strategies and that the user configured them for selection.
This functionality is based on the No-Free-Lunch Theorem for optimizations~\cite{Wolpert1997} and the idea of situation-dependent behavior of adaptation planning systems. 
Hence, the goal is to select the strategy that seems most promising for the current situation.
To do this, the framework uses the experience gained from previous executions of the strategies in similar situations.
However, which algorithm performs best in a new situation is not known a priori.
Therefore, the framework must test the available strategies and start a new round of learning for that situation.
A general definition of the algorithm selection problem can be found in~\cite{smith2009cross}.
In the following, we explain the general workflow of the Strategy Selection and refer to \cref{algo:framework:strategyselection}.
\begin{algorithm}
	\caption{Pseudocode workflow of the Strategy Selection component.}
	\label{algo:framework:strategyselection}
    \SetKwInOut{Input}{Input}
    \SetKwInOut{Data}{Data}
    \Input{Domain-Data-Model, current strategy, number of optimization attempts already performed, all observations for the current situation}
    \BlankLine
    
    strategy $\leftarrow$ current strategy\;
    \eIf{number of optimization attempts $<$ Domain-Data-Model.min\_optimization\_attempts}{
        return strategy\;
    }{
        exceed\_counter $\leftarrow$ 0\;
        \For{observation within Domain-Data-Model.window\_size}{
            \If{thresholds exceeded}{
                exceed\_counter++\;
            }
        }
        \If{exceed\_counter $>=$ Domain-Data-Model.threshold\_exceeds}{
            \eIf{all strategies already executed for this situation}{
                strategy $\leftarrow$ best performing strategy in history\;
            }{
                strategy $\leftarrow$ next strategy determined in Domain-Data-Model\;
            }
        }
    }
    return strategy\;
\end{algorithm}

Similar to the Situation Detection, this component also receives the Domain-Data-Model as input.
Additionally, the Coordination component sends the currently active adaptation planning strategy, the number of optimization attempts already performed for this strategy, and all available observations for the current situation.
These observations contain the performance measures of the adaptation planning strategy and form the basis for the decision logic.
First, the Strategy Selection sets the currently active strategy as the selected strategy since it assumes that no changes need to be made by default~(line~1).
Then, the component checks whether enough optimization attempts have been made to decide whether the strategy should be changed.
We decided to provide a fixed initial period during which multiple optimizations of the parameters are performed before considering a strategy infeasible for this situation.
If the actual number of optimization attempts has not reached the minimum number of optimization attempts defined in the Domain-Data-Model, it means that the Parameter Optimization component might need more time to optimize the parameters of this strategy, this component then returns the currently active strategy~(lines~2-3).
If the required number of optimization attempts has already been reached~(line~4), this component can select another strategy if the current strategy does not meet the performance expectations~(lines~5-8).
To do this, the component analyzes the performance of the strategy in the last observations with respect to a defined threshold and counts the number of times the threshold is exceeded.
The actual number of analyzed observations is determined using the \emph{window\_size} in the Domain-Data-Model.
The component provides two ways to define these threshold: (i)~hypervolume threshold and (ii)~individual value thresholds.
Full details of both methods are provided later in this section.
After the component determines the number of threshold violation in the last observations, it checks whether this number is above the predefined maximum allowed threshold violations~(line~9).
If this holds, the component proceeds and selects a new strategy~(line~10).
It then checks to see if all strategies for that situation have already been executed and if so, it selects the strategy that resulted in the best performance measurements~(line~11). 
Thus, the component computes the Hypervolume of performance measurements for each observation within the \emph{window\_size} and all strategies and selects the strategy that yields the highest average Hypervolume.
In the event that at least one strategy defined in the Domain-Data-Model was not executed for this situation, the Strategy Selection retrieves one of these strategies from the Domain-Data-Model~(line~13). 
This triggers a trial-and-error phase in this component, since the decision cannot be based on experience and the component is forced to try new combinations.
Finally, the component returns the selected strategy to the Coordination component~(line~14). 

The Strategy Selection component provides two possibilities to determine whether an algorithm meets the performance expectations or should be modified.
In both mechanisms, the component counts the number of threshold violations and compares them to the allowed threshold violations specified by the user. 
The first method the component offers so far is the Hypervolume threshold method which reduces the performance measures to a single score.
In this case, the component computes the Hypervolume metric~(c.f.~\cite{Wang2016}) and compares its value to a user-defined threshold. 
To calculate the Hypervolume, the user must specify reference values for each performance measure in the Domain-Data-Model.
These reference values can either be defined out of range for the performance measure in question, or set to a value within the range that should never be dropped below.
If the reference value is defined within the value range and the actual value falls below this value, the Hypervolume is defined as zero regardless of the other performance measures. 
However, the downside of this method is that it weights measures with a larger value range more heavily, so the user should apply a normalization mechanism before sending the performance measures to the framework.
Still, the advantage of this method is that the performance of the overall adaptation planning system is condensed into one metric and the user only needs to specify one threshold value.

The second possibility to determine whether to change the currently active adaptation planning strategy is to set individual value thresholds. 
This method requires the user to define individual thresholds for each performance measure of the Domain-Data-Model that should never be fallen below. 
Whenever one of the performance measures falls below this threshold, the Strategy Selection component counts this as a threshold violation, regardless of any possibly perfect performance of the other measures. 
This method allows the user to have more impact on the individual performance measures and value ranges of these measures are less important. 
The user can even rule out performance measures having an impact on the strategy selection by setting the threshold out of the value ranges.
Similar to the other components of the framework, the user can develop a customized version of the methods used in this component.
Additionally, the user can easily extent the functionality of this component due to its modular design.
For instance, Machine Learning techniques such as Random Forests~\cite{RandomForests} can be integrated to learn a model for the Strategy Selection.
This learning could use historical observation data to computes features as basis for the learned model.
As the framework detects new situations, the model should be retrained to also cover decisions for the new situation once sufficient observation data has been collected.

\subsubsection{Parameter Optimization}
The last component to be presented is the Parameter Optimization component depicted in~\cref{fig:framework:architecture}. 
The Coordination component invokes this component when a new strategy is determined, the situation changes, or the performance of the strategy decreases with respect to the performance measures of the use case. 
This component uses optimization techniques to determine the best performing parameter setting for the selected strategy. 
According to the No-Free-Lunch Theorem~\cite{Wolpert1997}, the choice of the optimization algorithm depends heavily on the use case being optimized. 
At the moment, this component uses Bayesian Optimization to optimize the parameter setting, but the desired technique can be easily replaced due to the modular nature of the framework.
The decision to use the Bayesian Optimization is based on our platooning coordination use case, as our study of situation-dependency showed that Bayesian Optimization is best s best for this use case~\cite{Lesch2021Towards}.

So far, the Parameter Optimization component applies Bayesian Optimization to determine a new set of parameters for the current strategy. 
For this computation, the component uses historical observation data of the same situation and strategy combination. 
The Coordination component is responsible for providing only relevant data to this component. 
If the situation-strategy combination has not changed since the last invocation of this component, the Bayesian Optimization integrates only the last observation into the optimization model to computer new parameters.
If either the situation or the selected strategy has changed since the last invocation, the optimization model must be re-trained using historical data of the new situation-strategy combination, if available. 
This allows the Parameter Optimization to react to the current situation and strategy and learn from previous decisions.

The Parameter Optimization component returns the new parameter set for the strategy to the Coordination component which forwards the adaptations to the use case.
The use case executes these adaptations and collects new observations, that is, performance data for the new parameter settings, and sends them to the framework. 
In the next round of execution by the framework, the optimization technique receives this performance data and performs the next optimization step to further optimize the strategy parameters.

When choosing which optimization technique to use, the user of the framework must keep in mind that the algorithm must be able to learn from previous decisions.
It should also be noted that the algorithm must process new incoming observations on the fly and does not need to be fully trained for every new observation.
Finally, the overhead to completely retrain the model when a new situation occurs or the selected strategy changes should be kept to a minimum.
In the best case, the user should choose an optimization technique whose optimization model can be extracted and reloaded when the component needs to handle a new situation and strategy combination.
This would limit the time required to completely re-train the model for each change in situation and strategy. 
For future extension of the framework, the general meta-heuristic search algorithm Stepwise Sampling Search~(S3)~\cite{Noorshams2015phd} could be tested for faster optimized parametrizations.

\subsection{Use Case-specific Adapter of the Framework}
\label{sec:framework:adapter}
All the components of the framework are designed to be generically applicable to a variety of use cases enabled by the Domain-Data-Model definition of use-case specific characteristics and an adapter that manages the connection between use case and framework as described in the following.
As mentioned earlier, the framework is modular and consists of components that users can adapt to the use case depending on their requirements. 
Nevertheless, all components are designed to handle any kind of data from a use case as long as the data and optimization goals are defined in the Domain-Data-Model.
This section briefly summarizes the required user actions to apply the framework for any use case. 

\begin{figure}[htb]
\begin{center}
\includegraphics[width=0.75\columnwidth]{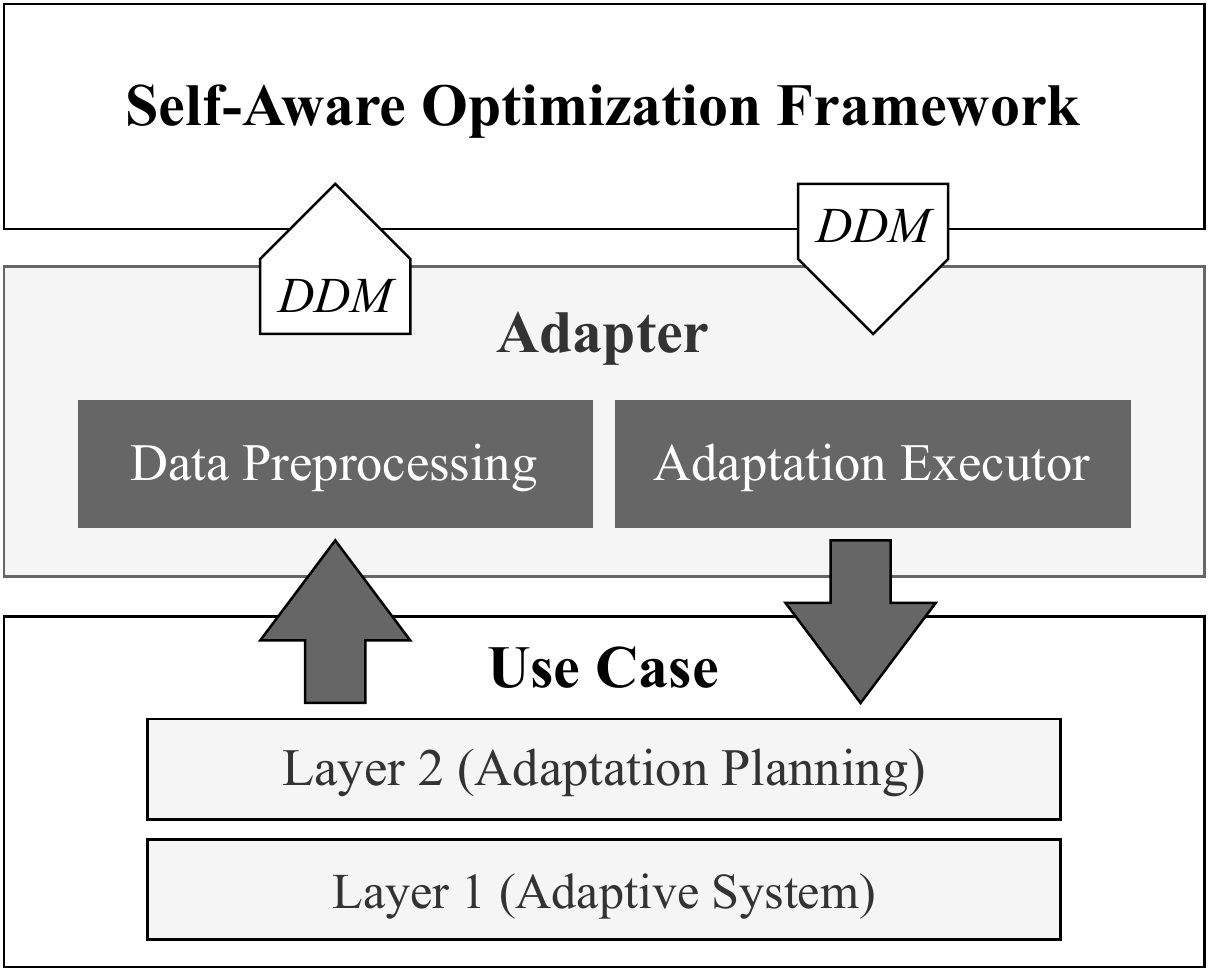}
\caption[Use case adapter for the self-aware optimization framework.]{Use case adapter for the generic self-aware optimization framework. The use case with its two layers adaptive system and adaptation planning are depicted at the bottom. It communicates with the Framework by sending observations and retrieving adaptation actions. Additional Data Preprocessing and Adaptation Executor components can provide a further abstraction level.}
\label{fig:framework:adapter}
\end{center}
\end{figure}

\cref{fig:framework:adapter} provides an overview of the architecture of the adapter required to connect the framework to any use case.
The self-aware optimization framework is shown at the top, while the use case consisting of the two lower levels~(see~\cref{sec:framework:systemmodel}) is shown at the bottom of the figure.
The center of the figure presents two adapter components which are used to connect the framework and the use case: (i)~Data Preprocessing and (ii)~Adaptation Executor.
The framework provides two interfaces that enable general applicability of the framework as they are implemented with REST APIs.
Further, the Domain-Data-Model defines the data sent through these APIs and provides all the necessary information to interpret the parameters, make adaptation decisions, and send adaptation actions. 
The API on the left receives the observation data, while the API on the right provides the possibility to retrieve adaptation decisions for the use case. 
We decided to further abstract the data handling from the use case and include an additional Data Preprocessing component.
This component receives raw monitoring data from the use case, preprocesses this data, and potentially calculates additional aggregate metrics that may be required to assess the performance of the use case.
Due to the REST API, the whole component can be replaced with a customized version to fit the desired use case.
The Adaptation Executor component, depicted on the center right of the figure, retrieves the adaptation decisions from the framework and converts these into specific adaptation actions for the use case.
This component also depends heavily on the use case and the user must customize the existing component to the requirements of the new use case. 
Since both adapter components handle data transfer to and from the framework, the implementation effort required to apply them to a new use case should be minimal.
If the use case already provides the possibilities to send monitoring data directly to the framework and retrieve and execute adaptation decisions, these adapter components may not be necessary.
However, we have chosen to provide a template for such components as they represent another level of abstraction to the use case and, thus, reduce the required computational effort in the use case.

\section{Evaluation}
\label{eval}
\subsection{Methodology}
\label{sec:eval:framework_method}
In this work, we use the platooning coordination use case as a running example of our self-aware optimization framework. 
In this context, we also evaluate our framework in this use case. 
We first define the applied scenarios, then summarize the testbed and specify the framework configuration for our evaluation before proposing our baseline approaches.

\begin{figure*}[htb]
    \centering
    \subfloat[Wednesday]{\includegraphics[width=0.4\textwidth]{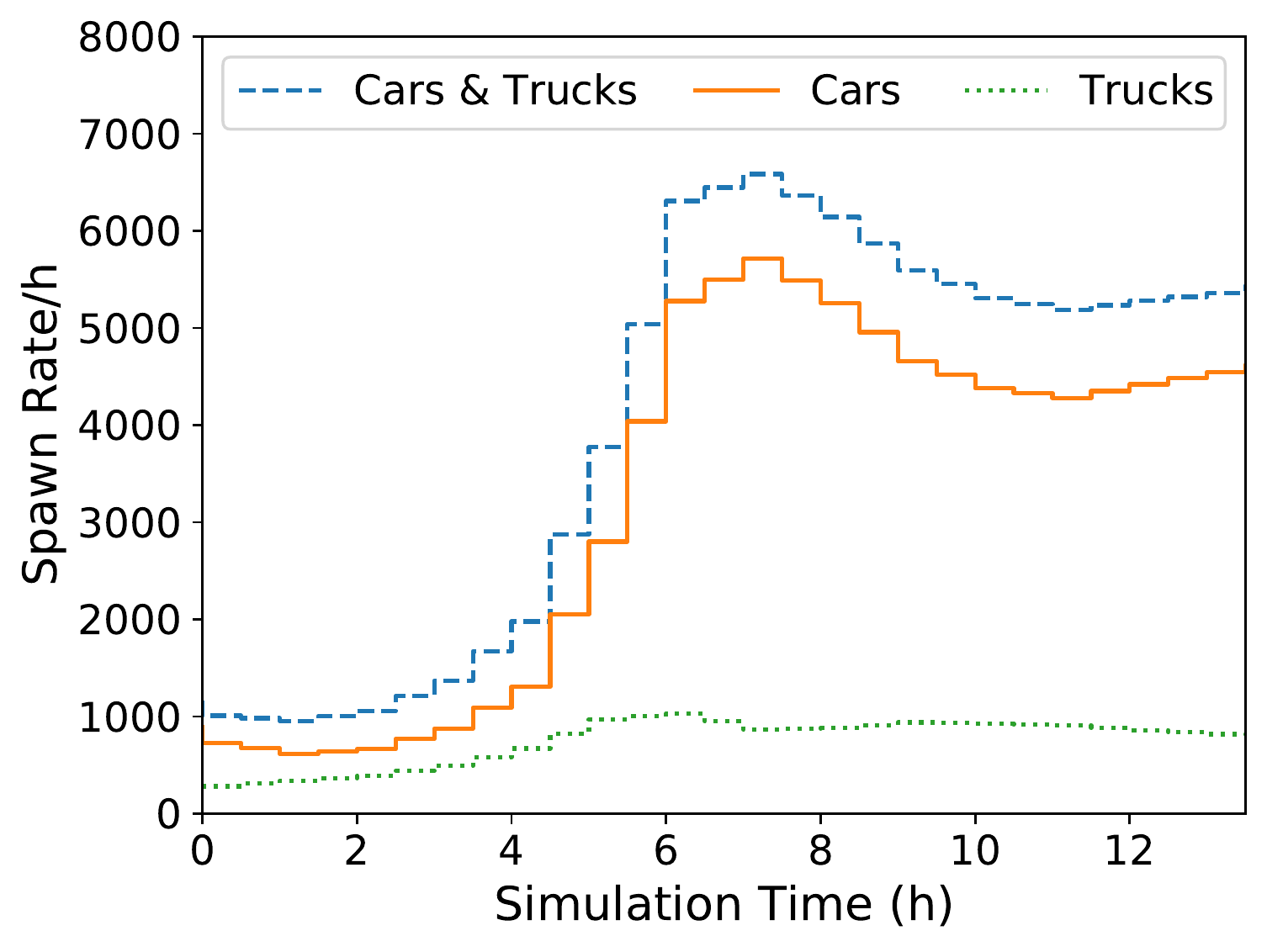}\label{fig:eval:framework:scenario_wed}}
    \hspace{2cm}
    \subfloat[Saturday]{\includegraphics[width=0.4\textwidth]{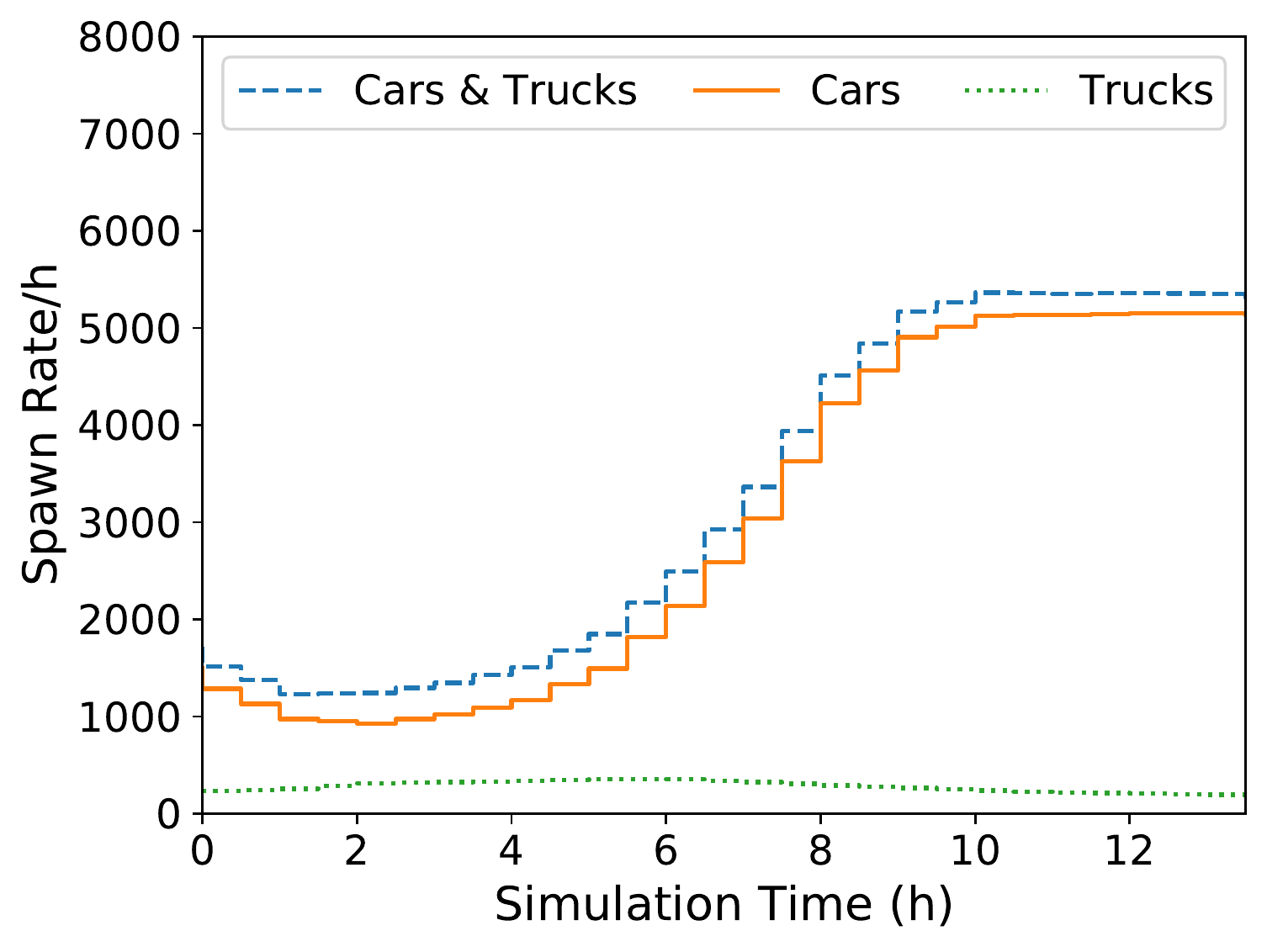}\label{fig:eval:framework:scenario_sat}}
    \caption
        [Considered traffic scenarios of the framework evaluation for Wednesday and Saturday.]
        {Considered traffic scenarios of the framework evaluation for Wednesday on the left and Saturday on the right. Total number of spawning vehicles is depicted as blue dashed line, cars are depicted as solid orange line, and trucks are depicted as dotted green line.}
    \label{fig:eval:framework:scenario}
\end{figure*}

We use a simulated road section of the German highway A8, which extends from the Stuttgart interchange to the Stuttgart-Degerloch exit. 
According to Süddeutsche Zeitung, this section is one of the busiest highway sections in Germany~\cite{a8befahrenWebsite}.
In addition to the realistic model of this highway section, we use real traffic data provided by the Federal Highway Research Institute of Germany~\cite{basttrafficstats} to define the vehicle spawn rates for our simulation.
After a detailed analysis of the traffic values for each day of the week with the goal of selecting two distinct days with individual traffic volume profiles, we selected Wednesday as the representative weekday, and Saturday as the representative weekend day.
\cref{fig:eval:framework:scenario} shows the traffic volume for the selected days between 12:00~AM and 2:00~PM.
As the simulation of such high traffic volume requires high computational power and a long computation time, we decided to simulate the first 14~hours of a day.
This time interval contains a typical traffic volume profile for weekdays and weekends and, therefore, provides a good balance between long runtime and comprehensive simulation.
We set the platooning percentage of all vehicles to 70\% as we assume that not every vehicle is capable of platooning or drivers choose not to participate. 
Furthermore, we set the maximum speed limit of cars to 120km/h, which corresponds to the actual speed limits on this section~\cite{a8tempoLimitWebsite}.
In our evaluation, we use two types of situation detection---OPTICS and rule-based situation detection---and two types of triggers for strategy selection---Hypervolume- and threshold-based triggers---which results in four simulations per traffic profile.
Since our approach involves Bayesian Optimization that incorporates randomness, we run three different random seeds in the traffic simulator SUMO for each simulation.

We perform our simulations in the cloud of the Chair of Computer Science II at the University of Würzburg.
This cloud consists of 18 hosts, each running RHEL-7-8.2003.0.el7.centos and oVirt Node 4.3.10 with KVM version 2.12.0.
The cloud contains one large ProLiant DL380 Gen9 host with two Intel(R) Xeon(R) CPU E5-2640 v3 @ 2.60~GHz CPU sockets and eight cores per socket.
The remaining hosts are ProLiant DL160 Gen9 type with two CPU sockets of type Intel(R) Xeon(R) CPU E5-2640 v3 @ 2.60~GHz, eight cores per socket, and two CPU threads per core.
We use three identical virtual machines for the simulations, which are deployed in our private cloud.
Each virtual machine has two CPU sockets, each with 4 cores running at 2.6~GHz and 32~GB available RAM.
We measure the simulation runtime of our scenarios, resulting in an average runtime of 9.5~days for the Wednesday scenarios and 9~days for Saturdays.
Since the traffic volume on Saturdays is lower than on Wednesdays, these simulations require less runtime.

We test two situation detection approaches and two triggers for strategy selection, which we now define.
For the situation detection, we use a rule-based approach as well as OPTICS, which we have already presented in \cref{sec:framework:architecture}.
As data input for the situation detection we use the amount of vehicles on the road.
We derived the rules for the rule-based approach by taking definitions for peak hours, medium, and low traffic volumes from the German city of Rostock~\cite{rostockNahverkehrWebsite}.
This study states that peak hours occur from Monday to Friday, which led us to the decision to use the highest traffic volume on Saturday as the upper limit for off-peak hours.
This results in the following rules: We consider the first situation with lowest traffic volume, where the maximum number of vehicles on the road section is 120.
We define the medium traffic volume from 121 to 280 vehicles and define the peak traffic volume above 280 vehicles on the road segment.
OPTICS requires the definition of the minimum number of points and the minimum cluster size, both of which we set to a value of 45.
We determined this value in a preliminary study with different parameters, which showed that this configuration is best suited for our use case.

Similar to the situation detection, we also evaluate two triggers for the strategy selection component: Hypervolume and individual thresholds.
Both methods incorporate the four objective metrics to assess the performance of the currently active strategy: (i)~throughput, (ii)~time loss, (iii)~platoon utilization, and (iv)~platoon time.
The Hypervolume requires the definition of a reference value which we set to -0.1, which is outside the range of values of all considered platoon metrics. 
We set the Hypervolume threshold to 0.3 and consider a time window size of five, in which the Hypervolume must fall below the threshold at least three times to trigger the strategy selection.
The threshold-based trigger requires the definition of an individual threshold per platoon metric, which we set as follows.
We set the throughput metric threshold to 0.5, since we assume that platooning coordination strategies have little impact on this metric. 
Furthermore, we set the threshold for the time loss metric to 0.9, since our preliminary study showed that the time loss metric was always above 0.85 for all runs, so we need a very strict threshold to have any effect at all.
We set the platoon utilization metric to 0.62, which is also close to the defined platoon percentage and should lead the system to high platoon utilization.
The threshold for the platoon time is 0.3, a comparatively low value that provides the framework with a large margin for testing different strategies.
The strategy selection component requires specifying the number of optimization cycles for each strategy as initial trial phase in which no other strategy can be selected. 
We set this value to ten. 
Finally, we specify the order in which the platooning coordination strategies are selected: Best-Distance, Best-Velocity, as well as Best-Distance-and-Lane. 
In our study regarding the situation-awareness of platooning coordination strategies~\cite{Lesch2021Towards} we analyzed that the Best-Velocity strategy is the most appropriate for this use case. 
This would mean that this strategy should to be tested first.
However, we decided to start with the Best-Distance strategy to force the framework to select another strategy.
\begin{table}[tb]
\small
\centering
\caption[Configuration of the framework and tested strategies, algorithms, and methods.]{Configuration of the framework and tested strategies, algorithms, and methods used in the evaluation. }
\label{tab:eval:framework:fw_conf}
\begin{adjustbox}{width=0.99\columnwidth}
\begin{tabular}{llp{3.5cm}}
    \toprule
    DDM Part & Parameter & Value\\
    \midrule
    Use Case        & Available strategies  & Best-Distance, Best-Velocity, Best-Distance-and-Lane\\
    Situation Detection & Algorithm         & RuleBased, OPTICS \\ 
    Strategy Selection & Method             & Hypervolume, threshold\\
                    & Min. opt. attempts    & 10\\
    Hypervolume     & Reference values         & -0.10\\
                    & Threshold             & 0.30 \\
                    & Time window size      & 5 \\
                    & Threshold exceeds     & 3 \\
    Thresholds      & Throughput            & 0.50\\
                    & Time loss             & 0.90 \\
                    & Platoon utilization   & 0.62 \\
                    & Platoon time          & 0.30\\\bottomrule
\end{tabular}
\end{adjustbox}
\end{table}

To evaluate the performance of our framework against a set of baseline approaches, we apply the Best-Distance, Best-Velocity, and a rule-based strategy to the two scenarios.
\cref{tab:eval:framework:baseline_conf} summarizes the configurations of our baseline strategies.
We derived the rule-based strategy from our study of the self-awareness of strategies~\cite{Lesch2021Towards}.
Since the Best-Velocity strategy performs by far the best in this study, we distinguish two cases in which we change the configuration depending on the number of vehicles on the road and the average car speed.
The rule-based strategy uses the first configuration when the number of vehicles is below 500 and the average car speed is above 125~km/h.
It applies the second configuration when the number of vehicles is higher than 500 and the average car speed is lower than 125~km/h. 
We also apply the same set of rules as fallback-mechanism in our framework when the applied situation detection cannot detect the current situation.
\begin{table}[htb]
\small
\centering
\caption[Configurations of the baseline approaches for evaluating the proposed framework.]{Configurations of the baseline approaches used in the evaluation.}
\label{tab:eval:framework:baseline_conf}
\begin{adjustbox}{width=0.99\columnwidth}
\begin{tabular}{p{2.5cm}cccc}
    \toprule
    Parameter Name & Best-Distance & Best-Velocity & Rules I & Rules II\\
    \midrule
    Advertising duration [m] & 10 & 10 & 10 & 5 \\ 
    Search distance front [m] & - & 600 & 600 & 400 \\ 
    Search distance back [m] & - & 250 & 250 & 200 \\ 
    Max. speed difference [km/h] & 35 & - & - & - \\ 
    Speed threshold lane 2 [km/h] & 100 & 100 & 100 & 100 \\ 
    Speed threshold lane 3 [km/h] & 130 & 130 & 130 & 130 \\ 
    Speed threshold lane 4 [km/h] & 160 & 160 & 160 & 160 \\ \bottomrule
\end{tabular}
\end{adjustbox}
\end{table}

\subsection{Evaluation of the Situation Detection Component}
\label{sec:eval:framework_sitdet}
\begin{figure*}[htb]
  \centering
  \subfloat[Ground truth for the situation detection.]{
  	\includegraphics[width=0.3\textwidth]{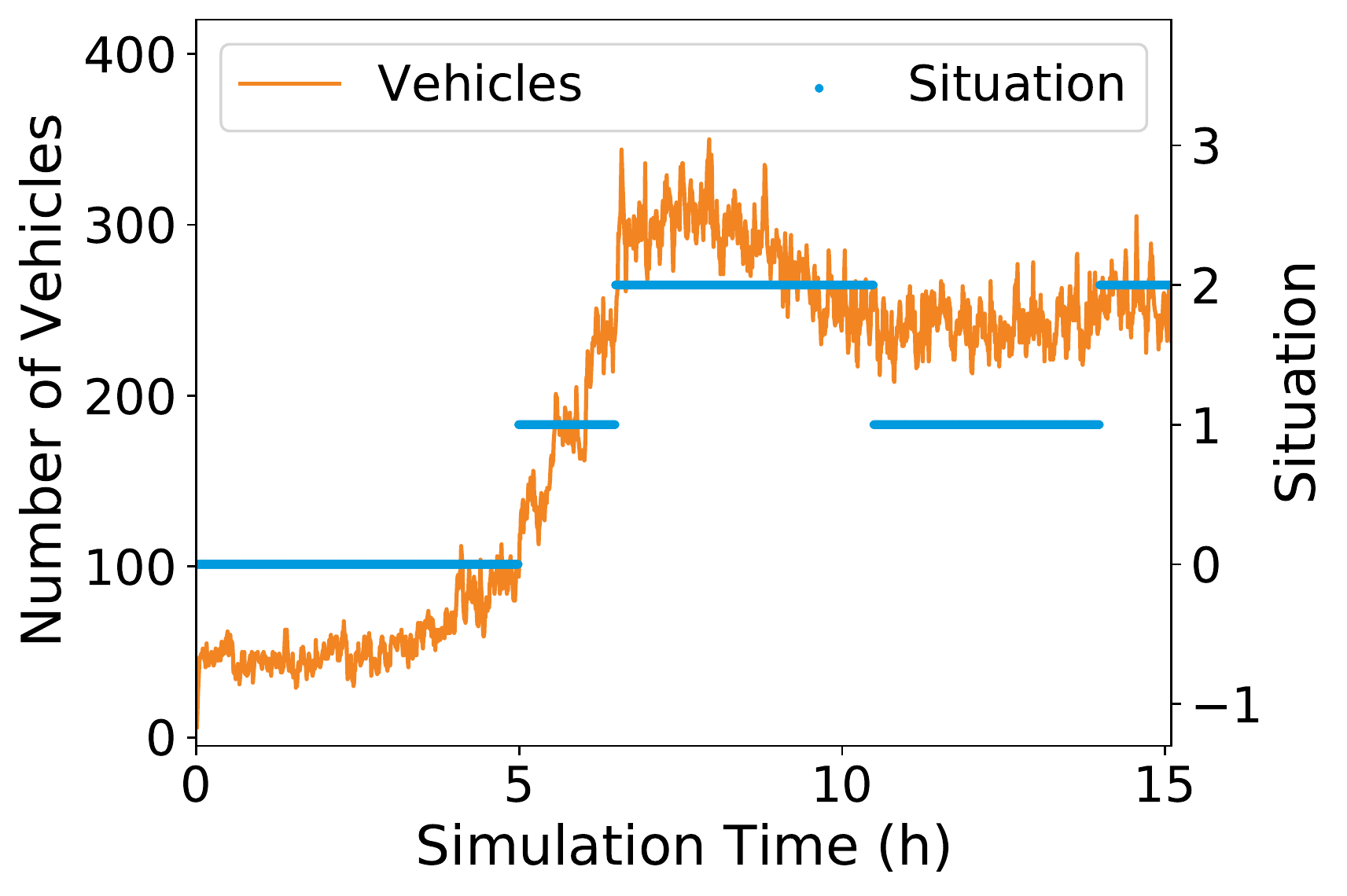}
  	\label{fig:framework_sitdet_wed_truth}
  }
  \hfill
  \subfloat[Detected situations when applying rule-based situation detection.]{
  	\includegraphics[width=0.3\textwidth]{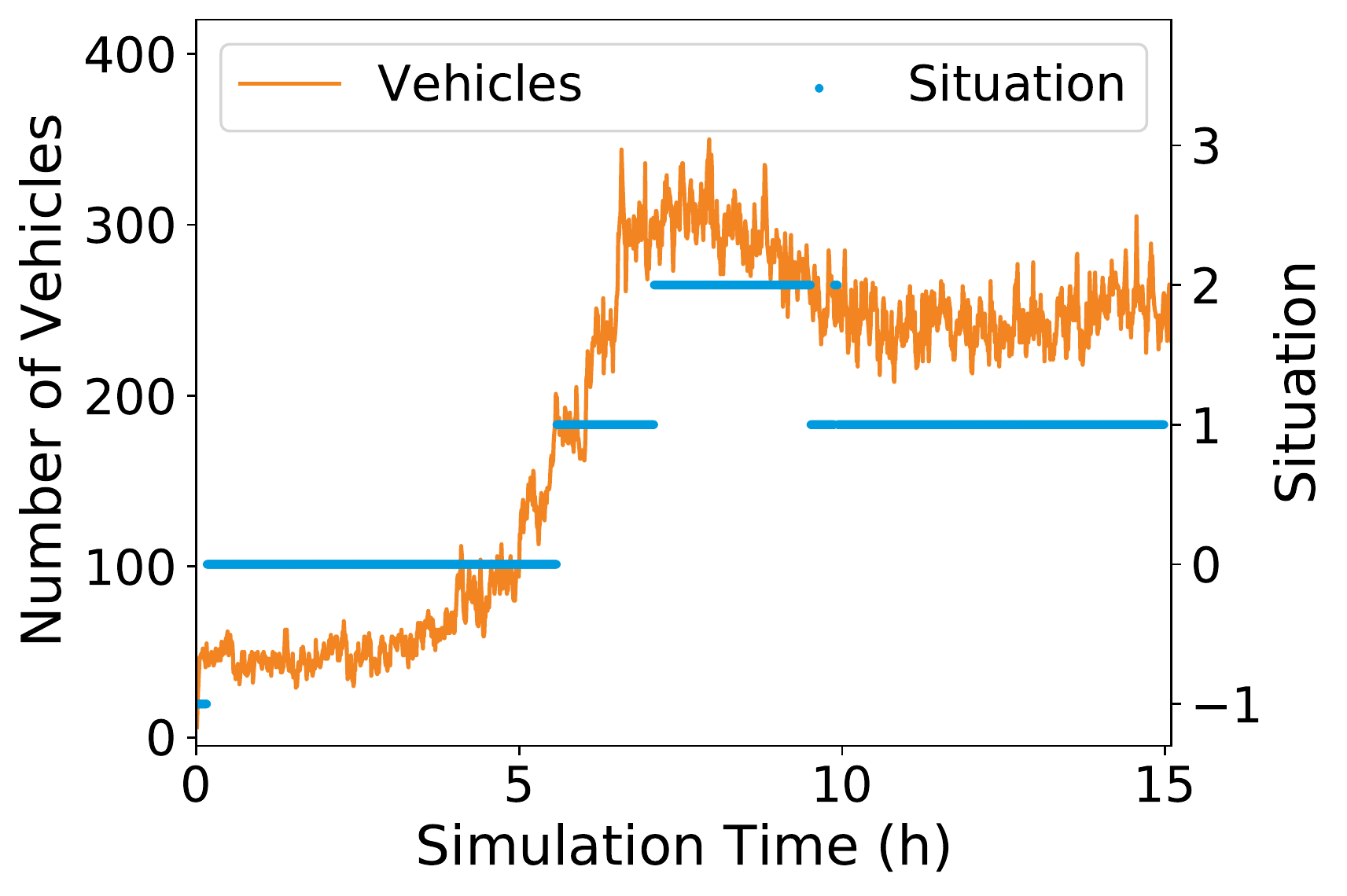}
  	\label{fig:framework_sitdet_wed_rules}
  }
  \hfill  
  \subfloat[Detected situations when applying OPTICS situation detection.]{
  	\includegraphics[width=0.3\textwidth]{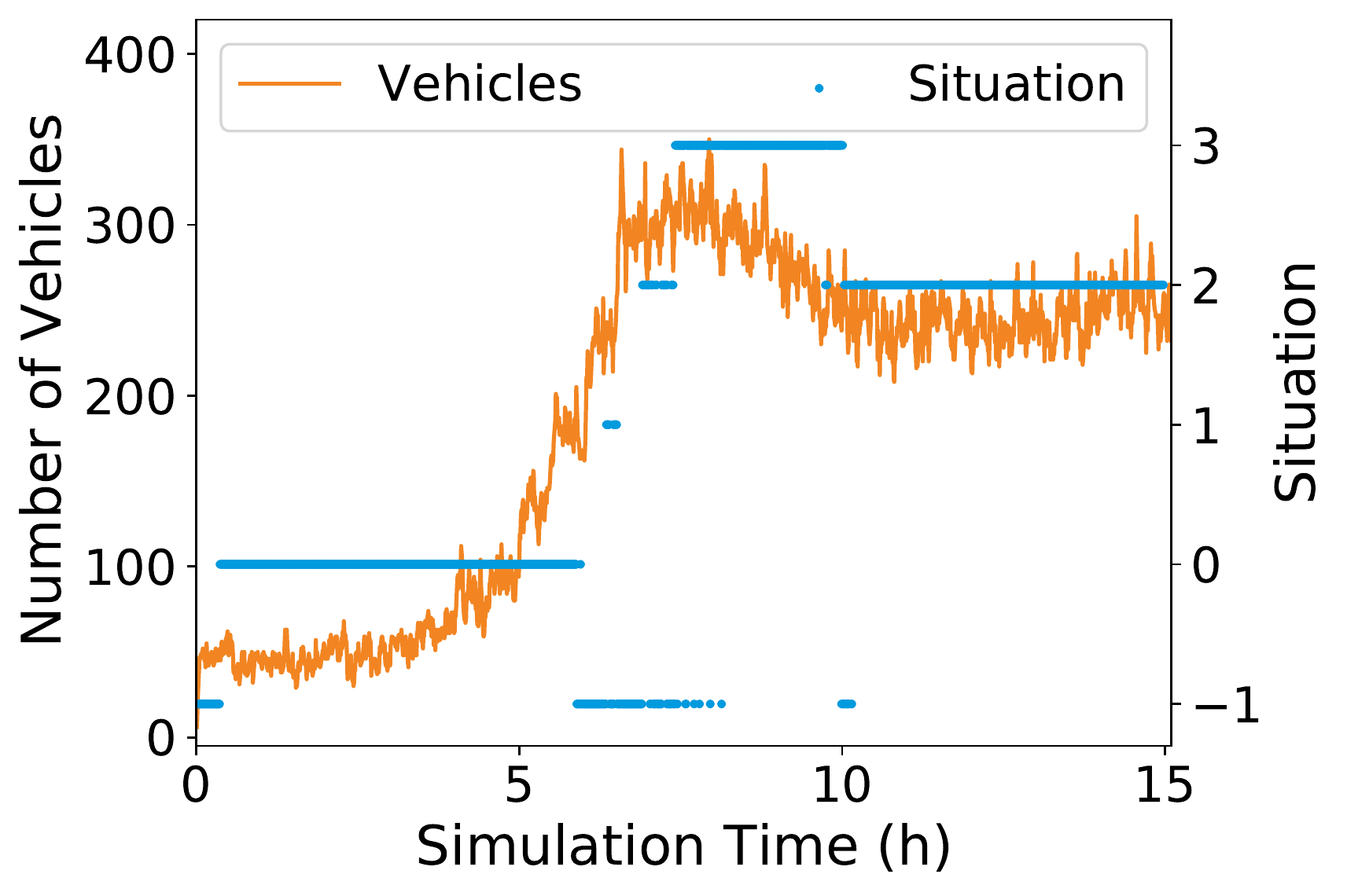}
  	\label{fig:framework_sitdet_wed_optics}
  }  
  \caption[Detected situations for Wednesday traffic data.]{Actual situations of the ground truth and detected situations of the rule-based and OPTICS approach for Wednesday traffic data. The orange line represents the vehicle spawn rate at a specific point in time. The blue dots represent the detected situation at the current point in time incorporating all previously observed data points.}
  \label{fig:framework_sitdet_wed}
\end{figure*}

In line with the workflow of our optimization framework, we start our evaluation with the situation detection component.
Keep in mind, that this component uses the current amount of vehicles on the road to identify a situation.
Therefore, we analyze the detected situations during the simulation for both scenarios and compare the rule-based and OPTICS approaches to the ground truth. 
\cref{fig:framework_sitdet_wed} shows the ground truth for situation detection and the results of the component applied to the Wednesday scenario.
The orange line represents the vehicle spawn rate, while the blue dots represent the cluster ID, that is, the detected situation, at a given time.
A feature of clustering algorithms such as OPTICS is that the identified clusters and observations assigned to them might change as new measurements considered. 
This can cause the cluster IDs for an observation to change over time, which is the motivation of our ongoing model learning approach in the situation detection component~(c.f. \cref{sec:framework:architecture}).
However, this behavior is not part of the illustration in \cref{fig:framework_sitdet_wed}.
The figure shows the cluster numbers assigned when the observation first occurred.
This represents the situation based on which the framework makes its decisions.
The figures~\ref{fig:framework_sitdet_wed_truth} and \ref{fig:framework_sitdet_wed_rules} show that the rule-based situation detection component is close to ground truth, as it identifies all three situations, but assigns fewer observations to the peak traffic cluster.
In addition, the rule-based approach does not detect the start of the second peak traffic cluster.
The good performance of this approach was expected since the rules were derived from the ground truth.
The situation detection using OPTICS, as shown in \cref{fig:framework_sitdet_wed_optics}, identifies the situations using clustering mechanisms and identifies four different situations, but is not able to cluster all observations denoted by cluster ID -1.
The four identified situations are less evenly distributed in terms of the observations they contain, as cluster number one contains only a few observations.
Nevertheless, this mechanism is able to distinguish different situations even if they do not completely consistent with the ground truth.

\begin{figure*}[htb]
  \centering
  \subfloat[Ground truth for the situation detection.]{
  	\includegraphics[width=0.3\textwidth]{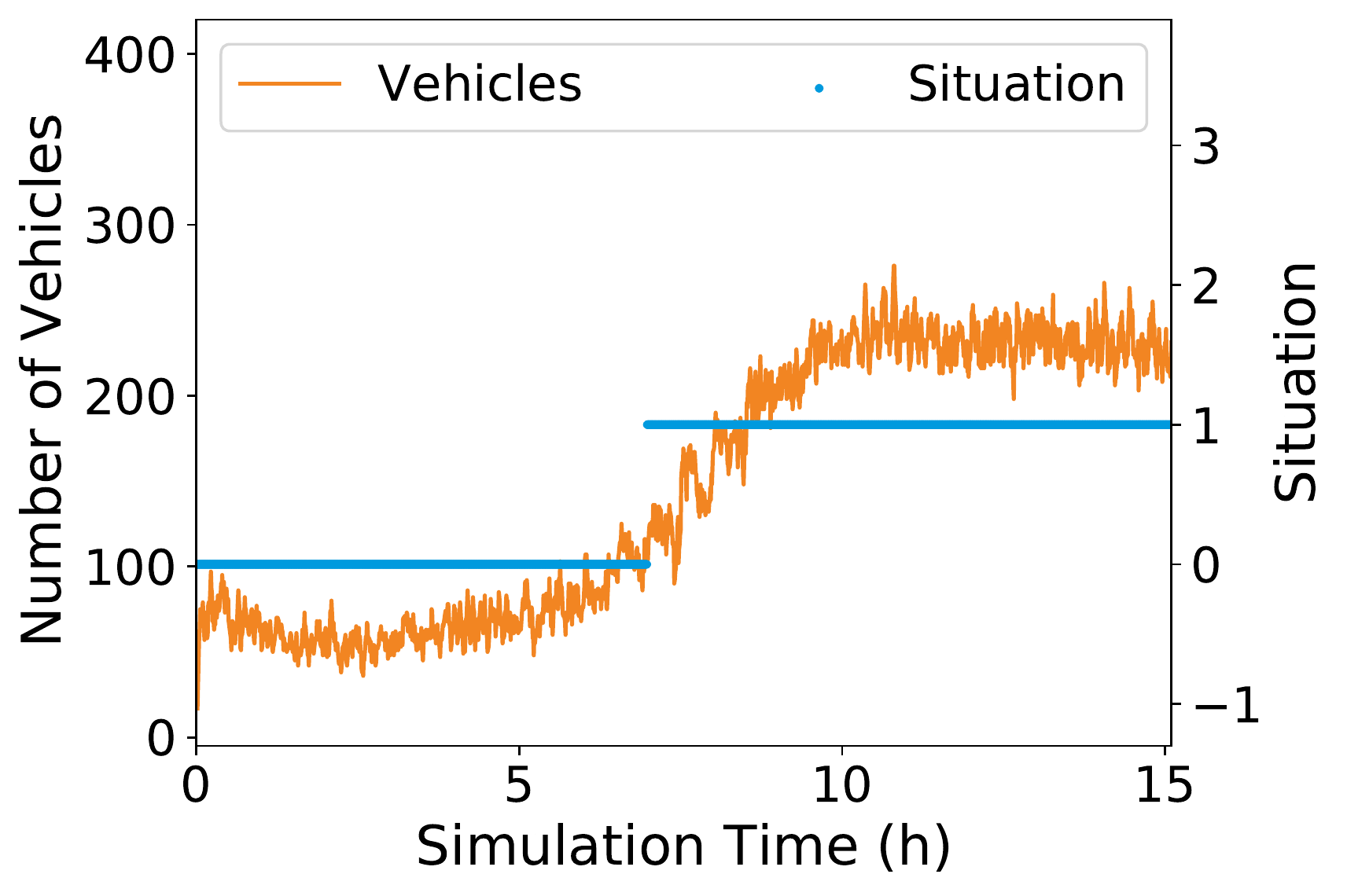}
  	\label{fig:framework_sitdet_sat_truth}
  }
  \hfill
  \subfloat[Detected situations when applying rule-based situation detection.]{
  	\includegraphics[width=0.3\textwidth]{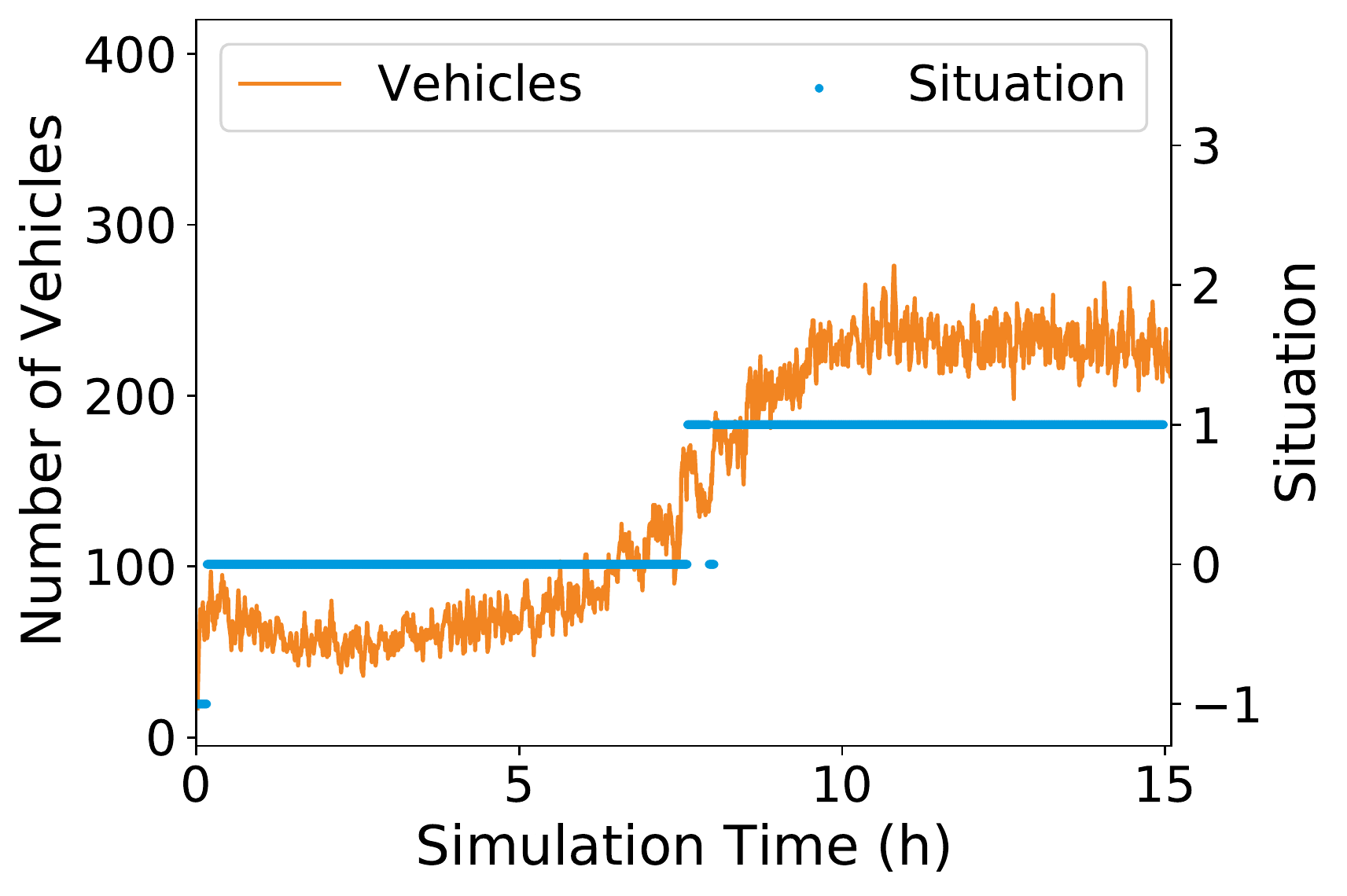}
  	\label{fig:framework_sitdet_sat_rules}
  }
  \hfill  
  \subfloat[Detected situations when applying OPTICS situation detection.]{
  	\includegraphics[width=0.3\textwidth]{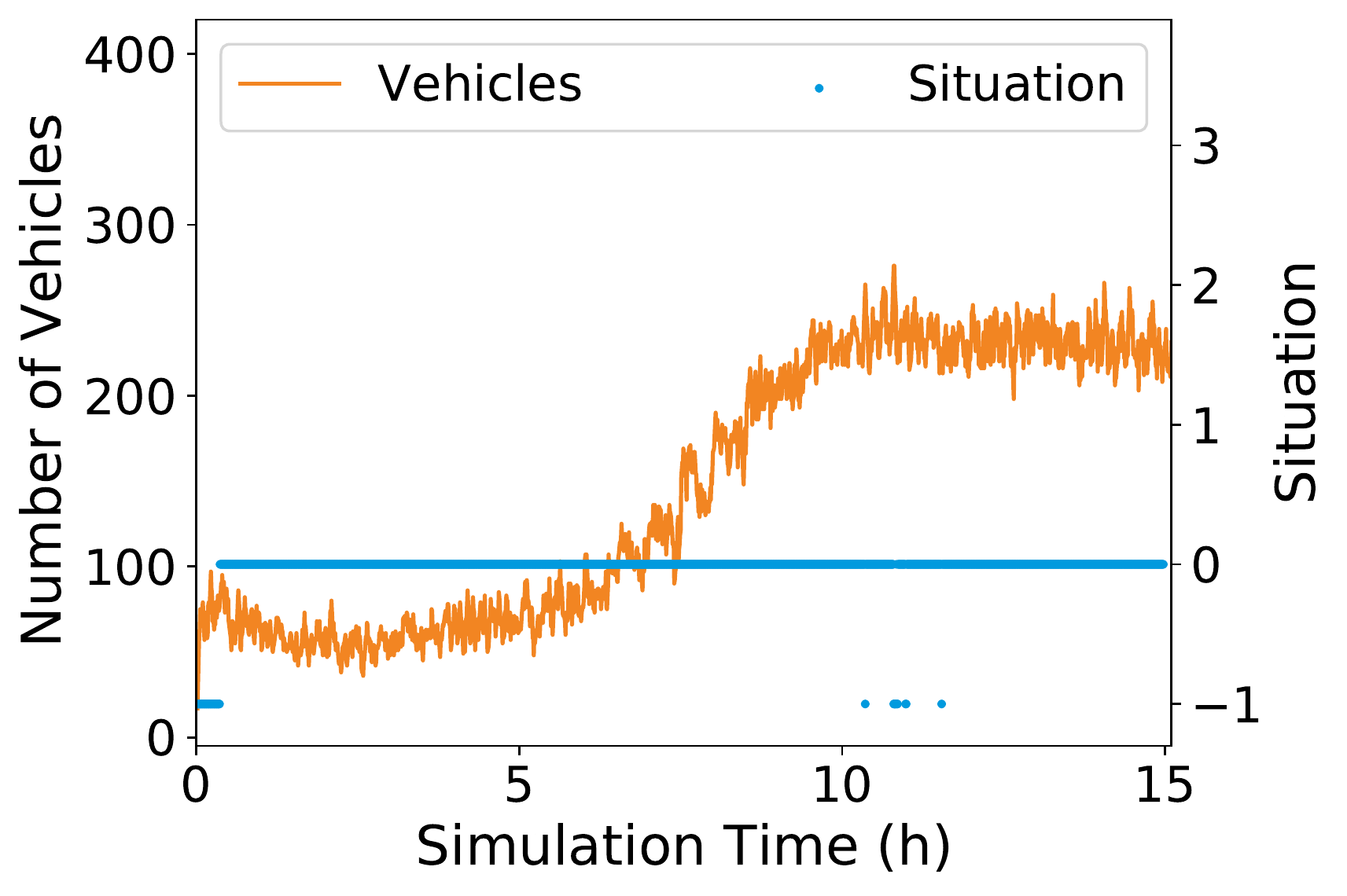}
  	\label{fig:framework_sitdet_sat_optics}
  }  
  \caption[Detected situations for Saturday traffic data.]{Actual situations of the ground truth and detected situations of the rule-based and OPTICS approach for Saturday traffic data. The orange line represents the vehicle spawn rate at a specific point in time.}
  \label{fig:framework_sitdet_sat}
\end{figure*}

The results of the situation detection component applied to the Saturday scenario are depicted in \cref{fig:framework_sitdet_sat}.
Again, the orange line represents the vehicle spawn rate and the blue dots represent the identified cluster ID.
Similarly to the Wednesday scenario, the rule-based approach is close to the ground truth, which is not surprising since the rules were derived from the ground truth.
However, the OPTICS approach shows a different behavior as it is not able to identify at least two different situations and combines all observations into one situation.
The poor performance of this approach could be due to an unfavorable parameter configuration resulting from our preliminary parameter study.
Another factor could be the lower number of vehicles on the road compared to the Wednesday scenario, which could lead to similar observation data.
Further evaluation using more extensive scenarios and additional parameter studies may provide more insight in the future.

In summary, this evaluation shows that the rule-based approach performs well against the defined ground truth for both scenarios.
The OPTICS approach identifies distinct situations in the Wednesday scenario, but only a single situation for the Saturday scenario. 
The ground truth derived rules work predictably well, but are a very rigid approach and do not provide flexibility for future changes. 
A rule set must be defined at design time using expert knowledge and will not be further adapted. 
On the other hand, the clustering approach OPTICS provides more flexibility, but does not find the situations defined in the ground truth as reliably. 
For the future, extended simulations with, for example, several days could reveal more potential for improvements.
In addition, rule learning methods could be used to adapt the rule-based situation detection during runtime.

\subsection{Evaluation of the Strategy Selection Component}
\label{sec:eval:framework_strsel}
\begin{figure*}[tb]
  \centering
  \subfloat[Selected Strategies when using the OPTICS situation detection and Hypervolume trigger.]{
  	\includegraphics[width=0.35\textwidth]{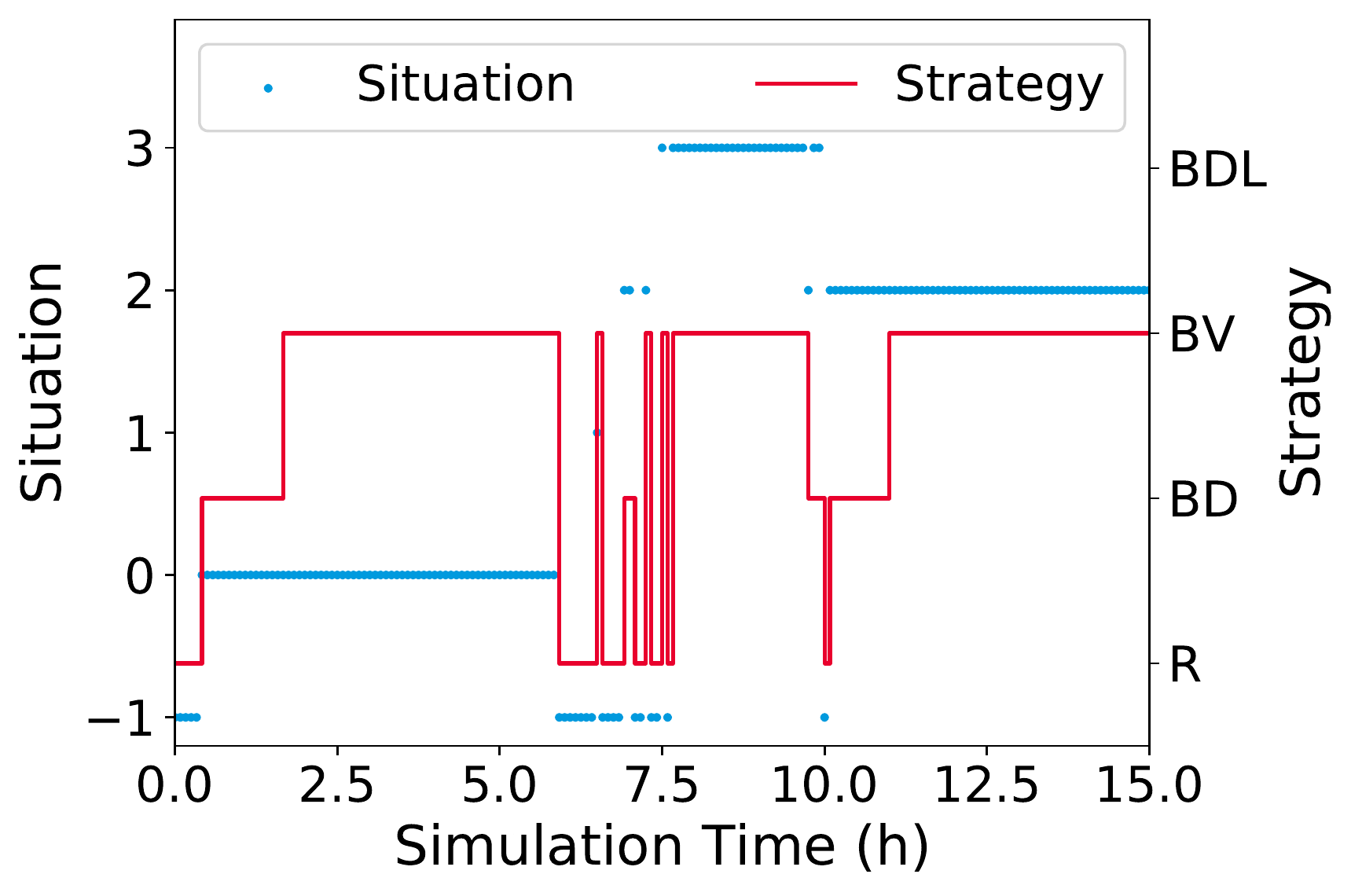}
  	\label{fig:eval:fraemwork_strsel_wed_hv_opt}
  }
  \hspace{2cm}
  \subfloat[Selected Strategies when using the OPTICS situation detection and individual threshold triggers.]{
  	\includegraphics[width=0.35\textwidth]{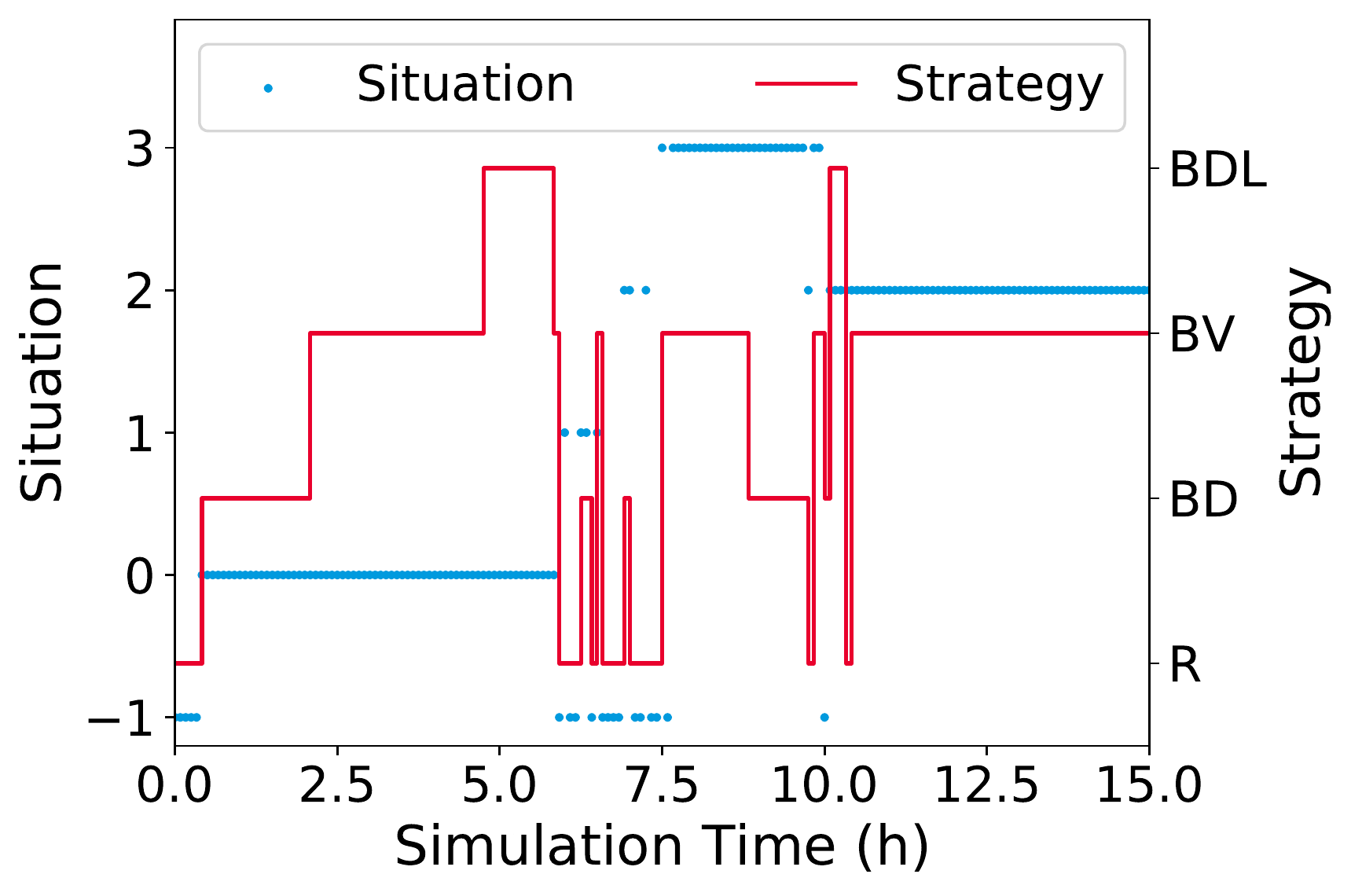}
  	\label{fig:eval:framework_strsel_wed_th_opt}
  }
  \caption[Strategy selection on Wednesday traffic data.]{Strategy selection on Wednesday traffic data. Blue points represent the detected situation at a specific point in time. The red line represents the selected adaptation planning strategy at a specific point in time. (R~=~\textit{Rules}, BD~=~\textit{BestDistance}, BV~=~\textit{BestVelocity}, and BDL~=~\textit{BestDistanceAndLane})}
  \label{fig:eval:framework_strsel_wed}
\end{figure*}

The second component of the framework, the strategy selection component, receives the detected situation and selects the most promising strategy to be applied in the adaptation planning system. 
In this section, we analyze the proper operation and performance of the strategy selection component.
Therefore, \cref{fig:eval:framework_strsel_wed} shows the selected strategies for the Wednesday scenario using OPTICS as the situation detection mechanism and the Hypervolume trigger in \cref{fig:eval:fraemwork_strsel_wed_hv_opt} as well as the individual thresholds as trigger in \cref{fig:eval:framework_strsel_wed_th_opt}.
We decided to use continuous line charts with vertical lines representing a strategy change to better visualize the changed strategies especially in cases where the selection changes back and forth frequently.
We base this evaluation solely on OPTICS, as it identifies different situations for the Wednesday scenario and is able to handle new situations not defined in a rule set.

The blue points represent the determined situation, while the red line illustrates the selected strategy at a certain point in time, that is, the height of the line represents the selected strategy. 
The left figure shows that the strategy selection component selects a strategy and switches to the next one if the performance metrics fall below the thresholds and the triggers activate the selection.
When using the Hypervolume trigger, the strategy selection remains at the Best-Velocity and does not switch to the Best-Distance-and-Lane within the first six simulation hours compared to the individual threshold trigger.
After this time, the observations are classified as noise by the situation detection, which causes the strategy selection to revert to the rule-based strategy. 
Whenever new situations occur, the strategy selection starts with the Best-Distance and tests its performance before switching to the Best-Velocity strategy.
The results show that the individual thresholds trigger the strategy selection more often than to the Hypervolume trigger because the selection component examines the Best-Distance-and-Lane twice. 
This is the intended behavior of the framework and tells us that it is working properly. 
Also, this may indicate that the individual thresholds are too restrictive and could be relaxed to avoid jitters between strategies.

\begin{figure*}[t] 
  \centering
  \subfloat[Selected Strategies when using the OPTICS situation detection and Hypervolume trigger.]{
  	\includegraphics[width=0.35\textwidth]{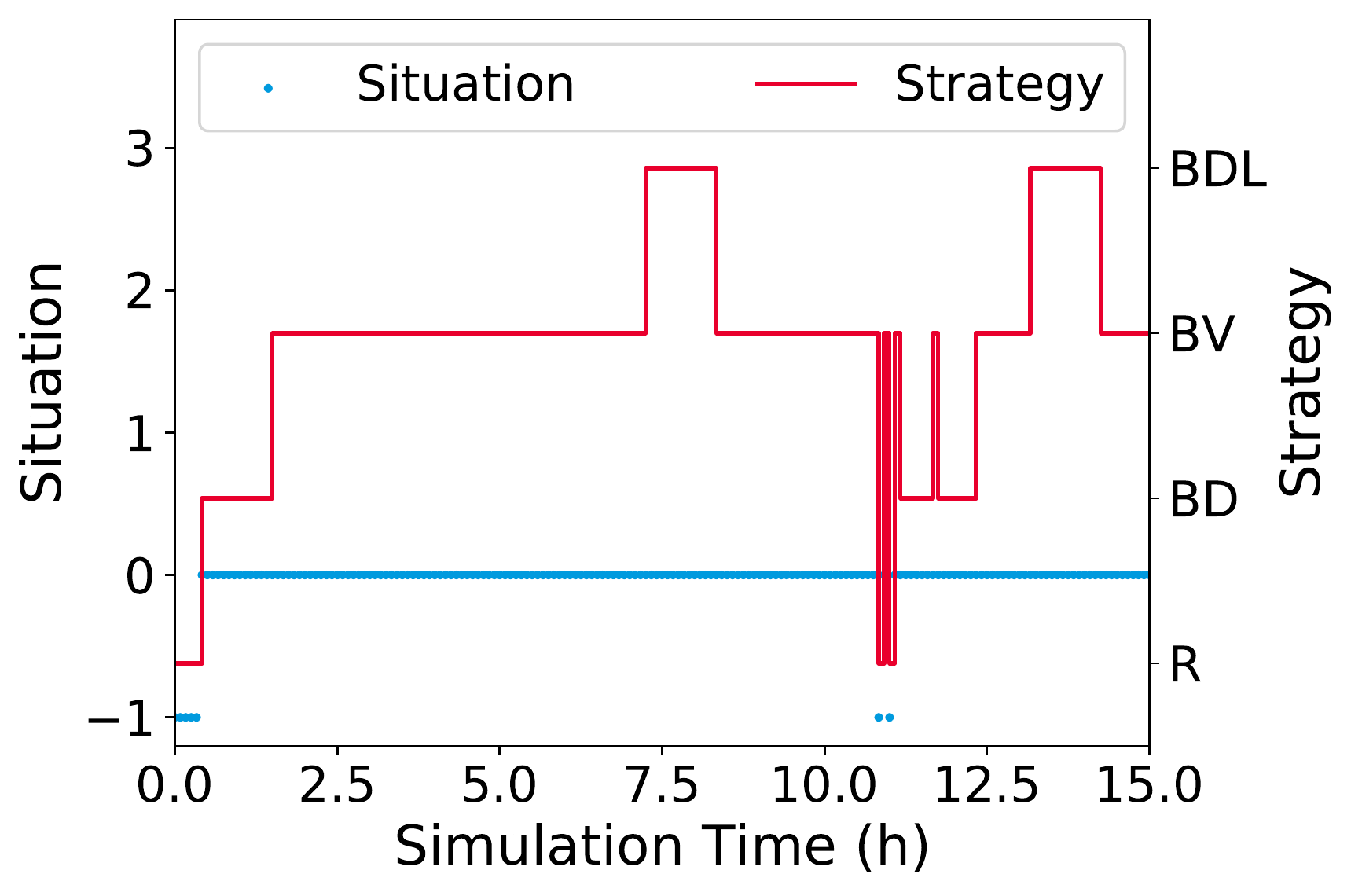}
  	\label{fig:eval:framework_strsel_sat_hv-opt}
  }
  \hspace{2cm}
  \subfloat[Selected Strategies when using the OPTICS situation detection and individual threshold triggers.]{
  	\includegraphics[width=0.35\textwidth]{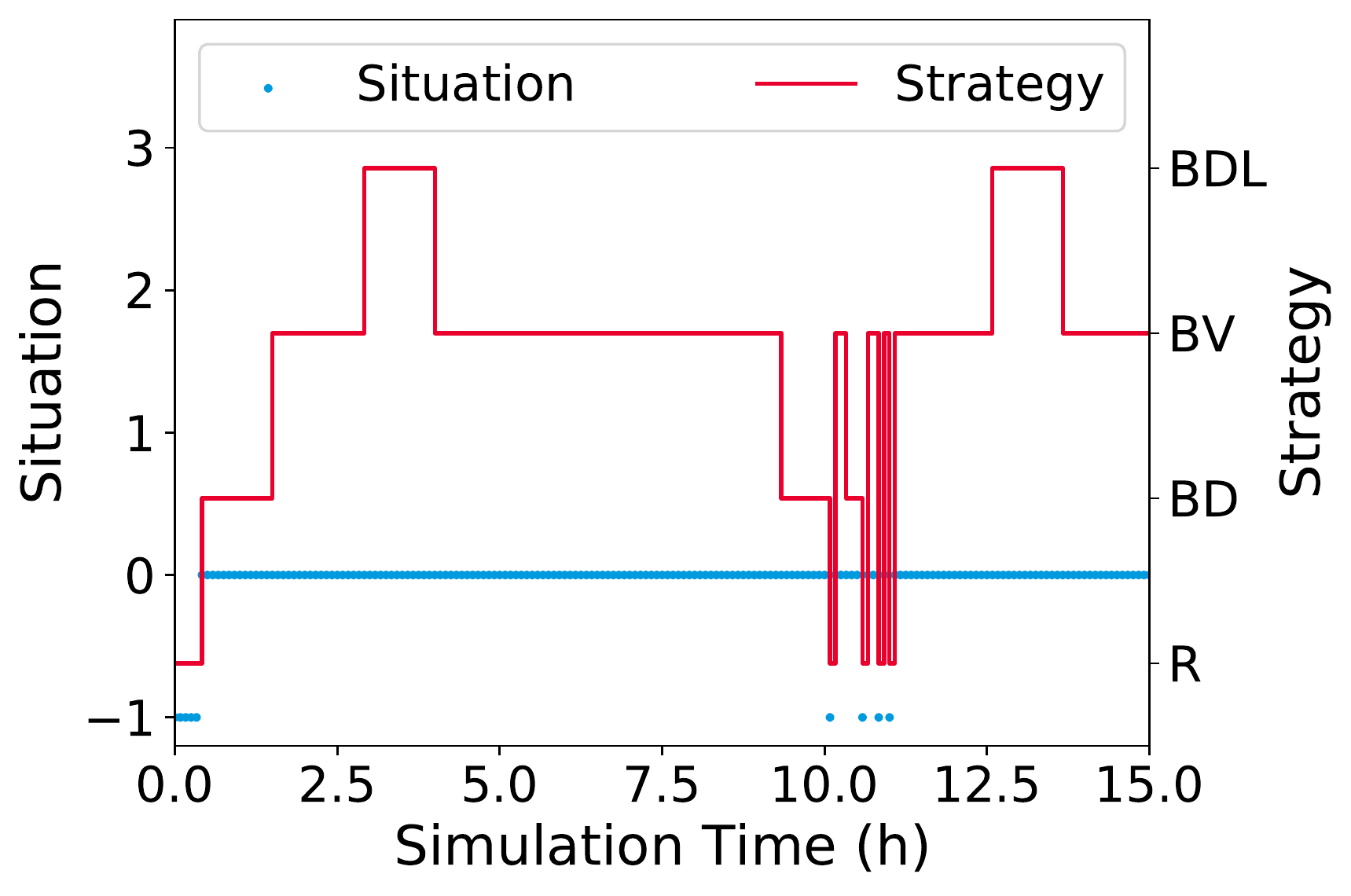}
  	\label{fig:eval:framework_strsel_sat_th-opt}
  }
  \hfill 
  \subfloat[Selected Strategies when using the rule-based situation detection and Hypervolume trigger.]{
  	\includegraphics[width=0.35\textwidth]{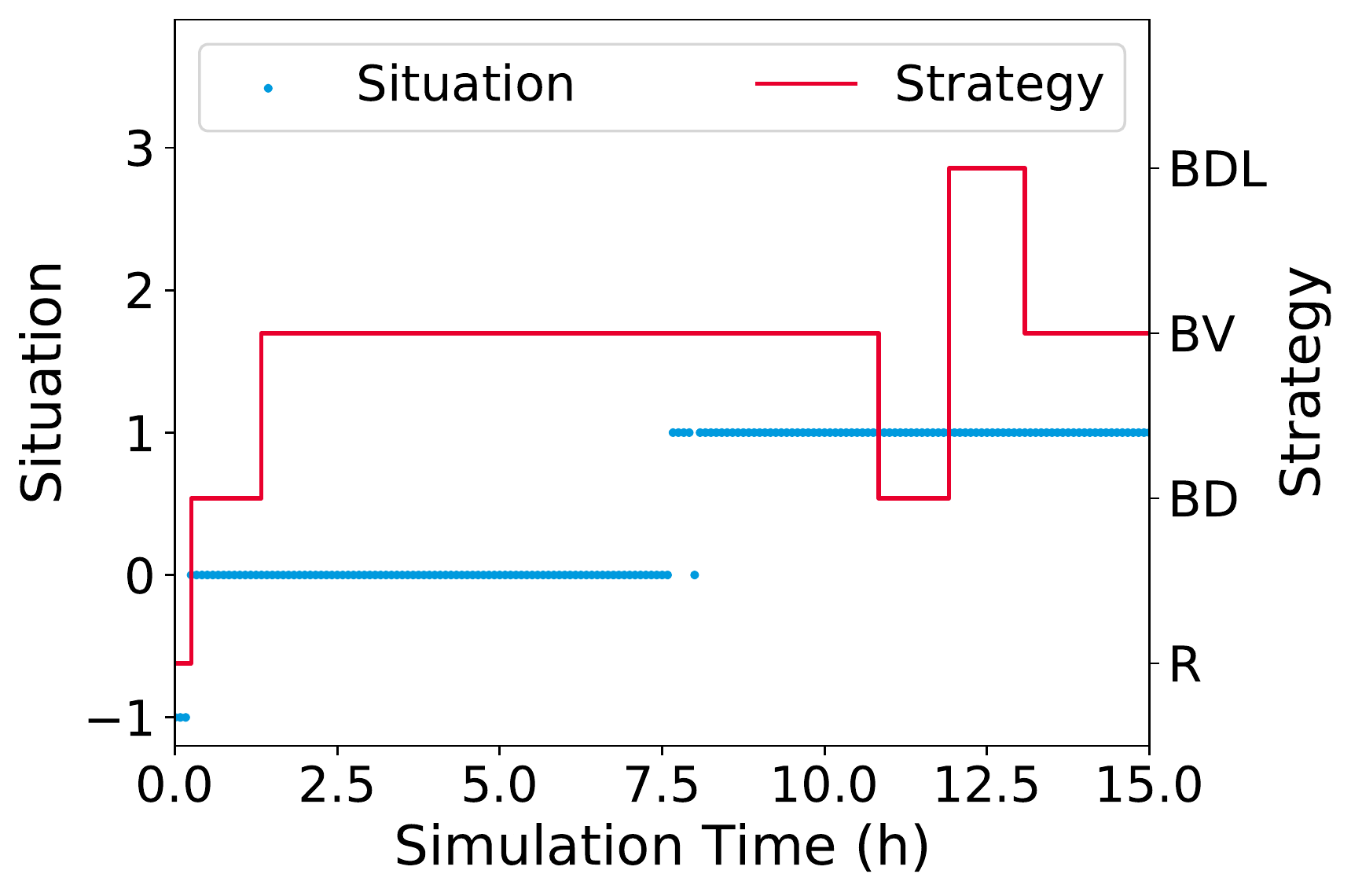}
  	\label{fig:eval:framework_strsel_sat_hv-rules}
  }
  \hspace{2cm}
  \subfloat[Selected Strategies when using the rule-based situation detection and individual threshold triggers.]{
  	\includegraphics[width=0.35\textwidth]{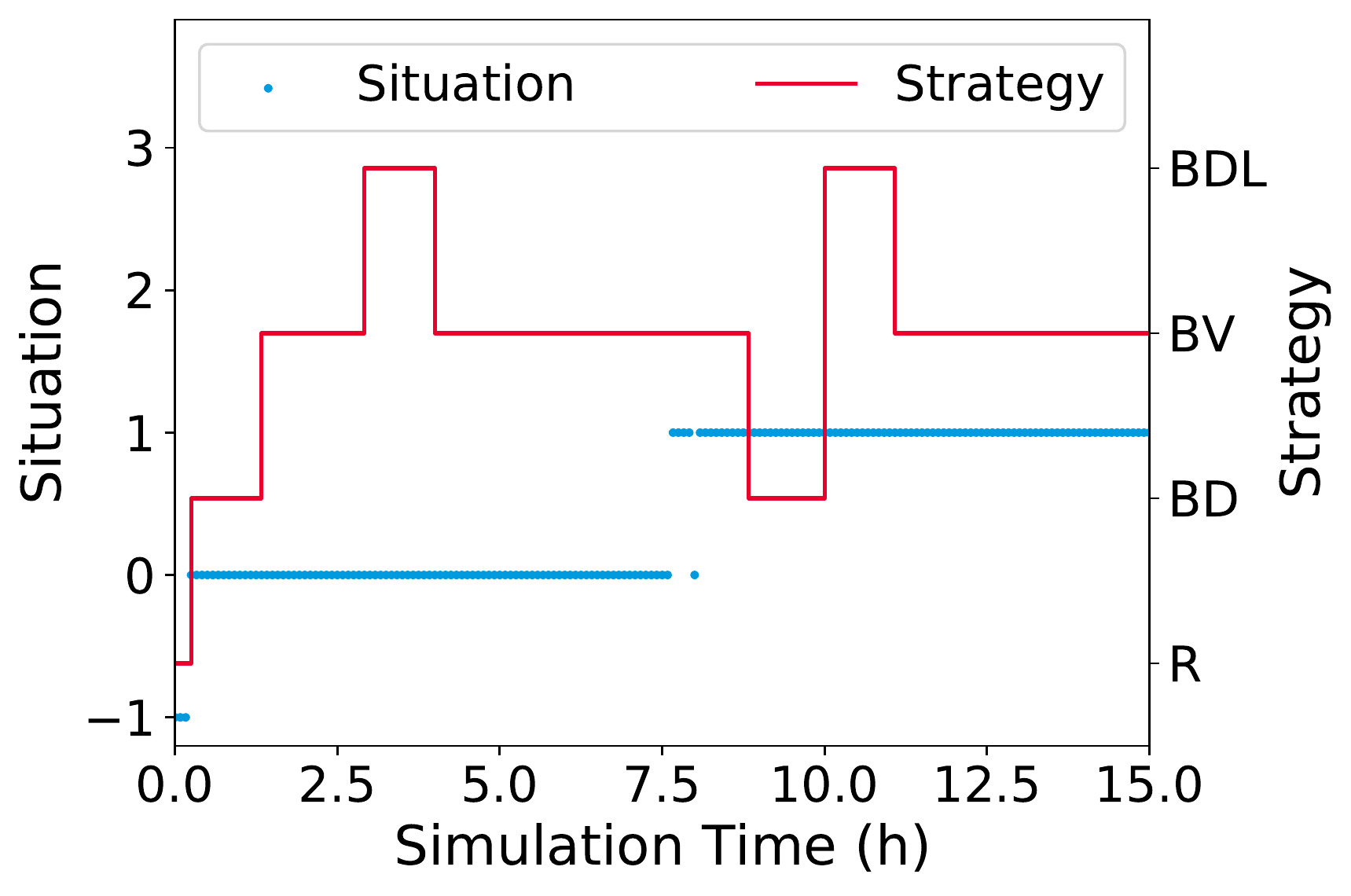}
  	\label{fig:eval:framework_strsel_sat_th-rules}
  }
  \caption[Evaluation of the strategy selection component.]{Strategy selection on Saturday traffic data. Blue points represent the detected situation at a specific point in time. The red line represents the selected adaptation planning strategy at a specific point in time. (R~=~\textit{Rules}, BD~=~\textit{BestDistance}, BV~=~\textit{BestVelocity}, and BDL~=~\textit{BestDistanceAndLane})}
  \label{fig:eval:framework_strsel_sat}
\end{figure*}

\cref{fig:eval:framework_strsel_sat} shows the results of the strategy selection component for the Saturday scenario using OPTICS and rule-based situation detection in combination with the Hypervolume and individual threshold triggers. 
The reason for using the rule-based situation detection in this evaluation is that OPTICS situation detection was not able to identify more than one situation for the Saturday scenario.
\cref{fig:eval:framework_strsel_sat_hv-opt} presents the OPTICS and Hypervolume evaluation, \cref{fig:eval:framework_strsel_sat_th-opt} presents the OPTICS and individual threshold evaluation, \cref{fig:eval:framework_strsel_sat_hv-rules} illustrates the rule-based and Hypervolume evaluation, and \cref{fig:eval:framework_strsel_sat_th-rules} shows the rule-based and individual threshold evaluation.
Again, the blue points represent the identified situation, and the red line represents the selected strategy at a given time. 
All figures show the desired exploratory behavior of the strategy selection when a new situation occurs due to the step-wise strategy change at the beginning. 
If a strategy performs well, it is not replaced and remains active until the triggers indicate a performance degradation. 
Since the OPTICS situation detection identifies only one situation and classifies some observations as noise, it shows a clear step-wise strategy change and a reversion to the rule-based strategy when the situation detection reveals noise. 
When using the rule-based situation detection, the strategy selection is more stable since no fallback mechanisms are required. 
However, \cref{fig:eval:framework_strsel_sat_hv-rules} shows an anomaly in the strategy selection behavior, as the detection of a new situation does not trigger a new exploration of strategies after around eight hours. 
A detailed analysis of this behavior led us to the conclusion that the detection of a situation change was not perfectly aligned with the strategy selection component and, hence, resulted in a lost situation change.
Thus, the currently active strategy, that is, the Best-Velocity, remains active until about eleven hours of simulation time. 
At this point, the Hypervolume trigger indicates a performance degradation of the current strategy and the strategy selection selects the Best-Distance strategy. 
However, it is discarded after the initial trial period and the strategy selection switches to the Best-Distance-and-Lane strategy. 
The same lost update of a new situation can be observed in \cref{fig:eval:framework_strsel_sat_th-rules}.
However, this figure shows a faster discarding of the currently active strategy, similar to the behavior in \cref{fig:eval:framework_strsel_sat_th-opt}.
This also indicates that the individual thresholds might be too restrictive and could be relaxed in the future to produce a more stable result.

In summary, this evaluation shows that both algorithm selection trigger methods work properly and activate the algorithm selection when the performance of the currently active strategy deteriorates.
While the Hypervolume threshold provides a more stable result, the individual thresholds appear to detect performance degradation earlier.
Therefore, the individual thresholds explore more possible strategies, but also result in higher jitter compared to the Hypervolume.
However, the definition of the individual thresholds can be adjusted in future evaluation studies to achieve a trade-off between detecting performance degradation quickly and reducing jitter.
All in all, both methods work properly and are capable of triggering the algorithm selection.

\subsection{Evaluation of the Parameter Optimization Component}
\label{sec:eval:framework_opt}
\begin{figure*}[htb]
\centering
  \subfloat[][Selected Strategies when using the rule-based situation detection and Hypervolume trigger.]{
  	\includegraphics[width=0.35\textwidth]{figures/framework_strsel_sat-hv-rules.pdf}
  	\label{fig:eval:framework_opt_sat_hv-strsel}
  }  
  \hspace{2cm}
  \subfloat[][Selected Strategies when using the OPTICS situation detection and individual threshold triggers.]{
  	\includegraphics[width=0.35\textwidth]{figures/framework_strsel_sat-th-opt.pdf}
  	\label{fig:eval:framework_opt_sat_th-strsel}
  }

  \subfloat[][Hypervolume score of the selected strategy when using the rule-based situation detection and Hypervolume trigger.]{
  	\includegraphics[width=0.35\textwidth]{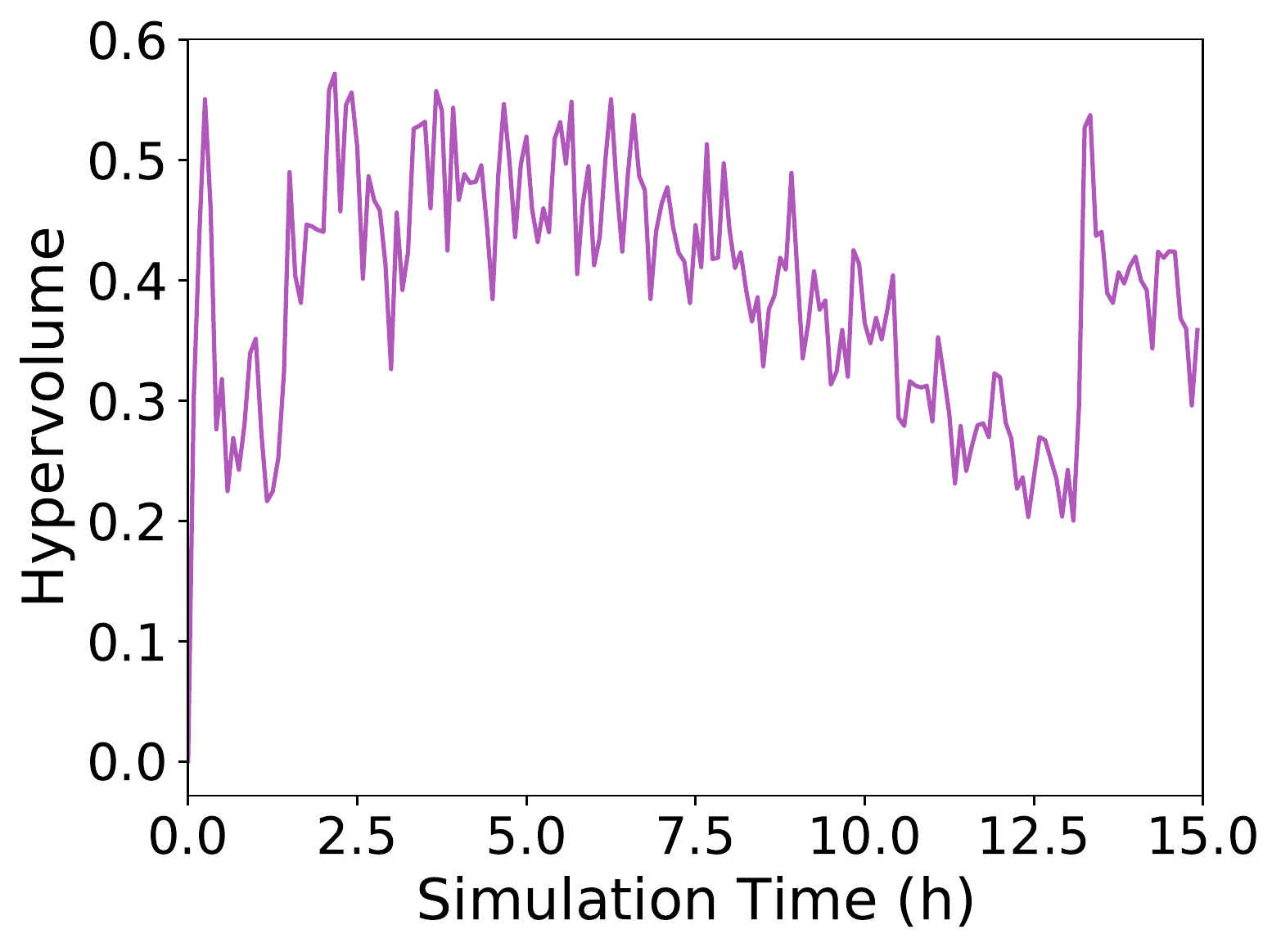}
  	\label{fig:eval:framework_opt_sat_hv-opt}
  }
  \hspace{2cm}
  \subfloat[][Hypervolume score of the selected strategy when using the OPTICS situation detection and individual threshold triggers.]{
  	\includegraphics[width=0.35\textwidth]{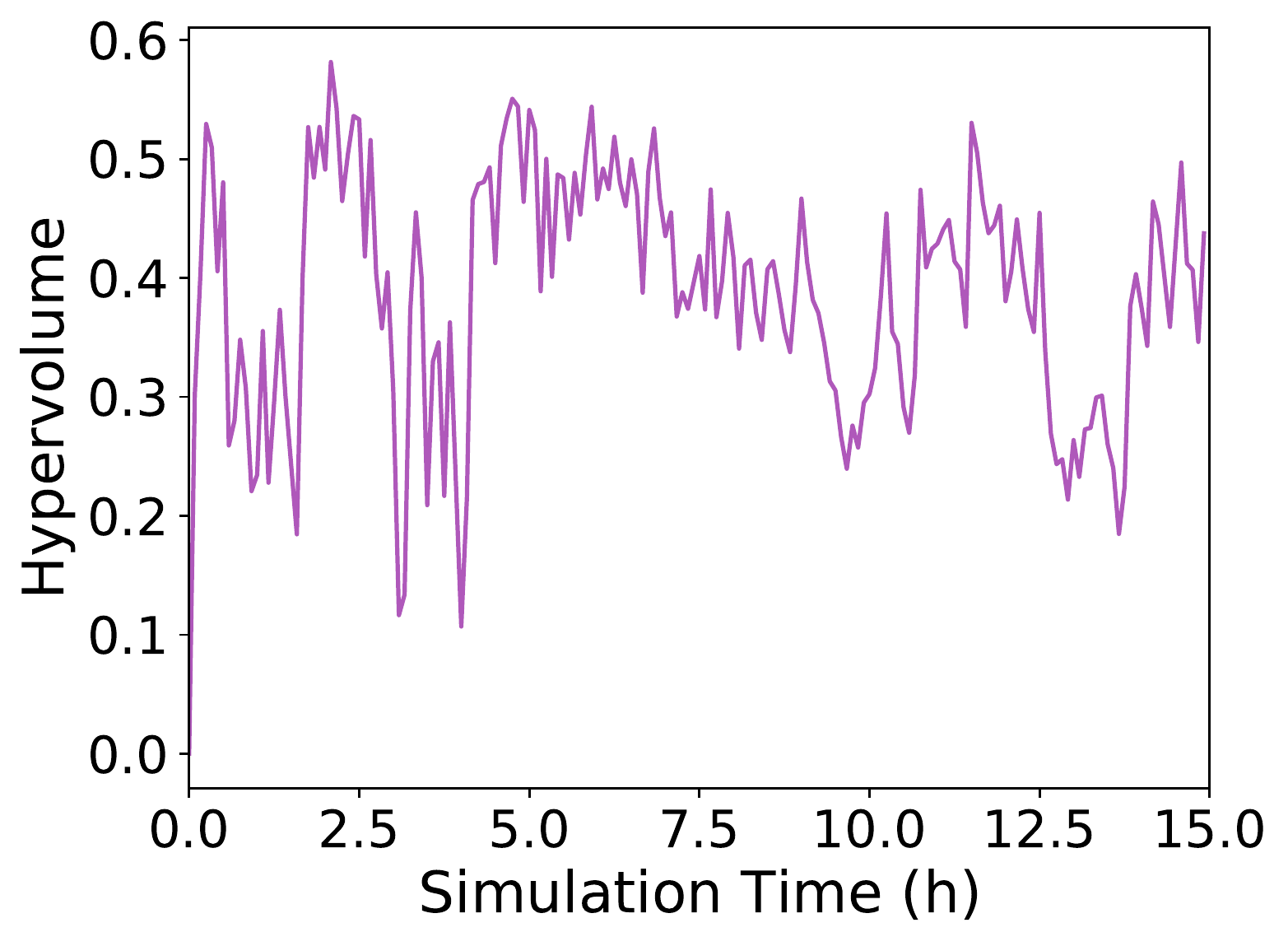}
  	\label{fig:eval:framework_opt_sat_th-opt}
  }
  \caption[Evaluation of the optimization component.]{Evaluation of the optimization component on the Saturday scenario. The left side represents configurations using the rule-based situation detection and Hypervolume triggers. The right side illustrates OPTICS situation detection and individual threshold triggers (R~=~\textit{Rules}, BD~=~\textit{BestDistance}, BV~=~\textit{BestVelocity}, and BDL~=~\textit{BestDistanceAndLane}).}
  \label{fig:eval:framework_opt_sat}
\end{figure*}

We evaluate the performance of our optimization component by analyzing the course of the Hypervolume metric used by this component to optimize the parameter configuration of the current adaptation planning strategy. 
The Hypervolume metric~(c.f.~\cite{Wang2016}) accumulates the platooning metrics into one objective metric that can be used by the single-objective Bayesian Optimization.
\cref{fig:eval:framework_opt_sat} shows evaluations of the Saturday scenario using rule-based situation detection and Hypervolume as trigger for the strategy selection component on the left~(\cref{fig:eval:framework_opt_sat_hv-strsel} and \cref{fig:eval:framework_opt_sat_hv-opt}).
The right side of the figure shows measurements for the Saturday scenario using OPTICS as situation detection mechanism and individual thresholds as triggers for strategy selection~(\cref{fig:eval:framework_opt_sat_th-strsel} and \cref{fig:eval:framework_opt_sat_th-opt}).
The top figures show the identified situations in blue in combination with the selected strategies in red.
The lower figures summarize the course of the Hypervolume metric, that is, the performance indicator of the platooning coordination strategy. 
The course of the Hypervolume metric appears to be very fluctuating for both configurations during the simulation time.
This was to be expected, since the optimization component needs some time to learn which parameter setting works well for which strategy and situation.
Therefore, it makes most sense to analyze time windows of the Hypervolume progression where the identified situation and strategy remain stable.
This is also a reason for choosing Saturday scenarios for this evaluation, as traffic volumes do not fluctuate as much as in Wednesday scenarios, which allows for longer time frames per situation and strategy.
When analyzing the first stable phase on the left between 2.5 and 7.5~hours of simulation time, the Hypervolume starts with a value of about 0.5~Hypervolume points and drops to 0.3~Hypervolume points.
Then, it stabilizes back to about 0.5~Hypervolume points, indicating that the optimization component has explored different parameter settings and stabilized to a well performing set of parameters.
As discussed earlier, the change in  the situation is lost at about 7.5~hours of simulation time, resulting in a sharply decreasing trend in the Hypervolume. 
This leads to the extended Hypervolume threshold that triggers the strategy selection at about 11~hours of simulation time.
The other configuration, depicted on the right of the figure, captures OPTICS and individual thresholds. 
In this evaluation, we can analyze the Hypervolume score for the simulation period starting at four hours up to eight hours of simulation time. 
The Hypervolume score shown on the bottom right starts at a low value of around 0.2 score points, but quickly increases to a value of 0.4 score points. 
This low start value is due to the recent strategy change from the Best-Distance-and-Lane strategy which was discarded in favor of the Best-Velocity strategy after its initial trial phase. 
After that, the Hypervolume score shows a slight increase to a value of about 0.58 score points, but then decreases again to values between 0.4 and 0.5 score points. 
This indicates, that the Optimization component finds better parameter settings for the selected strategy and then explores new parameter settings that unfortunately lead to worse Hypervolume values.
This triggers the strategy selection, and since all existing strategies have already been explored, the fallback rules take place. 

In summary, this evaluation shows us that the Optimization component has the potential to optimize the parameter settings of the adaptation planning strategies, as the Hypervolume score remains stable and shows slight increases in stable situations for situation and selected strategy.
However, negative effects also occur when the Optimization component explores new parameter settings, which may lead to worse results compared to the previous settings that performed well. 
This indicates that the stable phases of identified situations and selected strategies, that is, the time for the Optimization component to optimize the parameter settings, may be too short to find stable configurations with good performance. 
Extended evaluations over several days or even weeks could provide more insight into the required amount of experience for the Optimization component and increase the overall performance of this component.

\subsection{Evaluation of the Entire Framework}
\label{sec:eval:framework_entire}
In our final evaluation, we analyze the overall performance of the framework.
\begin{table*}[htb]
\small
\centering
\caption[Evaluation summary of the performance metrics of the framework for the Wednesday scenario.]{Evaluation summary of the average and standard deviation for performance metrics throughput, time loss, platoon utilization, and platoon time for the Wednesday scenario. The best values are shown in bold. (Hv = Hypervolume, Th = Threshold)}
\label{tab:eval:framework_wed}
\begin{tabular}{l c rr c rr c rr c rr}
\toprule

\textbf{Configuration} && \multicolumn{2}{c}{\textbf{Throughput}} && \multicolumn{2}{c}{\textbf{Time Loss}} && \multicolumn{2}{c}{\textbf{Platoon Utilization}} && \multicolumn{2}{c}{\textbf{Platoon Time}} \\ 

\cmidrule{3-4} \cmidrule{6-7} \cmidrule{9-10} \cmidrule{12-13}
                                && mean     & std   && mean     & std   && mean     & std   && mean     & std \\ \midrule
Best Distance                   && \textbf{0.9952}   & 0.0   && 0.8992   & 0.0   && 0.6251   & 0.0   && 0.4908   & 0.0 \\ 
Best Velocity                   && 0.9942   & 0.0   && \textbf{0.9199}   & 0.0   && 0.6973   & 0.0   && 0.6109   & 0.0 \\
Fallback Rules                  && 0.9950   & 0.0   && 0.9198   & 0.0   && \textbf{0.7176}   & 0.0   && \textbf{0.6518}   & 0.0 \\ \midrule
OPTICS \& Hv              && 0.9943   & 0.0003 && \textbf{0.9122}  & 0.0022 && \textbf{0.6690}  & 0.0030 && \textbf{0.5442}  & 0.0090 \\
Rule-based \& Hv                && \textbf{0.9946}   & 0.0004 && 0.9102  & 0.0011 && 0.6647  & 0.0039 && 0.5302  & 0.0076 \\
OPTICS \& Th              && 0.9945   & 0.0003 && 0.9110  & 0.0014 && 0.6566  & 0.0072 && 0.5275  & 0.0119 \\
Rule-based \& Th                && 0.9943   & 0.0003 && 0.9108  & 0.0003  && 0.6343  & 0.0109 && 0.5005  & 0.0083 \\ \bottomrule
\end{tabular}
\end{table*}
\begin{table*}[htb]
\small
\centering
\caption[Evaluation summary of the performance metrics of the framework for the Saturday scenario.]{Evaluation summary of the average and standard deviation for performance metrics throughput, time loss, platoon utilization, and platoon time for the Saturday scenario. The best values are shown in bold. (Hv = Hypervolume, Th = Threshold)}
\label{tab:eval:framework_sat}
\begin{tabular}{l c rr c rr c rr c rr}
\toprule

\textbf{Configuration} && \multicolumn{2}{c}{\textbf{Throughput}} && \multicolumn{2}{c}{\textbf{Time Loss}} && \multicolumn{2}{c}{\textbf{Platoon Utilization}} && \multicolumn{2}{c}{\textbf{Platoon Time}} \\ 

\cmidrule{3-4} \cmidrule{6-7} \cmidrule{9-10} \cmidrule{12-13}
                                && mean     & std           && mean     & std           && mean     & std       && mean     & std \\ \midrule
Best Distance                   &&  0.9945  & 0.0           && 0.9255  & 0.0            && 0.5999   & 0.0           && 0.4522   & 0.0 \\ 
Best Velocity                   && \textbf{0.9951}   & 0.0  &&  \textbf{0.9411}  & 0.0   && 0.6942   & 0.0          && 0.5833   & 0.0 \\
Fallback Rules                  && 0.9950  & 0.0        && 0.9401   & 0.0               &&  \textbf{0.7101}  & 0.0   && \textbf{0.6199}  & 0.0 \\ \midrule
OPTICS \& Hv              && 0.9949   & 0.0001            && 0.9309  & 0.0004 && 0.6360  & 0.0019 && 0.4918  & 0.0022 \\
Rule-based \& Hv                && \textbf{0.9950} & 0.0001                && 0.9297  & 0.0013 && 0.6367  & 0.0087 && 0.4880  & 0.0137 \\
OPTICS \& Th              && \textbf{0.9950}   & 0.0000            && 0.9323  & 0.0012 && \textbf{0.6511}  & 0.0065 && \textbf{0.5169}  & 0.0159 \\
Rule-based \& Th                && \textbf{0.9950}   & 0.0001            && \textbf{0.9333}  & 0.0024 && 0.5677  & 0.0504 && 0.4182  & 0.0520 \\ \bottomrule
\end{tabular}
\end{table*}

First, we compare the four defined configurations of the framework with the three baselines in terms of the four platooning metrics of throughput, time loss, platoon utilization, and platoon time. 
\cref{tab:eval:framework_wed} presents the mean and standard deviation results for these metrics for the Wednesday scenario and \ref{tab:eval:framework_sat} summarizes the results for the Saturday scenario for the three repetitions.
We highlight the best values of each platooning metric for the baseline group and the framework group in bold.
In both evaluation scenarios, the throughput metric results for all baselines and framework configurations are very close, with values between 0.9943 and 0.9952 and low standard deviations.
In the Wednesday scenario, the Best-Distance baseline and rule-based situation detection combined with Hypervolume thresholds perform best on the throughput metric with values of 0.9952 and 0.9946, respectively. 
In the Saturday scenario, all configurations of the framework perform equally well, while the Best-Velocity baseline performs best on the throughput metric with values of 0.9950 and 0.9951, respectively.
All applied configurations and baselines show higher diversity for the time loss metric, ranging from 0.8992 to 0.9122 for Wednesday and from 0.9255 to 0.9411 for Saturday.
Rule-based situation detection combined with individual thresholds performs best for this metric among all configurations tested, with a value of 0.9122 and 0.9333, but achieves a lower value compared to the Best-Velocity baseline, with a value of 0.9199 and 0.9411 for Wednesday and Saturday, respectively.
Results for the platoon utilization metric range from 0.6251 to 0.7176 and from 0.5999 to 0.7101 for Wednesday and Saturday, respectively. 
For this metric, the fallback rule baseline among the baselines and the OPTICS situation detection in combination with Hypervolume and individual thresholds perform best.
Finally, the results for the platoon time metric range from 0.4908 to 0.6518 and from 0.4182 to 0.6199 for Wednesday and Saturday, respectively.
Again, the fallback rules baseline performs best for both scenarios, and the OPTICS situation detection with Hypervolume and individual thresholds performs best among the framework configurations.
The combination of the close average values for all metrics and the small standard deviations does not suggest significant advantages for some configurations. 
However, this indicates that the framework performs comparably well when considering the results of the baseline, which was designed and configured with complete prior knowledge based on the preliminary situation-dependency study we published~\cite{Lesch2021Towards}.

\begin{figure}[htb]
\centering
  \includegraphics[width=0.8\columnwidth]{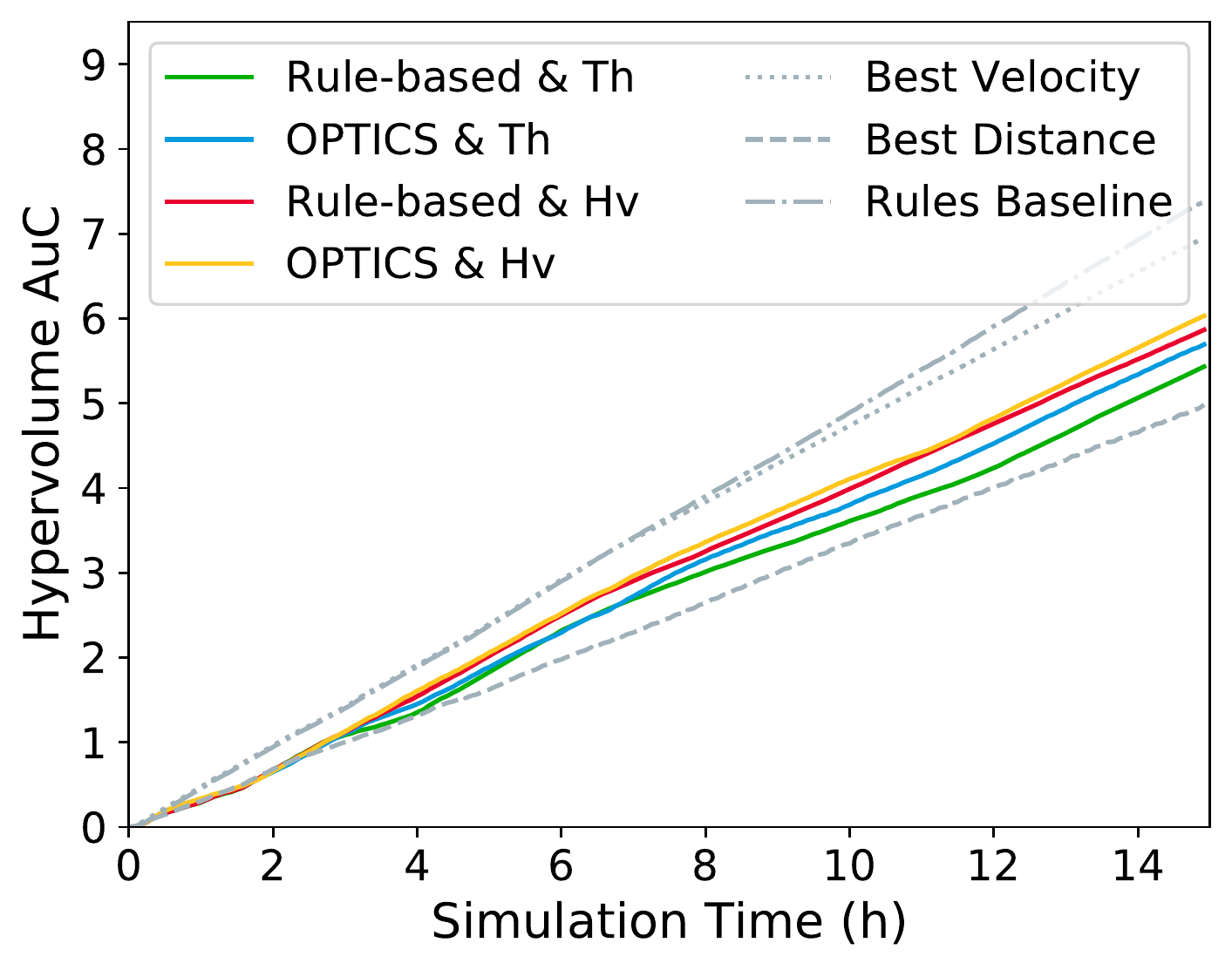}
  \caption[Mean area under curve over time for the Hypervolume score.]{Mean area under curve evaluation over time for the Hypervolume score of all tested configurations and the baselines on the Wednesday scenario. The different colors represent the tested configurations, the x-axis shows the simulation time, and the area under curve is depicted on the y-axis.}
  \label{fig:eval:framework_auc_wed}
\end{figure}

\begin{figure}[htb]
\centering
  \includegraphics[width=0.8\columnwidth]{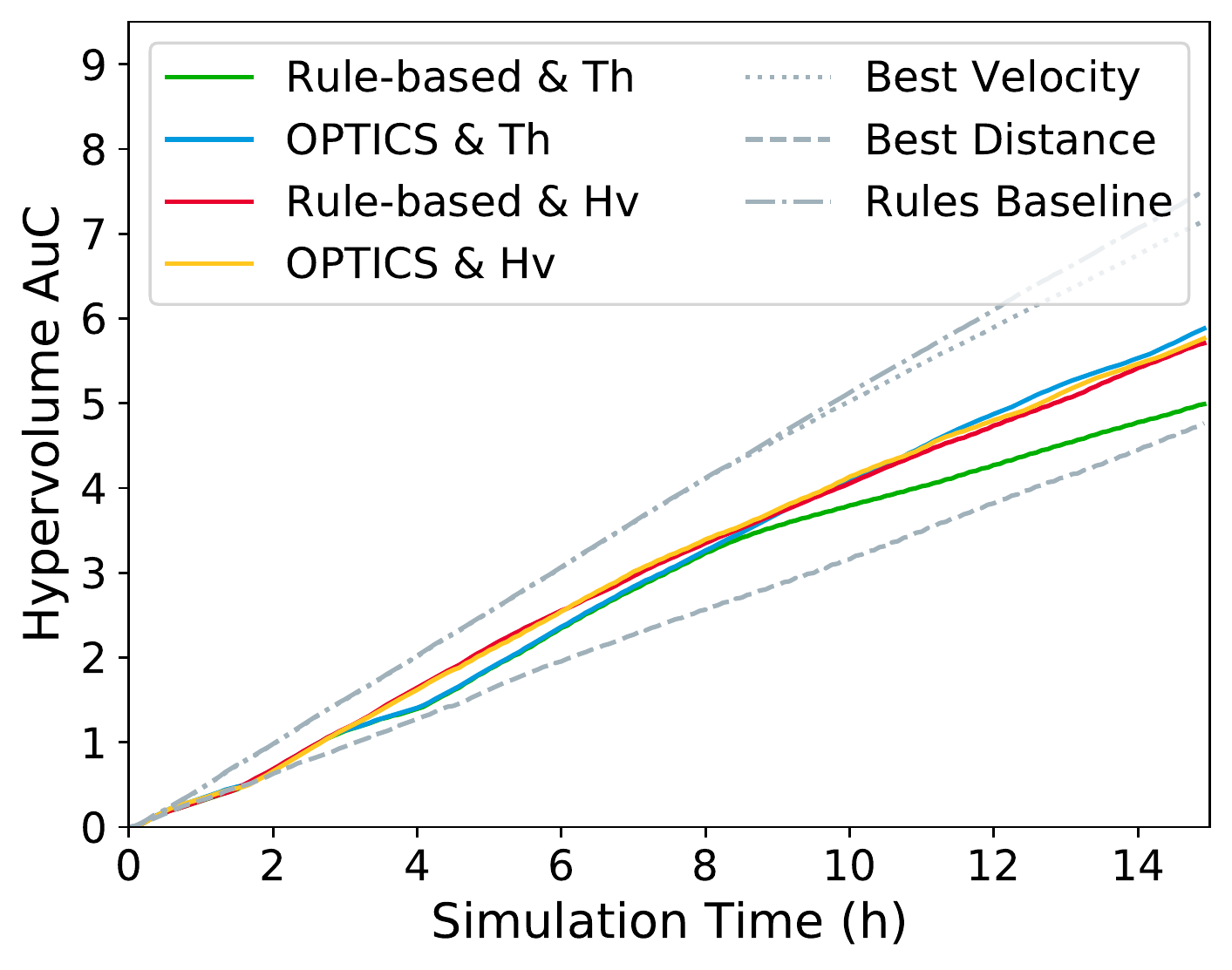}
  \caption[Mean area under curve over time for the Hypervolume score.]{Mean area under curve evaluation over time for the Hypervolume score of all tested configurations and the baselines on the Saturday scenario. The different colors represent the tested configurations, the x-axis shows the simulation time, and the area under curve is depicted on the y-axis.}
  \label{fig:eval:framework_auc_sat}
\end{figure}

In addition to evaluating the individual platooning metrics, we also analyze the progression of the performance over the simulation time.
Therefore, \cref{fig:eval:framework_auc_wed} and \cref{fig:eval:framework_auc_sat} present the mean Hypervolume area under curve over simulation time for all configurations and baseline strategies for Wednesday and Saturday.
The baseline strategies are depicted as gray lines with a dotted line for the Best-Velocity, a dashed line for Best-Distance and a dashed and dotted line for the rules baseline.
The colors represent the different configurations.
Both plots show a similar result: The Best-Velocity and rules baseline perform best, with a stable increasing gradient of the area under curve, while the Best-Distance baseline performs worst.
The curves of the framework configurations do not increase at a constant rate, but show more fluctuations in the gradient.
All lines are close to each other, but more noticeable differences appear as the simulation progresses.
The OPTICS and rule-base situation detection combined with the Hypervolume trigger perform best for Wednesday.
For the Saturday scenario, both configurations perform well again, but OPTICS in combination with individual thresholds outperforms them slightly from ten hours of simulation time. 
For both scenarios, the rule-based situation detection in combination with individual thresholds performs worst of all configurations.

It is not surprising that the Best-Velocity and the rules baseline perform best, since our use case study published in~\cite{Lesch2021Towards} extensively examined existing baseline strategies, their configuration, and their performance in various situations. 
Using this information, we then defined these baseline strategies to represent the best possible performance when complete knowledge of situations, strategies, and configuration was available at design time. 
However, such intensive studies are not feasible, especially in such dynamic, adaptive use cases.
Moreover, it is in the nature of the framework to perform worse than the gold standard, since it needs some time to explore possible strategies and configurations before it can learn and profit from earlier decisions.
The better performance of all framework configurations compared to the Best-Distance baseline shows that the framework is able to identify and select a strategy that works well.
This reduces the need of expert knowledge or extensive case studies for a use case and, hence, provides a valuable contribution to self-aware optimization.

\section{Conclusion}
\label{conclusion}
In today's world, circumstances, processes, and requirements for software systems are becoming increasingly complex.
In order to operate properly in such dynamic environments, software systems must adapt to these changes, which has led to the research area of Self-Adaptive Systems~(SAS).
Platooning is one example of adaptive systems in Intelligent Transportation Systems, which is the ability of vehicles to travel with close inter-vehicle distances.
This technology leads to an increase in road throughput and safety, which directly addresses the increased infrastructure needs due to increased traffic on the roads.
However, the No-Free-Lunch theorem states that the performance of one platooning coordination strategy is not necessarily transferable to other problems.
Moreover, especially in the field of SAS, the selection of the most appropriate strategy depends on the current situation of the system.
In this paper, we address the problem of self-aware optimization of adaptation planning strategies by designing a framework that includes situation detection, strategy selection, and parameter optimization of the selected strategies.
We apply rules and clustering techniques to identify the current situation, as well as Bayesian Optimization to tune the selected strategy's parameters. 
Further, we learn models of the system and its enviroment and reason on future decisions based on these models.
Finally, we apply the proposed framework on the platooning coordination case study and evaluate the performance of all components of the framework as well as the overall performance of the whole framework. 
In the future we plan to further enhance the components of the framework:
First, the coordination component processes the observations from the use case and triggers the other components. 
However, with increasing runtime of the framework, the amount of data collected from the use case increases. 
This leads to large data sets that do not necessarily contribute to good performance of the overall system, as the information may become outdated~\cite{Markovitch1988Role,Sukhov2020Prevention}.
Hence, it is useful to develop a strategy on how to discard or aggregate the increasing amount of data.
Further, the situation detection currently comprises of a rule-based and a clustering approach, but is not able to adapt the rule set with learned insights. 
Hence, a rule-learning mechanism could be applied to improve the rule base of the situation detection.
Currently, the strategy selection learns which strategy to choose based solely on all observations on the current situation. 
However, a global mechanism could provide benefits to the component by adjusting the order of strategies based on the performance of strategies previously experienced in all situations. 
This could reduce the trial-and-error phase for new situations and, thus, shorten the time to convergence.
The parameter optimization component currently provides the hypervolume metric and individual thresholds.
However, for other use cases, other techniques for multi-objective optimization could be useful, such as the concept of Pareto-optimality to provide the operator with a set of equally well performing configurations.
Further, approaches to reduce the search space for parameter tuning such as~\cite{pukhkaiev2018brise,Hutter2011Sequential} could speed up the component.
In general, we could apply forecasting techniques~\cite{ZuBaLeKrHeKoCu2019-ICAC-Autonomic} to anticipate future developments of the system and its environments to proactively plan adaptations.
In summary, we developed the framework using components, which allows for dynamic evolution of each component according to the individual requirements and best practices of the targeted use case.

\bibliographystyle{IEEEtran}
\bibliography{Bibliography.bib}

\begin{thebibliography}{10}
\providecommand{\url}[1]{#1}
\csname url@samestyle\endcsname
\providecommand{\newblock}{\relax}
\providecommand{\bibinfo}[2]{#2}
\providecommand{\BIBentrySTDinterwordspacing}{\spaceskip=0pt\relax}
\providecommand{\BIBentryALTinterwordstretchfactor}{4}
\providecommand{\BIBentryALTinterwordspacing}{\spaceskip=\fontdimen2\font plus
\BIBentryALTinterwordstretchfactor\fontdimen3\font minus
  \fontdimen4\font\relax}
\providecommand{\BIBforeignlanguage}[2]{{%
\expandafter\ifx\csname l@#1\endcsname\relax
\typeout{** WARNING: IEEEtran.bst: No hyphenation pattern has been}%
\typeout{** loaded for the language `#1'. Using the pattern for}%
\typeout{** the default language instead.}%
\else
\language=\csname l@#1\endcsname
\fi
#2}}
\providecommand{\BIBdecl}{\relax}
\BIBdecl

\bibitem{Cheng2009}
B.~H.~C. Cheng, R.~de~Lemos, H.~Giese, P.~Inverardi, and J.~Magee,
  \emph{{Software Engineering for Self-Adaptive Systems: A Research
  Roadmap}}.\hskip 1em plus 0.5em minus 0.4em\relax Springer Berlin Heidelberg,
  2009.

\bibitem{Krupitzer2015}
C.~Krupitzer, F.~M. Roth, S.~Vansyckel, G.~Schiele, and C.~Becker, ``{A survey
  on engineering approaches for self-adaptive systems},'' \emph{Pervasive and
  Mobile Computing}, vol.~17, pp. 184--206, 2015.

\bibitem{lesch2020toward}
V.~Lesch, ``{Toward a Framework for Self-Learning Adaptation Planning through
  Optimization},'' in \emph{Organic Computing: Doctoral Dissertation Colloquium
  2020}, S.~Tomforde and C.~Krupitzer, Eds.\hskip 1em plus 0.5em minus
  0.4em\relax Kassel University Press GmbH, jul 2020, pp. 17--31.

\bibitem{xie2014coping}
X.-F. Xie, S.~F. Smith, G.~J. Barlow, and T.-W. Chen, ``{Coping with real-world
  challenges in real-time urban traffic control},'' in \emph{Compendium of
  Papers of the 93rd Annual Meeting of the Transportation Research Board},
  2014, pp. 1--15.

\bibitem{robinson2010operating}
T.~Robinson, E.~Chan, and E.~Coelingh, ``{Operating Platoons On Public
  Motorways: An Introduction To The SARTRE Platooning Programme},'' in
  \emph{Proceedings of the 17th World Congress on Intelligent Transport
  Systems}, 2010.

\bibitem{alam2011fuel}
A.~Alam, ``{Fuel-efficient distributed control for heavy duty vehicle
  platooning},'' Ph.D. dissertation, KTH Royal Institute of Technology,
  Stockholm, 2011.

\bibitem{Sturm2020evaluation}
T.~Sturm, C.~Krupitzer, M.~Segata, and C.~Becker, ``{A Taxonomy of Optimization
  Factors for Platooning},'' \emph{IEEE Transactions on Intelligent
  Transportation Systems}, vol.~22, no.~10, pp. 6097--6114, 2021.

\bibitem{Wolpert1997}
D.~H. Wolpert and W.~G. Macready, ``{No free lunch theorems for
  optimization},'' \emph{IEEE Transactions on Evolutionary Computation},
  vol.~1, no.~1, pp. 67--82, 1997.

\bibitem{rice1976algorithm}
J.~R. Rice, ``{The Algorithm Selection Problem},'' in \emph{Advances in
  Computers}.\hskip 1em plus 0.5em minus 0.4em\relax Elsevier, 1976, vol.~15,
  pp. 65--118.

\bibitem{fredericks2019planning}
E.~M. Fredericks, I.~Gerostathopoulos, C.~Krupitzer, and T.~Vogel, ``{Planning
  as Optimization: Dynamically Discovering Optimal Configurations for Runtime
  Situations},'' in \emph{In Proceedings of the 13th International Conference
  on Self-Adaptive and Self-Organizing Systems}.\hskip 1em plus 0.5em minus
  0.4em\relax IEEE, 2019, pp. 1--10.

\bibitem{calinescu2020understanding}
R.~Calinescu, R.~Mirandola, D.~Perez-Palacin, and D.~Weyns, ``{Understanding
  Uncertainty in Self-adaptive Systems},'' in \emph{In Proceedings of IEEE
  International Conference on Autonomic Computing and Self-Organizing
  Systems}.\hskip 1em plus 0.5em minus 0.4em\relax IEEE, 2020, pp. 242--251.

\bibitem{endsley2017toward}
M.~R. Endsley, ``{Toward a Theory of Situation Awareness in Dynamic Systems},''
  in \emph{Human Factors: The Journal of Human Factors and Ergonomics
  Society}.\hskip 1em plus 0.5em minus 0.4em\relax Sage Journals, 2017,
  vol.~37, pp. 32--64.

\bibitem{liu2015situation}
W.~Liu, S.-W. Kim, S.~Pendleton, and M.~H. Ang, ``{Situation-aware decision
  making for autonomous driving on urban road using online POMDP},'' in
  \emph{2015 IEEE Intelligent Vehicles Symposium}.\hskip 1em plus 0.5em minus
  0.4em\relax IEEE, 2015, pp. 1126--1133.

\bibitem{rockl2007architecture}
M.~Rockl, P.~Robertson, K.~Frank, and T.~Strang, ``{An architecture for
  situation-aware driver assistance systems},'' in \emph{2007 IEEE 65th
  Vehicular Technology Conference-VTC2007-Spring}.\hskip 1em plus 0.5em minus
  0.4em\relax IEEE, 2007, pp. 2555--2559.

\bibitem{hardes2019dynamic}
T.~Hardes and C.~Sommer, ``{Dynamic Platoon Formation at Urban
  Intersections},'' in \emph{{Proceedings of the 44th IEEE Conference on Local
  Computer Networks}}, 2019, pp. 101--104.

\bibitem{porter2016losing}
B.~Porter and R.~Rodrigues~Filho, ``{Losing Control: The Case for Emergent
  Software Systems Using Autonomous Assembly, Perception, and Learning},'' in
  \emph{10th International Conference on Self-Adaptive and Self-Organizing
  Systems}.\hskip 1em plus 0.5em minus 0.4em\relax IEEE, 2016, pp. 40--49.

\bibitem{kang2020far}
S.~Kang, T.~Choi, and T.~P. Pavlic, ``{How far should I watch? Quantifying the
  effect of various observational capabilities on long-range situational
  awareness in multi-robot teams},'' in \emph{In Proceedings of the IEEE
  International Conference on Autonomic Computing and Self-Organizing
  Systems}.\hskip 1em plus 0.5em minus 0.4em\relax IEEE, 2020, pp. 146--152.

\bibitem{smith2009cross}
K.~A. Smith-Miles, ``{Cross-disciplinary perspectives on meta-learning for
  algorithm selection},'' \emph{ACM Computing Surveys}, vol.~41, no.~1, pp.
  1--25, 2009.

\bibitem{kerschke2019survey}
P.~Kerschke, H.~H. Hoos, F.~Neumann, and H.~Trautmann, ``{Automated Algorithm
  Selection: Survey and Perspectives},'' \emph{Evolutionary Computation},
  vol.~27, no.~1, pp. 3--45, 2019.

\bibitem{kerschke2019automated}
P.~Kerschke and H.~Trautmann, ``{Automated Algorithm Selection on Continuous
  Black-Box Problems by Combining Exploratory Landscape Analysis and Machine
  Learning},'' \emph{Evolutionary Computation}, vol.~27, no.~1, pp. 99--127, 03
  2019.

\bibitem{kotthoff2015improving}
L.~Kotthoff, P.~Kerschke, H.~Hoos, and H.~Trautmann, ``{Improving the State of
  the Art in Inexact TSP Solving Using Per-Instance Algorithm Selection},'' in
  \emph{Learning and Intelligent Optimization}, C.~Dhaenens, L.~Jourdan, and
  M.-E. Marmion, Eds.\hskip 1em plus 0.5em minus 0.4em\relax Cham: Springer
  International Publishing, 2015, pp. 202--217.

\bibitem{bischl2016aslib}
B.~Bischl, P.~Kerschke, L.~Kotthoff, M.~Lindauer, Y.~Malitsky, A.~Fréchette,
  H.~Hoos, F.~Hutter, K.~Leyton-Brown, K.~Tierney, and J.~Vanschoren, ``{ASlib:
  A benchmark library for algorithm selection},'' \emph{Artificial
  Intelligence}, vol. 237, pp. 41--58, 2016.

\bibitem{neumuller2012large}
C.~Neum{\"u}ller, A.~Scheibenpflug, S.~Wagner, A.~Beham, and M.~Affenzeller,
  ``{Large scale parameter meta-optimization of metaheuristic optimization
  algorithms with heuristiclab Hive},'' \emph{Actas del VIII Espa{\~n}ol sobre
  Metaheur{\'\i}sticas, Algoritmos Evolutivos y Bioinspirados}, 2012.

\bibitem{feurer2015initializing}
M.~Feurer, J.~T. Springenberg, and F.~Hutter, ``{Initializing Bayesian
  Hyperparameter Optimization via Meta-Learning},'' in \emph{Twenty-Ninth AAAI
  Conference on Artificial Intelligence}, 2015.

\bibitem{zhang2014empirical}
Y.~Zhang, M.~Harman, G.~Ochoa, G.~Ruhe, and S.~Brinkkemper, ``{An Empirical
  Study of Meta- and Hyper-Heuristic Search for Multi-Objective Release
  Planning},'' \emph{ACM Transactions on Software Engineering and Methodology},
  vol.~27, no.~03, 2018.

\bibitem{chis2013multi}
R.~Chis, M.~Vintan, and L.~Vintan, ``{Multi-objective DSE algorithms'
  evaluations on processor optimization},'' in \emph{In Proceedings of the 9th
  International Conference on Intelligent Computer Communication and
  Processing}.\hskip 1em plus 0.5em minus 0.4em\relax IEEE, 2013, pp. 27--33.

\bibitem{vinctan2015improving}
L.~Vin{\c{t}}an, R.~Chi{\c{s}}, M.~A. Ismail, and C.~Co{\c{t}}ofan{\u{a}},
  ``{Improving Computing Systems Automatic Multiobjective Optimization Through
  Meta-Optimization},'' \emph{IEEE Transactions on Computer-Aided Design of
  Integrated Circuits and Systems}, vol.~35, no.~7, pp. 1125--1129, 2015.

\bibitem{hardestowards}
T.~Hardes and C.~Sommer, ``{Towards Heterogeneous Communication Strategies for
  Urban Platooning at Intersections},'' in \emph{2019 IEEE Vehicular Networking
  Conference}.\hskip 1em plus 0.5em minus 0.4em\relax IEEE, 2019, pp. 1--8.

\bibitem{lewis2017towards}
P.~Lewis, K.~L. Bellman, C.~Landauer, L.~Esterle, K.~Glette, A.~Diaconescu, and
  H.~Giese, ``{Towards a framework for the levels and aspects of self-aware
  computing systems},'' in \emph{Self-Aware Computing Systems}.\hskip 1em plus
  0.5em minus 0.4em\relax Springer, 2017, pp. 51--85.

\bibitem{cox2005metacognition}
M.~T. Cox, ``{Metacognition in computation: A selected research review},''
  \emph{Artificial Intelligence}, vol. 169, pp. 104--141, 2005.

\bibitem{agarwal2009self}
A.~Agarwal, J.~Miller, J.~Eastep, D.~Wentziaff, and H.~Kasture, ``{Self-aware
  computing},'' Massachusetts Institute of Technology, Tech. Rep., 2009.

\bibitem{perrouin2012towards}
G.~Perrouin, B.~Morin, F.~Chauvel, F.~Fleurey, J.~Klein, Y.~Le~Traon,
  O.~Barais, and J.-M. J{\'e}z{\'e}quel, ``{Towards flexible evolution of
  Dynamically Adaptive Systems},'' in \emph{In Proceedings of the 34th
  International Conference on Software Engineering}.\hskip 1em plus 0.5em minus
  0.4em\relax IEEE, 2012, pp. 1353--1356.

\bibitem{gerostathopoulos2017strengthening}
I.~Gerostathopoulos, T.~Bures, P.~Hnetynka, A.~Hujecek, F.~Plasil, and
  D.~Skoda, ``{Strengthening Adaptation in Cyber-Physical Systems via
  Meta-Adaptation Strategies},'' \emph{ACM Transactions on Cyber-Physical
  Systems}, vol.~1, no.~3, pp. 1--25, 2017.

\bibitem{Kinneer2018uncertainty}
C.~Kinneer, Z.~Coker, J.~Wang, D.~Garlan, and C.~L. Goues, ``{Managing
  Uncertainty in Self-Adaptive Systems with Plan Reuse and Stochastic
  Search},'' in \emph{Proceedings of the 13th International Conference on
  Software Engineering for Adaptive and Self-Managing Systems}, 2018, pp.
  40--50.

\bibitem{sun2019reinbo}
X.~Sun, J.~Lin, and B.~Bischl, ``{ReinBo: Machine Learning pipeline search and
  configuration with Bayesian Optimization embedded Reinforcement Learning},''
  \emph{arXiv preprint arXiv:1904.05381}, 2019.

\bibitem{chai2019auto}
J.~Chai, J.~Chang, Y.~Zhao, and H.~Liu, ``{An Auto-ML Framework Based on GBDT
  for Lifelong Learning},'' \emph{arXiv preprint arXiv:1908.11033}, 2019.

\bibitem{thornton2013auto}
C.~Thornton, F.~Hutter, H.~H. Hoos, and K.~Leyton-Brown, ``{Auto-WEKA: Combined
  selection and hyperparameter optimization of classification algorithms},'' in
  \emph{Proceedings of the 19th ACM SIGKDD international conference on
  Knowledge discovery and data mining}, 2013, pp. 847--855.

\bibitem{li2017hyperband}
L.~Li, K.~Jamieson, G.~DeSalvo, A.~Rostamizadeh, and A.~Talwalkar,
  ``{Hyperband: A novel bandit-based approach to hyperparameter
  optimization},'' \emph{The Journal of Machine Learning Research}, vol.~18,
  no.~1, pp. 6765--6816, 2017.

\bibitem{lesch2020comparison}
V.~Lesch, C.~Krupitzer, K.~Stubenrauch, N.~Keil, C.~Becker, S.~Kounev, and
  M.~Segata, ``{A Comparison of Mechanisms for Compensating Negative Impacts of
  System Integration},'' \emph{Future Generation Computer Systems}, vol. 116,
  pp. 117--131, March 2021.

\bibitem{LeKrTo2019}
V.~Lesch, C.~Krupitzer, and S.~Tomforde, ``{Multi-objective Optimisation in
  Hybrid Collaborating Adaptive Systems},'' in \emph{Proceedings of the 7th
  edition in the Series on Autonomously Learning and Optimising Systems,
  co-located with 32nd GI/ITG ARCS 2019}.\hskip 1em plus 0.5em minus
  0.4em\relax Gesellschaft fuer Informatik (GI), may 2019.

\bibitem{camara2017self}
J.~C{\'a}mara, K.~L. Bellman, J.~O. Kephart, M.~Autili, N.~Bencomo,
  A.~Diaconescu, H.~Giese, S.~G{\"o}tz, P.~Inverardi, S.~Kounev \emph{et~al.},
  ``{Self-Aware Computing Systems: Related Concepts and Research Areas},'' in
  \emph{Self-Aware Computing Systems}.\hskip 1em plus 0.5em minus 0.4em\relax
  Springer, 2017, pp. 17--49.

\bibitem{kramer2007self}
J.~Kramer and J.~Magee, ``{Self-Managed Systems: an Architectural Challenge},''
  in \emph{In Proceedings of Future of Software Engineering}.\hskip 1em plus
  0.5em minus 0.4em\relax IEEE, 2007, pp. 259--268.

\bibitem{kounev2017notion}
S.~Kounev, P.~Lewis, K.~L. Bellman, N.~Bencomo, J.~Camara, A.~Diaconescu,
  L.~Esterle, K.~Geihs, H.~Giese, S.~G{\"o}tz \emph{et~al.}, ``{The Notion of
  Self-aware Computing},'' in \emph{Self-Aware Computing Systems}.\hskip 1em
  plus 0.5em minus 0.4em\relax Springer, 2017, pp. 3--16.

\bibitem{Kephart2003}
J.~O. Kephart and D.~M. Chess, ``{The Vision of Autonomic Computing},''
  \emph{IEEE Computer}, vol.~36, no.~1, pp. 41--50, 2003.

\bibitem{tomforde2011observation}
S.~Tomforde, H.~Prothmann, J.~Branke, J.~H{\"a}hner, M.~Mnif,
  C.~M{\"u}ller-Schloer, U.~Richter, and H.~Schmeck, ``{Observation and Control
  of Organic Systems},'' in \emph{Organic Computing--A Paradigm Shift for
  Complex Systems}.\hskip 1em plus 0.5em minus 0.4em\relax Springer, 2011, pp.
  325--338.

\bibitem{Wang2016}
S.~Wang, S.~Ali, T.~Yue, Y.~Li, and M.~Liaaen, ``A practical guide to select
  quality indicators for assessing pareto-based search algorithms in
  search-based software engineering,'' in \emph{Proceedings of the 38th
  International Conference on Software Engineering}, 2016, pp. 631--642.

\bibitem{nguyen2012computational}
S.~Nguyen, M.~Zhang, M.~Johnston, and K.~C. Tan, ``{A Computational Study of
  Representations in Genetic Programming to Evolve Dispatching Rules for the
  Job Shop Scheduling Problem},'' \emph{IEEE Transactions on Evolutionary
  Computation}, vol.~17, no.~5, pp. 621--639, 2012.

\bibitem{chand2018use}
S.~Chand, Q.~Huynh, H.~Singh, T.~Ray, and M.~Wagner, ``{On the Use of Genetic
  Programming to Evolve Priority Rules for Resource Constrained Project
  Scheduling Problems},'' \emph{Information Sciences}, vol. 432, pp. 146--163,
  2018.

\bibitem{ghandar2009computational}
A.~Ghandar, Z.~Michalewicz, M.~Schmidt, T.-D. To, and R.~Zurbrugg,
  ``{Computational Intelligence for Evolving Trading Rules},'' \emph{IEEE
  Transactions on Evolutionary Computation}, vol.~13, no.~1, pp. 71--86, 2009.

\bibitem{alelyani2013feature}
S.~Alelyani, J.~Tang, and H.~Liu, ``{Feature Selection for Clustering: A
  Review},'' \emph{{Data Clustering: Algorithms and Applications}}, vol.~29,
  no. 110-121, 2014.

\bibitem{tibshirani2001estimating}
R.~Tibshirani, G.~Walther, and T.~Hastie, ``{Estimating the number of clusters
  in a data set via the gap statistic},'' \emph{Journal of the Royal
  Statistical Society: Series B (Statistical Methodology)}, vol.~63, no.~2, pp.
  411--423, 2001.

\bibitem{elbowMethodBlogWebsite}
R.~Gove, ``{Using the elbow method to determine the optimal number of clusters
  for k-means clustering - bl.ocks.org},''
  \url{https://bl.ocks.org/rpgove/0060ff3b656618e9136b}, accessed: 2021-03-12.

\bibitem{RandomForests}
M.~Guillame-Bert, S.~Bruch, J.~Gordon, and J.~Pfeifer, ``{Introducing
  TensorFlow Decision Forests},''
  \url{https://blog.tensorflow.org/2021/05/introducing-tensorflow-decision-forests.html},
  Nov 2021, [Online; Accessed 2. Nov. 2021].

\bibitem{Lesch2021Towards}
V.~Lesch, T.~Noack, J.~Hefter, S.~Kounev, and C.~Krupitzer, ``{Towards
  Situation-Aware Meta-Optimization of Adaptation Planning Strategies},'' in
  \emph{Proceedings of the 2nd IEEE International Conference on Autonomic
  Computing and Self-Organizing Systems (ACSOS 2021)}.\hskip 1em plus 0.5em
  minus 0.4em\relax IEEE, 2021, best paper candidate.

\bibitem{Noorshams2015phd}
Q.~Noorshams, ``{Modeling and Prediction of I/O Performance in Virtualized
  Environments},'' Ph.D. dissertation, Karlsruhe Institute of Technology (KIT),
  Karlsruhe, Germany, February 2015.

\bibitem{a8befahrenWebsite}
dpa/lsw, ``{Verkehr - Stuttgart - Meistbefahrener Autobahnabschnitt:
  Unfallzahlen verdoppelt - Wirtschaft - SZ.de},''
  \url{https://www.sueddeutsche.de/wirtschaft/verkehr-stuttgart-meistbefahrener-autobahnabschnitt-unfallzahlen-verdoppelt-dpa.urn-newsml-dpa-com-20090101-170806-99-537657},
  last Accessed: 2021-10-06.

\bibitem{basttrafficstats}
``{bast (Bundesanstalt für Straßenwesen) - Automatische Zählstellen 2018},''
  \url{https://www.bast.de/BASt_2017/DE/Verkehrstechnik/Fachthemen/v2-verkehrszaehlung/Daten/2018_1/Jawe2018.html},
  last Accessed: 2021-11-12.

\bibitem{a8tempoLimitWebsite}
J.~Breithut, ``{A8 zwischen Stuttgart und Leonberg: Polizei stellt
  Autobahn-Blitzer wieder auf},''
  \url{https://www.stuttgarter-nachrichten.de/inhalt.a8-zwischen-stuttgart-und-leonberg-polizei-stellt-autobahn-blitzer-wieder-auf.631561fb-8f7f-4881-a4cc-74eac1f4a158.html},
  last Accessed: 2021-10-06.

\bibitem{rostockNahverkehrWebsite}
P.~T.~V. AG, ``{Regionaler Nahverkehrsplan Mittleres Mecklenburg/Rostock},''
  \url{https://www.planungsverband-rostock.de/wp-content/uploads/2018/07/NVP\%5F\%5Fbersicht.pdf},
  last Accessed: 2021-10-07.

\bibitem{Markovitch1988Role}
S.~Markovitch and P.~D. SCOTT, ``The role of forgetting in learning,'' in
  \emph{Machine Learning Proceedings 1988}, J.~Laird, Ed.\hskip 1em plus 0.5em
  minus 0.4em\relax San Francisco (CA): Morgan Kaufmann, 1988, pp. 459--465.

\bibitem{Sukhov2020Prevention}
S.~Sukhov, M.~Leontev, A.~Miheev, and K.~Sviatov, ``Prevention of catastrophic
  interference and imposing active forgetting with generative methods,''
  \emph{Neurocomputing}, vol. 400, pp. 73--85, 2020.

\bibitem{pukhkaiev2018brise}
D.~Pukhkaiev and S.~G{\"o}tz, ``{BRISE: energy-efficient benchmark
  reduction},'' in \emph{Proceedings of the 6th International Workshop on Green
  and Sustainable Software}, 2018, pp. 23--30.

\bibitem{Hutter2011Sequential}
F.~Hutter, H.~H. Hoos, and K.~Leyton-Brown, ``{Sequential Model-Based
  Optimization for General Algorithm Configuration},'' in \emph{Learning and
  Intelligent Optimization}, C.~A.~C. Coello, Ed.\hskip 1em plus 0.5em minus
  0.4em\relax Berlin, Heidelberg: Springer Berlin Heidelberg, 2011, pp.
  507--523.

\bibitem{ZuBaLeKrHeKoCu2019-ICAC-Autonomic}
M.~Z{\"{u}}fle, A.~Bauer, V.~Lesch, C.~Krupitzer, N.~Herbst, S.~Kounev, and
  V.~Curtef, ``{Autonomic Forecasting Method Selection: Examination and Ways
  Ahead},'' in \emph{Proceedings of the 16th {IEEE} International Conference on
  Autonomic Computing}.\hskip 1em plus 0.5em minus 0.4em\relax IEEE, June 2019.

\end{thebibliography}

\end{document}